\documentclass[]{iopart}
\usepackage{iopams}
\usepackage{graphicx}
\usepackage{bm}
\expandafter\let\csname equation*\endcsname\relax
\expandafter\let\csname endequation*\endcsname\relax
\usepackage{amsmath}
\usepackage{color,soul}

\newcommand{\vxi}{{\boldsymbol \xi} }
\newcommand{\vsigma}{{\boldsymbol \sigma} }

\newcommand{\vh}{\mathbf{h}}

\newcommand{\vS}{\mathbf{S}}
\newcommand{\vOm}{\mathbf{\Omega}}

\newcommand{\beq}{\begin{equation}}
\newcommand{\eeq}{\end{equation}}
\newcommand{\bes}{\begin{subequations}}
\newcommand{\ees}{\end{subequations}}
\newcommand{\bea}{\begin{eqnarray}}
\newcommand{\eea}{\end{eqnarray}}
\newcommand{\ba}{\begin{array}}
\newcommand{\ea}{\end{array}}
\newcommand{\beqn}{\begin{eqnarray*}}
\newcommand{\eeqn}{\end{eqnarray*}}

\newcommand{\f}[2]{\frac{#1}{#2}}

\newcommand{\om}{\omega}

\newcommand{\la}{\langle}
\newcommand{\ra}{\rangle}

\newcommand{\dg}{\dagger}

\def\nn{\nonumber}

\newlength{\sizeonefig}
\newlength{\sizetwofig}
\begin{document}

\title{The Theory of Spin Noise Spectroscopy: A Review} 

\author{Nikolai A. Sinitsyn$^1$ and Yuriy V. Pershin$^2$}

\address{$^1$Theoretical Division, Los Alamos National Laboratory, Los Alamos, NM 87545, USA}
\address{$^2$Department of Physics and Astronomy and Smart State Center for Experimental Nanoscale Physics, University of South Carolina, Columbia, South Carolina 29208, USA}
\ead{nsinitsyn@lanl.gov}
\ead{pershin@physics.sc.edu}

\vspace{10pt}
\begin{indented}
\item[] \today 
\end{indented}

\begin{abstract}
Direct measurements of spin fluctuations  are becoming the mainstream approach for studies of complex
condensed matter, molecular, nuclear, and atomic systems. This review covers recent progress in the field of
optical Spin Noise Spectroscopy (SNS) with an additional goal to establish an introduction into its theoretical foundations.
Various theoretical techniques that have been  recently used to interpret results of SNS measurements are explained alongside
with examples of their applications. 
\end{abstract}

\pacs{72.70.+m, 72.25.Rb, 78.67.Hc}
%
%
\submitto{\RPP}
%
%

\vspace{1cm}

\ioptwocol

\tableofcontents
\maketitle

\section{Introduction}

\begin{figure}[b]
\centerline{\includegraphics[width=0.75\columnwidth]{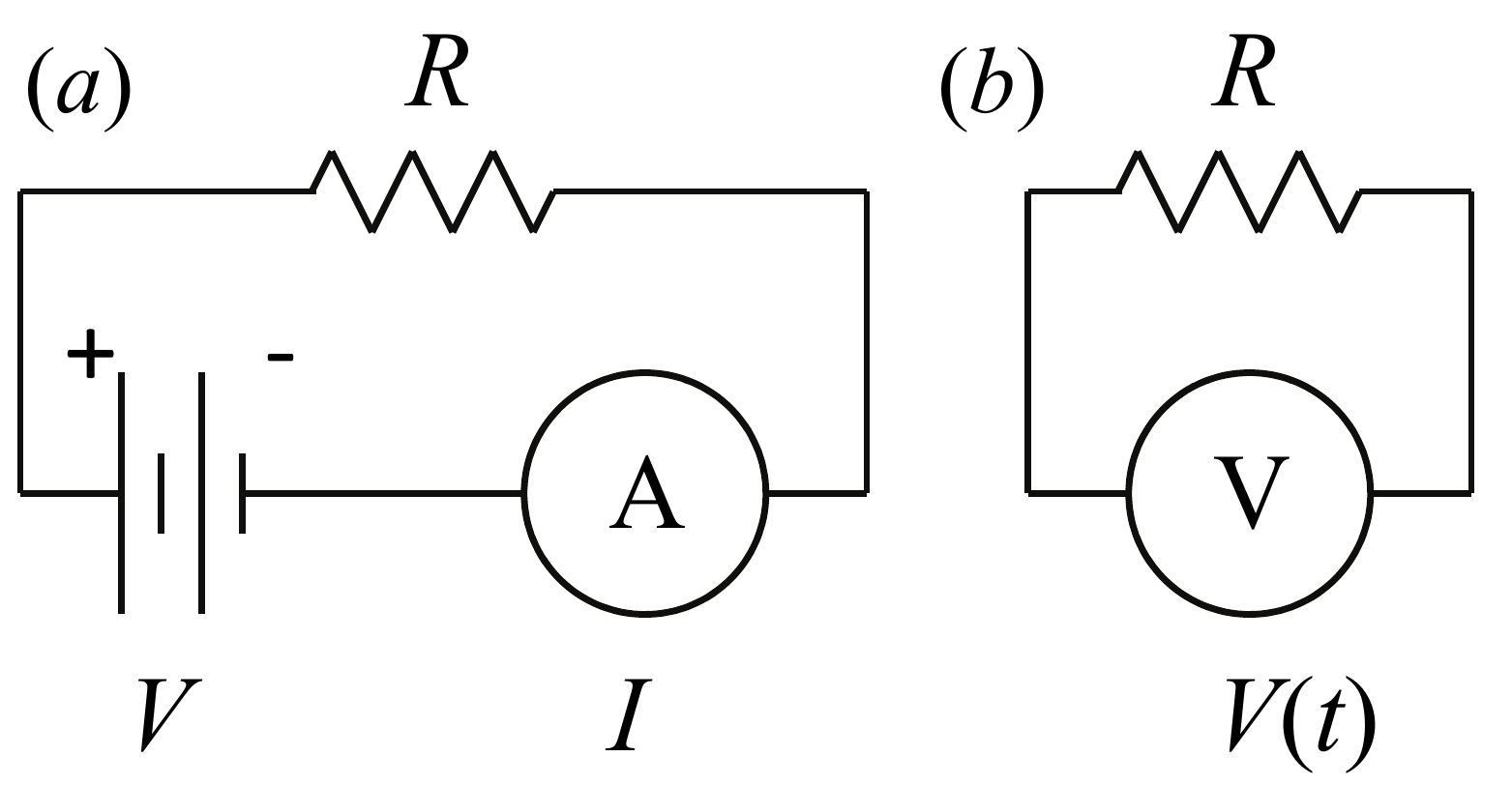}}
\caption{Circuits for (a) standard and (b) noise measurements of the resistance $R$. Here, A is the ammeter measuring the average current $I$ and V is the voltmeter measuring voltage fluctuations $V(t)$. The constant voltage $V$ across the battery in ($a$) is known.
}
\label{fig1_1}
\end{figure}

In experimental measurements, the noise  is not necessarily a
complication. From noise, one can often extract  valuable
information regarding its sources.
Consider, for example, the circuit in Fig.~\ref{fig1_1}(a) containing a resistor and battery.
The standard way of measuring the resistance $R$ is by application of a constant voltage $V$ to the resistor and the measurement of average electric current, $I$. The resistance is then given by
$R=V/I$. The noise-based measurement is a less invasive approach.
It was shown as early as in 1928 \cite{Johnson28a,Nyquist28a} that
the resistance  can be found from voltage
fluctuations at the thermodynamic equilibrium (the Johnson-Nyquist
noise \cite{Johnson28a,Nyquist28a})  without application of any
external voltage (see Fig.~\ref{fig1_1}(b) circuit). Indeed, the fluctuation dissipation theorem states
that the voltage variance (mean square) per Hertz of the bandwidth, is
given by  \cite{ft2,ft3}
\begin{equation}
 \langle V^2\rangle=4k_B TR,	
 \label{noise1}
 \end{equation}
 where $k_B$ is the Boltzmann constant, $T$ is temperature, and the
 average is taken over many repeated measurements. Equation~(\ref{noise1})
can be used to find the resistance $R$ from measured voltage fluctuations and
known temperature.

The above example shows that by studying  equilibrium
fluctuations we are able to obtain a single characteristic of a
circuit component. However, we can achieve much more. By measuring higher
correlators of noise (e.g., averages of higher powers of $V$) or by measuring the
noise in a nonequilibrium regime, one can avoid restrictions imposed by
the fluctuation-dissipation theorem and obtain the most detailed description
of electron interactions \cite{shot-noise-rev,nazarov-book,levitov-rev}.
For example, let's consider the case when the applied voltage induces
a current, $I$, with mean and variance, $\langle I \rangle$ and
var$(I) = \langle I^2 \rangle - \langle I \rangle^2$, respectively. The variance can be split
into the equilibrium and nonequilibrium parts, var($I$)=var($I$)$_{eq}$+var($I$)$_{neq}$.
The nonequilibrium part carries considerable
information about the physics of electrons. In particular, if
electrons combine into quasiparticles of charge $e^*$, then the ratio of the
nonequilibrium variance to mean is proportional to $e^*$, i.e. var($I$)$_{neq}$/
$\langle I  \rangle \sim e^*$ \cite{shot-noise-rev}.  This observation was used in practice
to obtain the long-sought experimental proof of the fact that the
quasiparticles in the fractional quantum Hall effect have fractional
charges \cite{4,5}.

Unfortunately, the charge current noise has been very hard to study
because of multiple parasitic effects induced, e.g., by contacts with leads, measurement devices, etc.
The proof of the fractional charge of anyons is one of only a few transformative experimental results in the field of electron counting statistics, while theoretical developments were much more promising. Among them are predictions of new types of quantum phase transitions in rare fluctuations \cite{fcs-qpt}, approaches to measure the  quantum entanglement entropy \cite{klich1} and qubit states \cite{8}, and results in the area of full counting statistics of strongly correlated electron states \cite{9,nazarov-rev}.

\begin{figure}[b]
\centerline{\includegraphics[width=0.75\columnwidth]{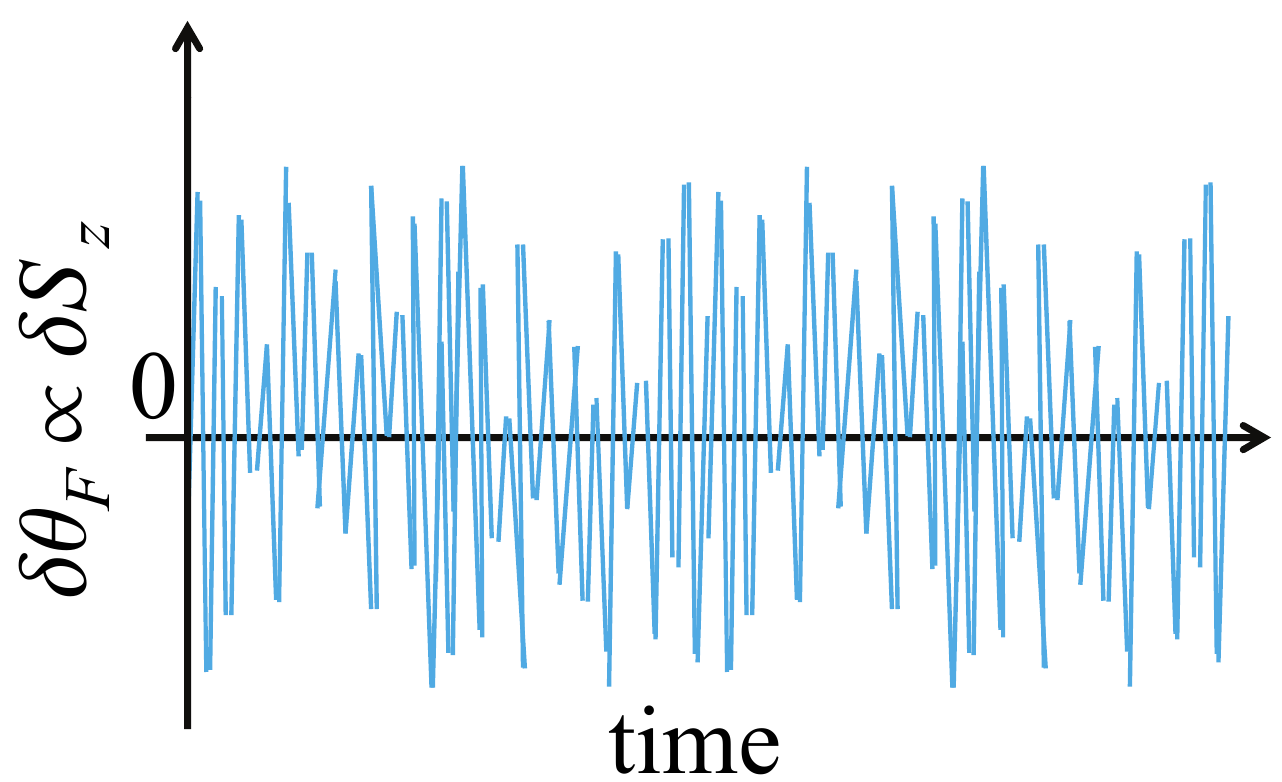}}
\caption{ Faraday rotation noise signal in real time.
}
\label{noise-fig1}
\end{figure}

Spin noise spectroscopy (SNS) is an alternative route
to obtain information about dynamics of various systems based on their {\it spin fluctuations}.
This method was introduced in
the pioneering work of Alexandrov and Zapasskii as early as in 1981 \cite{aleksandrov1981magnetic}.
However, it took over two decades of instrumentation progress including
development of real-time fast Fourier transformation (FFT) spectrum
analyzers and ultrafast digitizers before this technique had become
established as a powerful method to study spin dynamics in atomic
gases \cite{noise-nat}, conduction electrons \cite{Oestreich05}, and localized states in semiconductors \cite{Crooker10}. It should be emphasized that currently,
there is a large variety of available optical and non-optical spin noise measurement techniques targeting different applications. Some of the experimental methods, such as the resonance force microscopy  \cite{cantilever-science} and NV-center magnetometry \cite{NV2,NV3,nV1}, have been introduced  only recently and may experience considerable development in the near future.

\begin{figure}[t]
\centerline{\includegraphics[width=0.9\columnwidth]{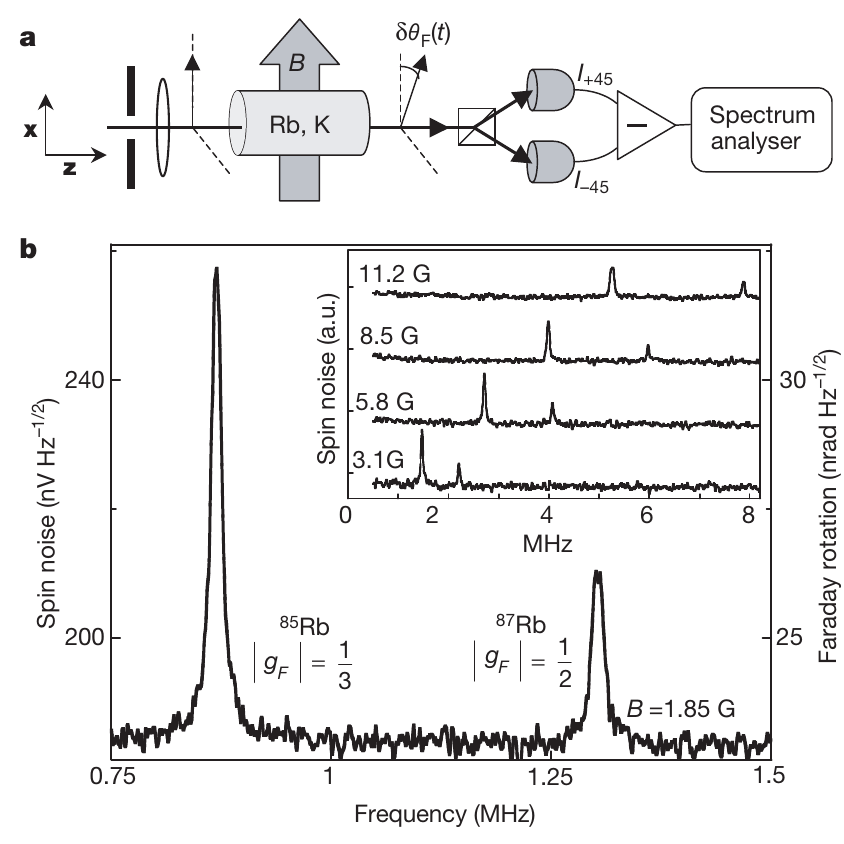}}
\caption{SNS of  Rb vapor \cite{noise-nat}: (a)
The experimental setup: a linearly polarized laser beam ($\sim$1-100 $\mu$m diameter) passes through a small volume of gas.
The outgoing beam experiences measurable Faraday rotation of polarization, $\delta \theta_F$,
proportional to the  instantaneous gas magnetization in the volume. (b) The noise power spectrum from Rb vapor showing two peaks from $^{85}$Rb and $^{87}$Rb at corresponding Larmor frequencies. For more details, see Ref. \cite{noise-nat}. Reprinted with permission from Macmillan
Publishers Ltd: Nature 431, pp. 49–52, 2004, Copyright \copyright (2004) \cite{noise-nat}.
}
\label{noise-fig}
\end{figure}

 In this review, we will focus  on the frequently used optical approach
based on the Faraday rotation spectroscopy, which has a relatively broad range of applications and has been the topic of many theoretical and experimental studies during the last decade.  To illustrate the main idea of this technique, let us consider an ensemble of spins in a small mesoscopic volume of an atomic gas or a semiconductor that contains $N$ atoms or electrons with spins in a constant magnetic field $B$. Statistical fluctuations of $N$ paramagnetic spins should generate spin fluctuations of the order of $\sqrt{N} $, even in a zero magnetic field at equilibrium \cite{noise-nat}. Optically, such fluctuations are measured  by detecting the Faraday rotation (Fig.~\ref{noise-fig}(a))  of the polarization axis of a linearly polarized beam ($\sim$1-100 $\mu$m diameter) passing through a sample
\footnote{Note that the relative fluctuations $\sim \sqrt{N}/N=1/\sqrt{N}$ increase with decreasing the beam diameter.}.
A typical experimental setup is shown in Fig.~\ref{noise-fig1}.

The polarization rotation angle is proportional to the local instantaneous magnetization of the illuminated region and its fluctuations can be traced with sub-nanosecond resolution
in time \cite{high-band,Starosielec08a,gaas2,Hubner13a},
thus providing considerable statistical data about local
magnetization dynamics. The spin noise power spectrum is found as FFT of the autocorrelation function $\la S_z(t) S_z(0) \ra$ (for an example, see Fig.~\ref{noise-fig}(b)). The position and shape of the spin noise
peaks are used to extract physical parameters of a particular system.
Importantly, the frequency of the measurement beam can be detuned from  system resonances  so that dissipative effects during measurements are minimized. Therefore, spin fluctuations can be studied  in spin systems at the thermodynamic equilibrium without inducing undesired extrinsic excitations.
Potentially, this technique can be applied to any material that demonstrates measurable Faraday or Kerr rotations induced by partial spin polarization. These include most known semiconductors, topological insulators, unconventional superconductors, as well as atomic gases.

The research on SNS is highly interdisciplinary with diverse applications in the areas of atomic gases, conduction electrons, spintronics, optoelectronics, semiconductor quantum dots, spin glasses, and  micromagnetics. By itself, SNS essentially differs from the traditional methods used in materials science.  Moreover, the very type of data that SNS deals with is unusual for
condensed matter physics. SNS works with {\it  stochastic signals} that
often have to be processed with a rate of gigabytes per second, sometimes during days or longer \cite{gaas4,instr-rev}.

 Several reviews have been already written about  SNS
with the focus on experimental developments \cite{alexandrov-rev, Zapasskii13b,OestreichPhysE,Oestreich-rise,Oestreich-rev3,glazov-rev}. The goal of the present review is different. Despite diverse applications, there are established theoretical methods and results that can be useful for theorists entering or working in this field.  We wrote our review  to provide an introduction to  SNS from the theoretical point of view  minimizing the discussion of instrumentation. Thus, our main attention is focused on the theoretical justification of the technique, theoretical methods to study spin noise, and the problems  that  will likely benefit from advances in the theory of SNS in the future.

Our review is organized as follows. In Sec.~2, we define various spin correlators and discuss their elementary properties. In Sec.~3, we review available experimental results from the theoretical perspective, illustrating theoretical problems and providing examples of elementary theoretical calculations. Moreover, in order to place the optical SNS into a broader context, subsection~3.5 contains a brief review of several alternative spin noise characterization techniques.
In Sec.~4, we provide quantum theory of spin correlators in the weak measurement framework. Section~5 contains an introduction to the theory of measurement based on the Faraday rotation effect. Section~6 is devoted to sum rules that are applicable practically to all spin systems studied by  SNS. Sections 7-9 review  theoretical  methods  that are frequently used in the research on SNS. Section~10 reviews possible future research directions and extensions of SNS.

Different sections of this review are written at different levels of complexity.
Sections~2-3 set the framework of problems for the rest of the review and provide introduction to the  SNS at a nonspecialist level. Sections~4-7 contain introductions to different  theoretical aspects of the optical SNS with many worked out examples. Although there are occasional cross-references between  these sections, they  are independent of each other and can be read in an arbitrary order. Subsections~8.1 and 8.2 are written in the same style but they should be more accessible to readers  with prior experience in spin drift-diffusion equation and Kubo formula.
In sections~8.3-10, we return to the ``review" mode in order to cover a large variety of other, more specified, theoretical  ideas without attempting to provide a detailed introduction.


\section{Spin Correlators: Definition and Basic Properties}

\subsection{2nd Order Correlator and Noise Power Spectrum}

The simplest characteristic that describes correlations of a signal in SNS is the spin-spin correlator
\begin{equation}
g_2(t_2,t_1)\equiv \la S_z(t_2) S_z(t_1) \ra,
\label{c2-220}
\end{equation}
where $S_z(t)$ is the time-dependent spin polarization in the observation region and $z$ is the measurement axis.
Generally, this correlator is a function of two time arguments, however, if measurements are performed in the steady state then (due to the translation invariance in time) this correlator depends only on the time difference $t$:
\begin{equation}
g_2(t)=\la S_z(t) S_z(0) \ra.
\label{c2-22}
\end{equation}

In many (but not all \cite{real-time-SNS}) experimental setups, it is much more convenient to process signals in the frequency domain.
Let's consider first the straightforward way to calculate the correlator (\ref{c2-22}) from an experimentally measured ``trajectory" $S_z(t)$. In order to obtain (\ref{c2-22}), the total measurement time interval should be divided into relatively long time intervals $T_m \gg \tau_s$, where $\tau_s$ is the characteristic spin relaxation time. Each time interval of size $T_m$ also consists of $N_m \gg 1$ elementary intervals of size $dt \ll \tau_s$. Naively, one may think that in order to obtain (\ref{c2-22}), one should take all pairs of time moments, $t_i$ and $t_{i+j}$, within the same time interval of size $T_m$, and calculate all possible products of measurement results $S_z(t_{i+j})S_z(t_i)$. Then, the average over $i$
provides $\la S_z(t_j) S_z(0) \ra$. Additional averaging over the
 large time intervals of size $T_m$ further increases the precision of calculations.
This straightforward process, however, is not optimal as
$N_m$ measurement points in the interval $T_m$ require the
calculation of  $\sim N_m^2$ products $S_z(t_{j+i})S_z(t_{i})$ in order to estimate correlator (\ref{c2-22}) for all different $t$.
For measurements performed with, e.g.,  a nanosecond resolution,
 continuous data processing imposes very strict
hardware requirements that are difficult to satisfy.


The widely used approach is based on the Fourier transform of the recorded trajectory during the measurement interval $T_m$
\begin{equation}
a(\omega) =\frac{1}{\sqrt{T_m}} \int\limits_0^{T_m} \textnormal{d}t e^{i\omega t} S_z(t).
\label{aom}
\end{equation}
A typical FFT algorithm (e.g., the Cooley-Tukey FFT algorithm) requires $t\sim N_m {\log}(N_m) \ll N_m^2$ steps to perform. Note that since the measured trajectory $S_z(t)$ is real, $a(-\omega)=a^*(\omega)$. For a system in the steady state, $\la a(\omega)\ra=0$ at
 $\omega \ne 0$, where the average is considered over repeated time intervals of size $T_m$ under  identical experimental conditions~\footnote{Generally, $S_z(t)$ may have a constant component, so that $\la S_z(t)\ra=0$ is not required.}.

 The most accessible physically interesting characteristic in the frequency domain is the {\it  noise power spectrum},  defined as
\begin{equation}
C_2(\omega) = \langle |a(\omega)|^2 \rangle.
\label{noiseP1wk}
\end{equation}
One can easily prove that, due to the translation invariance in time,
$$
\langle a(\omega) a(\omega') \rangle=0, \quad {\rm for} \,\, \omega \ne -\omega'.
$$
It is crucial for fast signal processing that $C_2(\omega)$ depends only on a single argument, so that its calculation for all independent values of $\omega$ scales as $\sim N_m$ for a discretized signal, as desired, so the slowest step in determining $C_2(\omega)$ is application of the FFT algorithm.

Another important property of the noise power spectrum is the additivity, namely, the contributions from independent noise sources add to each other. Indeed, suppose that the measured Fourier transform of the signal is given by $a(\omega) = a_{\rm ph}(\omega) +\zeta(\omega)$, where $a_{\rm ph} (\omega)$ is  the physical spin noise signal and $\zeta(\omega)$ is an uncorrelated from $a_{\rm ph} (\omega)$ background noise that originates, e.g., from an amplifier or represents the photon shot noise. In such a case $\langle a_{\rm ph}(\omega) \zeta(\omega) \rangle = 0$, so
\beq
C_2(\omega) = \la  | a_{\rm ph} (\omega)|^2 \ra +  \la  | \zeta (\omega)|^2 \ra.
\label{backgr}
\eeq
It is possible  then to get rid of the background part by measuring it
separately. The most popular way to do this is to measure the spectrum in a large external magnetic field applied in the transverse
direction to the measurement beam. This field induces fast spin
precessions that  move the physical part of the noise power to very
high frequencies. By measuring the background level in such a field, the
physical noise is obtained by subtracting $\la  | \zeta (\omega)|^2
\ra$ from (\ref{backgr}).

The noise power spectrum is also often easier to interpret than the spin-spin
correlator in  real time. However, if correlators are needed in the
time domain, there is the following important theorem that relates the
noise power spectrum to the time correlator $g_2(t)$:

\underline{\it Wiener-Khinchine theorem} \cite{kogan-book}: In the steady state, the noise power spectrum and the spin-spin correlator are related by the Fourier transform:
\begin{equation}
C_2(\omega)=2 \int\limits_0^{\infty} \textnormal{d}t \cos(\omega t) \la S_z(t) S_z(0) \ra.
\label{c2-1}
\end{equation}
This theorem requires several conditions, such as the requirement that the measurement time is much longer than the spin relaxation time ($T_m \gg \tau_s$), which are discussed in more detail, e.g., in \cite{kogan-book,pecselin-book}.
Apart from these details, the proof is straightforward: Substituting
the definition of the Fourier transformed trajectory (\ref{aom}) into (\ref{noiseP1wk}), and using the property $g_2(t_1,t_2) = g_2(t_1-t_2)$ and the fact that, since $S_z(t)$ is real, $a^*(\omega)=a(-\omega)$,  we  find

\begin{eqnarray}
C_2(\omega) = \frac{1}{T_m} \lim \limits_{T_m \rightarrow \infty} \int\limits_{-T_m/2}^{T_m/2} \textnormal{d}t_1 \int\limits_{-T_m/2}^{T_m/2} \textnormal{d}t_2 \, e^{i\omega (t_1-t_2)} \nonumber \\ \;\;\;\;\;\;\;\;\;\;  \times \la S_z(t_2) S_z(t_1) \ra =  \int\limits_{-\infty}^{\infty } \textnormal{d}t \, e^{i\omega t} \la S_z(t) S_z(0) \ra. \label{proof1}
\end{eqnarray}
Taking into account that $S_z(t)$ is the real valued signal, i.e. it is not an operator, we can use the property $ \la S_z(-t) S_z(0) \ra =  \la S_z(t) S_z(0) \ra$, which demonstrates that the last integral in (\ref{proof1}) coincides with (\ref{c2-1}). This proves the theorem.

The optical SNS is not restricted to
measurements of a single spin component or to fluctuations in
homogeneous spin systems. In such situations, one can consider
cross-correlations of spin densities with different indexes, e.g.,
\begin{equation}
C_{AB}(\om)=\int\limits_{-\infty}^{\infty}\textnormal{d}t\, e^{i\omega t}  \la S_A(t)S_B(0) \ra,
\label{cross-corr}
\end{equation}
where A and B are indexes that may correspond to different atomic
species in an interacting gas mixture or  different spin
projections probed by noncollinear beams. Generally, the cross-correlator $C_{AB}(\om)$ is not
positive definite. Moreover, it can have both real and imaginary
components.

An example of the experimentally measured noise power spectrum is shown, e.g., in Fig.~\ref{noise-fig}(b). Typically, the spectrum consists of one or several peaks. Two types of peaks are most frequently encountered: the Lorentzian shaped peaks
\beq
C_{2}^L(\omega) \sim \frac{1}{(\omega-\omega_L)^2+\Gamma^2},
\label{lor-c2-1}
\eeq
and Gaussian shaped peaks
\beq
C_{2}^G(\omega) \sim e^{-\frac{\omega^2}{2\Gamma^2}},
\label{gauss-c2-1}
\eeq
where $\Gamma$ is a characteristic spin relaxation rate.
Performing the inverse Fourier transform, one can verify that, in the time domain, the Lorentzian shape corresponds to the exponentially damped spin precession
\beq
g_{2}^L(t) \sim e^{-\Gamma |t|} \cos(\omega_L t).
\label{exprel1}
\eeq
The Fourier transform of the Gaussian function remains the Gaussian one:
\beq
g_{2}^G(t) \sim e^{-\Gamma^2 t^2/2}.
\label{gauss-rel1}
\eeq
The exponential relaxation  (\ref{exprel1}) is usually associated with processes involving numerous fast mutually-uncorrelated microscopic interactions, such as atomic collisions, that contribute to the observed relaxation.
The relaxation according to the Gaussian law (\ref{gauss-rel1}) usually indicates the presence of dephasing processes, as in the case of an ensemble of similar systems with a Gaussian distribution of time-independent parameters.

\subsection{Higher Order Correlators}
A spin system's $n$-th order time-correlator is the average of a product of
$n$ results of the spin polarization measurements  taken at
different moments of time. The information content of 2nd order
correlators is intrinsically limited. Actually, this is not surprising as the noise power spectrum  tells us only
the weight of each frequency in the dynamics of spin
polarization without providing any information regarding correlations among different frequencies.
The complete information about an interacting spin system is contained in the full
set of all-order correlators \cite{2Dspectrum-mukamel}. Hence, the {\it higher order} SNS
is very interesting direction for future research.

 Higher order  correlators depend on more than one frequency, therefore, they are usually represented by multidimensional data, e.g. in the form of 2D or 3D density plots, that
contain additional and, possibly, considerably larger amount of
information about the system than the noise power spectrum  \cite{c4-1}. This information can be particularly useful when the studied spin
system experiences fluctuations at different time scales, such as in the case of a
qubit interacting with a slowly changing configuration of nuclear spins and simultaneously fast fluctuating spin-orbit fields due to interactions with phonons.
In other words,  higher order correlators are sensitive to the homogeneous broadening even in a strongly inhomogeneously broadened system,
i.e. they can be used to separate the effect of a static disorder from the intrinsic spin dynamics.

Another interesting property of higher order correlators is their intrinsic ``quantumness": Quantum measurements at intermediate time moments generally break the unitarity of evolution during the observed time interval. This fact may strongly influence higher order correlators, as it was demonstrated in a recent experiment \cite{alex-c3}, which extracted the quantum life time from an inhomogeneously broadened spectrum of a quantum dot spin qubit by using quantum properties of  the 3rd order correlator of spin projection operators.


To define higher order cumulants in the frequency domain,  let us introduce the normalized spin polarization
\begin{equation}
\delta S_z(t) = S_z(t)-\la S_z\ra,
\label{deltas}
\end{equation}
and consider its Fourier transform $a(\omega)$.  Similarly to the 2nd order correlator, in the steady state, only the products of $a(\omega_i)$ with $\sum_i \omega_i=0$ remain non-zero after the averaging. Hence, the next nearest nontrivial correlator of $a(\omega)$ is the 3rd order one  given by
\begin{equation}
C_3(\omega_1,\omega_2) = \langle a(\omega_1) a(\omega_2) a^* (\omega_1+\omega_2) \rangle.
\label{c3-1}
\end{equation}
A specific property of $C_3$ is that it is zero in a system with
time-reversal symmetry because spin variables are odd under the time reversal. Note also that the 3rd order correlator (\ref{c3-1}) is generally complex-valued.

 In the steady state, the 4th order cumulant is generally a
 function of three independent frequencies. Its definition depends on
 whether the sum of any two of these frequencies is zero or not. If the latter
 is the case, the 4th cumulant in the frequency domain can be defined as
 \begin{equation}
 C_4 (\omega_1,\omega_2,\omega_3) = \la   a(\omega_1) a(\omega_2) a(\omega_3) a^*(\omega_1+\omega_2+\omega_3) \rangle .
 \label{c4-1}
 \end{equation}
In the case of $ \quad \omega_1 =-\omega_3 \ne \omega_2$, the 4th order cumulant is defined as
\begin{equation}
 C_4 (\omega_1,\omega_2) \equiv \langle |a(\omega_1)|^2 |a(\omega_2)^2| \rangle -  \langle |a(\omega_1)|^2  \rangle \langle |a(\omega_2)^2| \rangle.
 \label{c4-11}
 \end{equation}
 Note that (\ref{c4-11}) cannot be obtained as a special case of (\ref{c4-1}).  The choice of the form of cumulants at equal values of some frequency parameters  is dictated by the requirement that such cumulants should be zero for Gaussian fluctuations of $a(\omega)$, so that the higher order cumulants do not duplicate the information that can be obtained from the lower ones.
 The correlators that depend on two frequencies, such as the 3rd order correlator (\ref{c3-1}) or the correlator (\ref{c4-11}), are often called the {\it bispectra}. A bispectrum  indicates how spin noise components at two different frequencies ``talk" to each other. For example, if $C_4(\omega_1,\omega_2)$ is negative, one can conclude the presence of anti-correlations, i.e. the observation of a strong signal at one frequency means that another frequency is likely suppressed, etc. An illustrative example of the experimental measurement of the correlator (\ref{c4-11}) can be found in \cite{fuxiang-c4}. Moreover, for $\omega_1=\omega_2=-\omega_3 \equiv \omega$,  the 4th cumulant is given by:
\begin{equation}
 C_4 (\omega) \equiv  \langle | a(\omega)|^4  \rangle -  2\langle |a(\omega)|^2  \rangle^2.
\label{c4-2}
 \end{equation}

Finally, there is a generalization of the correlator (\ref{c4-11}) called the ``noise
of susceptibility'' \cite{susceptibility-noise,forth-corr-maj}. In the correlator
(\ref{c4-2}), the product of 4th order spin variables is averaged over
repeated time intervals of duration $T_m$.  Instead, one can
measure the noise powers, $|a(\omega_1)|^2$ and $|a_T(\omega_2)|^2$,  of {\it different} time intervals  separated
by a given duration $T$, e.g., $(t,t+T_m)$ and $(t+T,t+T+T_m)$. One can  then  consider their
product averaged over all such pairs of intervals with different
$t$, separated by time $T$:
\begin{eqnarray}
\chi^{(4)} (\omega_1,\omega_2|T) = \nonumber \\
\quad =\la |a(\omega_1)|^2
|a_{T}(\omega_2)|^2\ra -\la |a(\omega_1)|^2 \ra \la |a_T(\omega_2)|^2 \ra.
\label{chi4}
\end{eqnarray}

\section{Survey of Experimental Results and Systems}
\label{exp}



\subsection{Hot Atomic Vapors}

Hot atomic vapors are usually evaporated gases of atoms, such as K, Rb, Cs, above the room temperature (e.g., at 400K).
Atoms in a hot vapor are relatively energetic, so that
the average occupancy of states in these vapors is small ($\ll 1$).
In this regard, hot atomic vapors are different from ultra-cold gases, in which atoms tend
to occupy the lowest energy levels.

Hot atomic vapors have broad applications in magnetometry and isotope
separation techniques \cite{magnetometry-book,isotope-book}. As spin
fluctuations limit the precision of these applications,
future advances in these fields will likely depend on our
understanding of  spin noise  and our ability to control it. In fact, strategies to suppress unwanted spin fluctuations
by a feedback control have been recently demonstrated \cite{atomic1,atomic2}.
We also mention that a random number generator based on the atomic spin noise was recently proposed  \cite{random-number-gener}.

\begin{figure}
\centerline{\includegraphics[width=0.9\columnwidth]{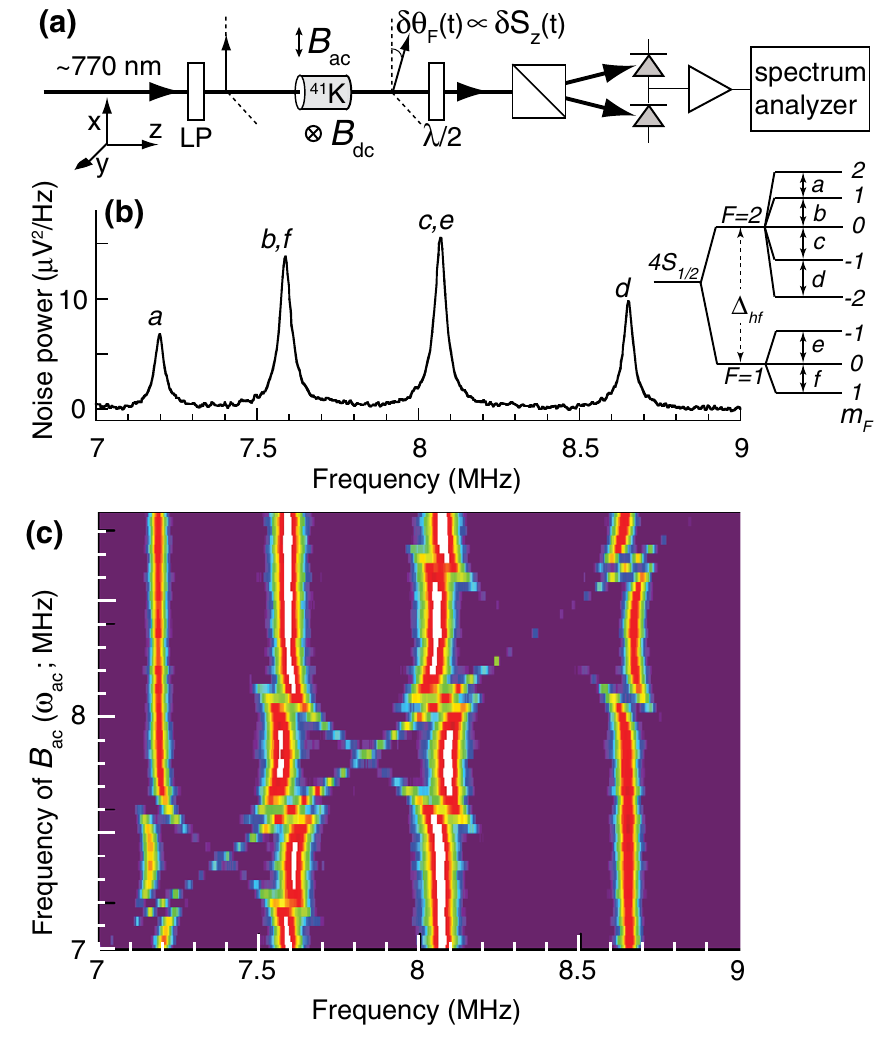}}
\caption{Spin noise spectroscopy of $^{41}$K vapour subjected to a dc and weak ac
magnetic fields. (a) The measurement setup. (b)  The equilibrium noise power
    spectrum of $^{41}$K atomic vapor ($B_{ac}=0$). (c) The density plot of the noise power
  spectrum vs frequency of an ac-magnetic field applied in the transverse to
  the measurement beam direction. Simultaneous response of more than
  one peak to such external perturbation reveals cross-correlations
  between different spin resonances.
  Reprinted figure with permission from P. Glasenapp, et al.,
  Physical Review Letters 113, p. 156601, 2014 \cite{crooker-noneq1}.
  Copyright \copyright (2014) by the American Physical Society.}
\label{ac-SNS}
\end{figure}

\subsubsection{Equilibrium SNS.}  The nowadays interest in spontaneous spin fluctuations at the thermodynamic equilibrium originates from
the pioneering  experiments on Rb and K atomic vapors \cite{noise-nat}. These experiments clearly
resolved spin noise power peaks (Fig.~\ref{noise-fig}(b)) and explored the strategies for
digital signal processing. The noise power peaks were associated with resonances of
hyperfine coupled electronic spins.  This work has clearly demonstrated that spin interactions can be studied
at the thermodynamics equilibrium without externally exciting a system.

Consider the atomic vapor of $^{41}$K that was investigated by SNS in Refs.~\cite{crooker-06pra2,crooker-06pra}.
Figure~\ref{ac-SNS} shows the experimental setup and the equilibrium noise power spectrum of this vapor measured in the applied magnetic field.
The spectrum consists of four Lorentzian peaks centered at different frequencies.

\begin{figure}
\begin{center}
\includegraphics[width=0.65\columnwidth]{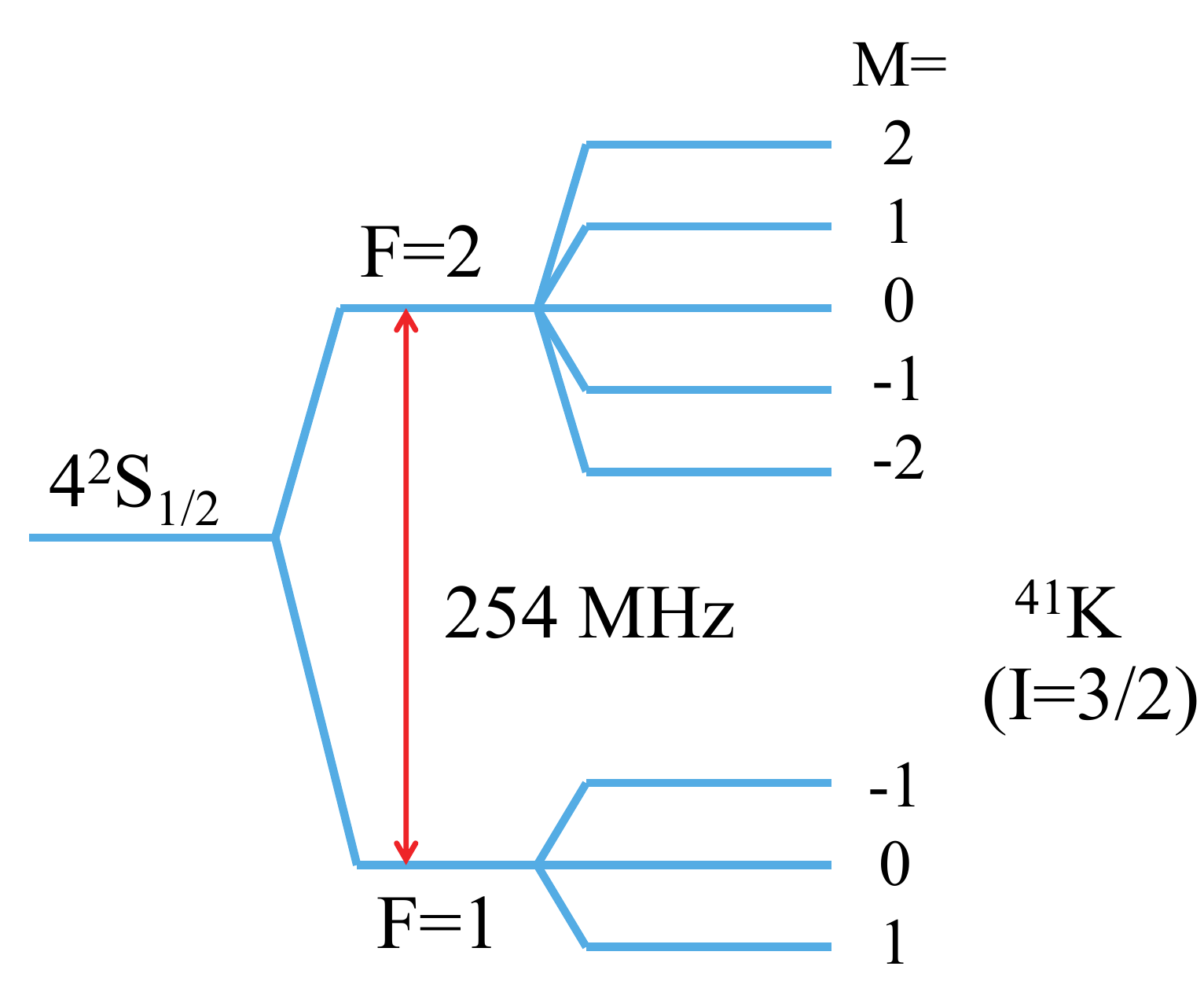}
\end{center}
\caption{4$^2$S$_{1/2}$ level diagram for $^{41}K$ atom in a constant magnetic field.}
\label{ls}
\end{figure}

In $^{41}$K atoms, the uncompensated electronic spin with $S=1/2$ couples relatively strongly with a nuclear spin of $I=3/2$. The eigenstates of the exchange Hamiltonian $\hat{H}_0=\hat{\bf S}\cdot \hat{\bf I}$ split into two multiplets with the total angular momentum $F=1$ and $F=2$. The hyperfine splitting between the two multiplets is about 254 MHz, which is much higher than the frequency range of the spectrum shown in Fig.~\ref{ac-SNS}(b). The external magnetic field ${\bf B}=B\hat{y}$  modifies the Hamiltonian: $\hat{H}\approx \hat{H}_0 + g{\bf B}\cdot \hat{\bf S} $. Figure~\ref{ls} shows that the magnetic field ${\bf B}$ splits each multiplet into energy levels characterized by the projection of the total angular momentum $M$ on the field axis. Different pairs of nearest levels generally
experience a different size of splitting. We also note that the temperature of the atomic gas $T\sim 400K$ is orders of magnitude higher than the splitting between any pair of energy levels in Fig.~\ref{ls}. Therefore, one can consider that these levels are occupied with equal probabilities.

Let us assume that the measurement axis is transverse to the magnetic field direction as shown in Fig.~\ref{ac-SNS}(a). The optical beam couples to the electron spin  operator $\hat{S}_z$ that has nonzero matrix elements only between the pairs of states with $\Delta M =\pm  1$, where $M$ is the angular momentum projection on the magnetic field axis. The spin-spin correlator in real time can then be estimated as
\bea
\nn \la S_z(t)S_z(0) \ra &\sim &\sum_{i}  \la E_i| e^{i \hat{H}t}  \hat{S}_z  e^{-i \hat{H}t} \hat{S}_z |E_i\ra = \\
 &=&\sum_{i; j=i\pm 1} e^{i(E_j-E_i)t -\gamma_{ij} t  }|\la E_i|\hat{S}_z|E_j\ra|^2,  \label{SNPS2}
\eea
where $E_j$ are the energies of the levels and index $i$ runs through all states. Note that in Eq.~(\ref{SNPS2}), for each resonance with the energy difference $E_j-E_i$, we introduced a phenomenological relaxation rate $\gamma_{ij}$.

Equation~(\ref{SNPS2}) shows that each pair of adjacent energy levels within each multiplet in Fig.~\ref{ls} contributes to the correlator with a damped oscillatory term $\sim e^{-\gamma_{ij}t} \cos([E_i-E_j]t)$.
After taking the Fourier transform, each such a term produces a Lorentzian peak in the noise power spectrum centered at $\omega = |E_i-E_j|$ (for $\omega>0$) with the width determined by the relaxation rate $\gamma_{ij}$.
Hence, the positions of these peaks can be used to find characteristic splittings between the energy levels, while their widths can be used to extract the relaxation rates. Moreover, Eq.~(\ref{SNPS2}) shows that the amplitudes of the peaks contain information about the matrix elements $\la E_i|\hat{S}_z|E_j\ra$.

Finally, we note that energy splittings between the levels with $M=0,1$ for $F=1$ and $F=2$, as well as between the levels with $M=-1,0$ for $F=1$ and $F=2$, are degenerate. This explains the only four peaks observed in Fig.~\ref{ac-SNS}(b) instead of six peaks as Fig.~\ref{ls} suggests.
Consequently, each of the two central peaks in Fig.~\ref{ac-SNS}(b) is the result of the overlap of two peaks from different resonances. This partly explains the relatively high amplitudes of the central peaks.

Relative heights of the peaks and
their behavior in external fields in relation to the matrix
elements of the measured electronic spin operator at different moments of time
were discussed in \cite{crooker-06pra2,crooker-06pra}. In
\cite{Katsoprinakis07}, an SNS-based approach to high precision relaxation time
measurements was developed.
In \cite{PhysRevA.84.043851}, the effect of the probe beam frequency on the spin
noise spectrum was explored. Peculiarities of near the resonance response were
attributed to collective effects.  Atomic gases were also used
as a simple testbed to improve SNS instrumentation
\cite{PhysRevLett.104.013601} and to demonstrate extensions of
this technique \cite{fuxiang-c4,crooker-13prl}.

Experiments with hot atomic gases are relatively simple. They do not require  cryogenic
equipment. Statistical filtering of physical spin correlators from the background noise in hot atomic gases,
such as $^{41}$K, can be achieved within few seconds. Moreover, the
electronic quantum states of different atomic isotopes have  been extensively studied by atomic physicists
throughout the 20th century and can be considered very well
understood. Due to this simplicity,  a methodology to teach the basics
of SNS in undergraduate laboratories was proposed in \cite{education}.
Hot atomic vapors are also
ideal for testing new measurement approaches and producing artificially correlated spin systems. Below, we discuss a few such
examples.

\subsubsection{Two-Color Spectroscopy.}
Electron spins cannot be generally considered identical even in the
same observation region. They can be localized near different
impurities, inside different quantum dots, etc. There are numerous
natural and engineered systems in which interactions between ``spins
of different kind" are important. Examples include the Kondo-lattices in correlated-electron materials \cite{heavy-fermions1, heavy-fermions2}, decoherence of solid-state spin qubits by a nuclear spin bath \cite{nuclear-bath1, nuclear-bath2, nuclear-bath3, Li12}, ferromagnetism in diluted magnetic semiconductors \cite{DMS1, DMS2}, and spin-exchange pumping of noble gas nuclei for medical imaging \cite{exchange-pumping}.


\begin{figure}
\centerline{\includegraphics[width=0.99\columnwidth]{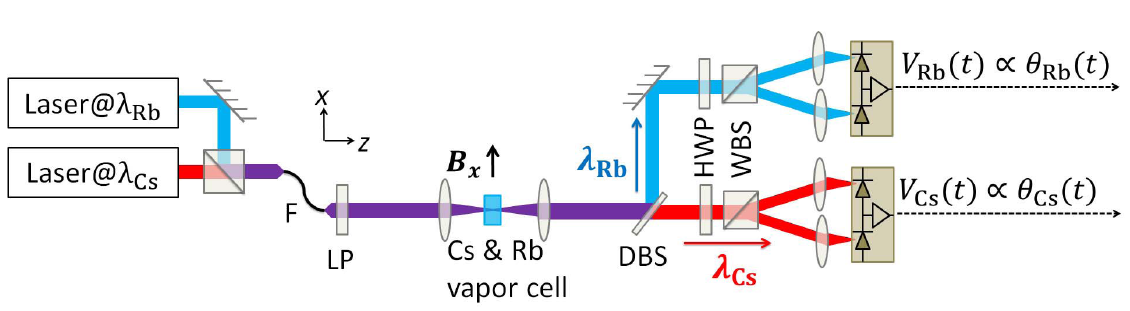}}
\caption{Two probe beams measure fluctuations of spins of
  different atomic species. Cross-correlations between
spin fluctuations of different spin species provide information
regarding interactions between these different types of spins.
  Reprinted with permission from Macmillan Publishers Ltd:
  Scientific Reports 5, article number: 9573, 2015, Copyright \copyright (2015) \cite{two-color-2}.}
\label{two-color}
\end{figure}


In \cite{two-color-2},  cross-correlations between spins of different kind were effectively
characterized by multi-probe SNS. The idea of this experiment is shown schematically in
Fig.~\ref{two-color}, wherein two atomic vapors, Cs and Rb,   interact
by the direct spin exchange at the thermal equilibrium. In this experiment, the spin fluctuations
from Cs and Rb are independently detected, and signatures of
spin interactions are observed in the cross-correlator (\ref{cross-corr}) of these two spin
noise signals. The authors of \cite{two-color-2} have demonstrated
that the spin relaxation and spin exchange rates can be measured
separately without ever perturbing the gas mixture from
the thermodynamic equilibrium.

\subsubsection{Beyond Equilibrium and Linear Response.}
One of the central results in statistical physics -- the fluctuation-dissipation theorem -- tells us that the information provided by the 2nd order correlators taken at the thermodynamic equilibrium is equivalent to the information that can be obtained by measuring a macroscopic linear response characteristic, such as the spin susceptibility. In this sense, the 2nd order correlator of spin fluctuations at the thermodynamic equilibrium does not provide any new information compared to the information that can be obtained, e.g., by the well developed pump-probe technique \cite{Young02a}, at least in principle.

However, the 2nd order correlator measured at {\it nonequilibrium} conditions is generally independent of the linear response characteristics.
The authors of \cite{crooker-noneq1} have explored the spin noise in the presence of a weak
radio frequency magnetic field, which was transverse both to the measurement and constant field axes.
This work demonstrated the possibility of a {\it multidimensional} SNS \cite{2Dspectrum-mukamel,2Dspectrum-2}, in which
the noise  power spectrum is represented as a function of both intrinsic and
driving field frequencies in a 2D density plot (as an example see Fig.~\ref{ac-SNS}(c)).  Such multidimensional
data show underlying patterns of correlations between different
frequencies and reveal numerous effects at the fluctuation level beyond the equilibrium and linear
response.

The multidimensional plots, such as Fig.~\ref{ac-SNS}(c), can be interpreted within the so called ``dressed state" Hamiltonian
\beq
\hat{H}_{\rm d}=\hat{H}+\hat{H}_{\rm EM}+\hat{V}
\label{ham-d}
\eeq
that describes the interaction of an atom with a coherent state of the electromagnetic field. Here $\hat{H}$ is the unperturbed atomic Hamiltonian, $\hat{H}_{\rm EM}$ is the secondary quantized Hamiltonian of the magnetic ac-field
\beq
\hat{H}_{\rm EM}=\omega_{ac} \hat{a}^{\dagger} \hat{a},
\eeq
where $\hat{a}^{\dg}$ and $\hat{a}$ are the creation and annihilation operators of the photons corresponding to this field mode, and
\beq
\hat{V}=\sum_{i} \tilde{\mu}_{i}B_{ac} (|E_i \ra \la E_{i+1}  | (\hat{a}+\hat{a}^{\dagger})+\textnormal{h.c.})
\label{ham-v}
\eeq
is the term describing the coupling of the radio frequency field of the amplitude $B_{ac}$ to the atomic spin states.

The electron spin-spin correlator of such a  system can be calculated similarly to Eq.~(\ref{SNPS2})
with the only difference that the index $i$ runs now throughout the eigenstates of the total Hamiltonian  (\ref{ham-d}).
For a weak ac-field, the interaction term (\ref{ham-v}) can be treated as a perturbation. In the first order in $\hat{V}$, the off-resonance values of the ac-field do not influence the spectrum.  However, if $\omega_{ac}$ is in resonance with any of the peaks then the corresponding peak splits into the so-called Mollow triplet \cite{Mollow69a} of three peaks. Moreover, if such a resonance shares a quantum level with another resonance, the latter also splits in the so-called
Autler-Townes doublet \cite{Autler55a} of two peaks. The distance between such splitted peaks is linear in the ac-field amplitude, $B_{ac}$. Higher ac-field amplitude leads to clearly observable nonlinear effects in the noise power spectrum, such as the appearance of the spin noise power peak centered at $\omega=\omega_{ac}$, additional splittings of other peaks, shifts of their positions, etc.  \cite{crooker-noneq1}.

The theory of dressed states is very well developed and cannot be reviewed here in detail. Fortunately, fairly complete and comprehensive introductions into this topic already exist (we refer to Refs.~\cite{budker-book,dressed-state-book} for further information). It should be emphasized that the importance of the work  \cite{crooker-noneq1} stems from the fact that for the first time  the rich physics of dressed states was observed at the fluctuation level without applications of pump laser pulses \cite{crooker-noneq1}.

The experimental demonstration of spontaneous spin fluctuations under steady nonequilibrium
conditions in atomic vapors \cite{crooker-noneq1} was followed by several theoretical studies
suggesting a broad range of applications for  {\it non-equilibrium} SNS.
In particular, it was proposed to use an ac-magnetic driving to
identify spin relaxation rates when the spin noise power spectrum is inhomogeneously broadened
(the quantum dots case) \cite{PhysRevB.91.155301}. In the case of electronic transport,
several effects (related to spin noise) were predicted both in the linear response regime
\cite{sinitsyn-13prl,sinitsyn-she} and far from equilibrium
\cite{golub-noneq}. The optical pumping can also be used to create nonequilibrium conditions. Spin noise of optically
induced exitons in semiconductors was explored theoretically in \cite{PhysRevB.90.085303}.

\subsubsection{Quantum Effects in Atomic Spin Noise.}
Squeezed spin states can be less susceptible to quantum
projection noise and thus can be used to increase the sensitivity of
magnetometers \cite{PhysRevA.50.67}. Experimental studies of quantum
projection noise and other spin fluctuations in such artificially
correlated atomic systems were reported in
\cite{projective-SNS,mitchel-12,mitchel-10,mitchel-11,polzik-01nat}.
The magnitude of artificially introduced correlations makes possible
observing their higher order spin cumulants experimentally \cite{Dellis13}.

We anticipate numerous applications of the optical SNS in atomic Bose-Einstein condensates and other ultracold atomic and molecular systems.
Currently, however, we are still at the very early stage in investigations of the  spontaneous equilibrium spin fluctuations in these systems using the optical
SNS. In fact, the proof-of-principle experiment was reported \cite{ultracold-noise}. This work communicates time-dependent spin fluctuations
in a cold atomic Fermi gas paying attention also to quantum measurement effects.

\subsection{Conduction Electrons in GaAs}

The first proof-of-principle measurements of the noise power spectrum of
conduction electron spins in GaAs revealed a much weaker useful signal in
comparison to atomic gases \cite{Oestreich05}. Nevertheless,
considerable improvements at the instrumentation and software levels \cite{gaas4,noise-gaas,ultrafast-sns} have
quickly enabled numerous SNS applications:


\begin{figure}
\centerline{\includegraphics[width=0.9\columnwidth]{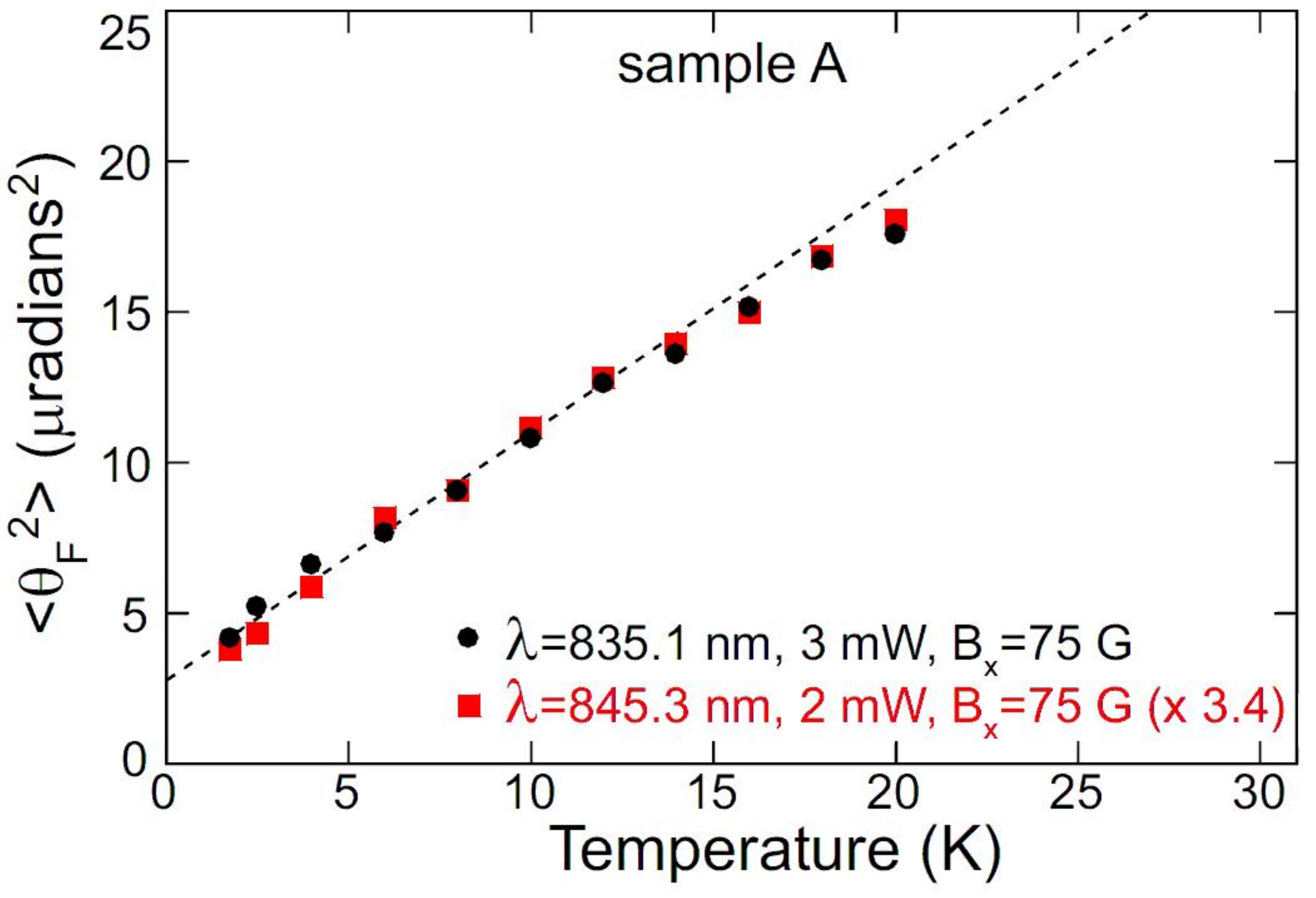}}
\caption{The integrated noise power $P_{\rm int}=\int C_2(\omega) \textnormal{d}\omega$
  of electronic spins in GaAs as a function of temperature. Linear
  scaling is in good agreement with predictions of the Fermi liquid
  theory. A finite offset, however, indicates either localization
  effects or the presence of other spin noise sources in the sample.
  Reprinted figure with permission from S. A. Crooker et al.,
  Physical Review B 79, p. 035208, 2009 \cite{noise-gaas}.
  Copyright \copyright (2009) by the American Physical Society.
  }
\label{noise-T}
\end{figure}

\begin{itemize}

\item The temperature-dependent spin relaxation rate and the Lande $g$-factor at the
thermodynamic equilibrium were studied in \cite{noise-gaas}. The dependence of the spin noise power spectrum on the
doping concentration level was explored in \cite{gaas3,gaas5}. The influence of optical
excitations on  spin noise in semiconductors was investigated in
\cite{Huang2011}.

\item Measurements of  spin noise with {\it  picosecond } time resolution
were reported in \cite{high-band}. Such a {\it high
bandwidth spectroscopy} was used to observe  inhomogeneous broadening of
the spin noise peak of conduction electrons in a relatively large
magnetic field \cite{high-band}.

\item The inverse
Faraday effect \cite{inverse-faraday}, i.e. the appearance of an effective magnetic
field induced by a circularly polarized light, was observed at the spin
noise level \cite{inverseFR-SNS}. 

\item The observation of a slow evolution of the noise power spectrum of
electrons interacting with polarized nuclear spins was reported in
\cite{impurity-noise-3}.

\end{itemize}

Spin dynamics of conduction electrons in $n$-doped materials has been relatively well
understood prior to the appearance of SNS.
However, some features of  spin noise in GaAs are still obscure. One such
observation \cite{noise-gaas} concerns the behavior of the
integrated noise power, $P_{\rm int}=\int P(\omega) \textnormal{d}\omega$, as function of
temperature $T$. The Fermi liquid theory predicts its linear
scaling $P_{\rm int}\sim T$ at low temperatures. However, the experiment \cite{noise-gaas} showed rather a linear scaling
with an offset, $P_{\rm int}= a+bT$, where $a$
and $b$ are constants (Fig.~\ref{noise-T}). The origin of this offset is not quite
understood. It can be due to the conduction electron localization or presence of deep
donor bound states.


\begin{figure}
\centerline{\includegraphics[width=0.99\columnwidth]{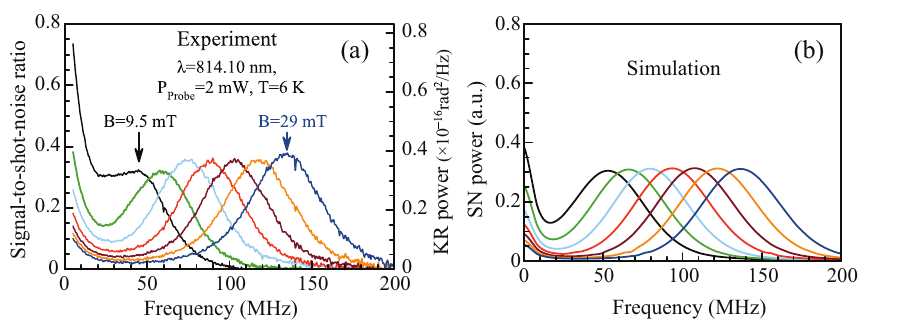}}
\caption{Spin noise from a single GaAs quantum well embedded
into a microcavity \cite{Poltavtsev14}. (a) Measured noise
  power spectrum of conduction electrons. (b) Predictions of the
  theoretical model in \cite{Poltavtsev14} that includes exciton effects.
  Reprinted figure with permission from S. V. Poltavtsev et al.,
  Physical Review B 89, p. 081304, 2014 \cite{Poltavtsev14}.
  Copyright \copyright (2014) by the American Physical Society.
  }
\label{zero-peak1}
\end{figure}


Another not fully explained feature of the noise power
spectrum is a small peak centered at zero frequency that persists even
in presence of an in-plane magnetic field (Fig.~\ref{zero-peak1}). The
magnetic field shifts most of
the noise power to the major Lorentzian peak centered at the Larmor frequency but
a small fraction of the noise power remains peaked at
$\omega=0$. Different explanations of this phenomenon have been
proposed. For example, a
small ellipticity of the probe beam polarization can lead to the
inverse Faraday effect \cite{inverse-faraday}, which does lead to a zero frequency
peak \cite{impurity-noise-3}. Another possible explanation is the presence of localized
states, which contribution to the spin noise power is usually centered
at zero frequency at weak external fields \cite{Crooker10}. There
has also been a proposal to relate this effect to excitons \cite{Poltavtsev14}.


An important future goal is to extend  SNS to conduction electrons
in other materials. New measurement schemes have been proposed to increase the useful
signal-to-background ratio \cite{real-time-SNS,greilich-13,squeezed-15}. A big step in this
direction was an experiment on placing a 2D electron system in a
microcavity that considerably enhanced  coupling of the probe beam
to spins of conduction electrons \cite{Poltavtsev14}. This geometry
increased the useful signal by orders of
magnitude and allowed studies of spin noise from only a few hundreds of
electrons instead of millions in early experiments. Such micro-cavities can also be used to produce  strongly correlated  spin-photon states, which can be revealed by  spin noise measurements \cite{cavity-noise}.





\subsection{Quantum Dots}

Prior to the advances in SNS, the experimental studies of decoherence and
relaxation of quantum dot spin qubits
could not verify numerous theoretical predictions.
Earliest applications of the
SNS to hole-doped InGaAs quantum dots showed a relatively strong
useful signal per spin
\cite{Crooker10,ingaas-2}. Subsequent studies of the spin noise power
spectrum of quantum dots  have not only verified some of the theories
but also revealed some unexpected behavior. It was clearly demonstrated that
 SNS is capable of resolving  important questions
in materials science and uncover new physical phenomena.

\begin{figure}
\centerline{\includegraphics[width=0.99\columnwidth]{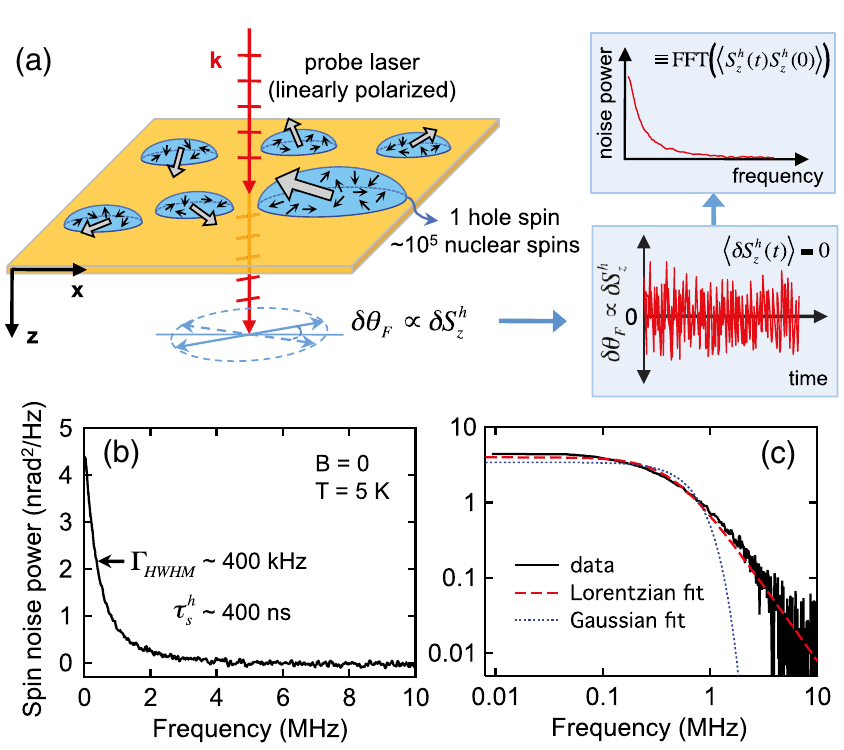}}
\caption{Spin noise from holes coupled to nuclear spin baths in (In, Ga)AS quantum dots \cite{Li12}.
(a) Experimental setup, (b) typical spin
noise power spectrum of the resident holes at a low temperature
(5 K) and zero applied magnetic field ($B=0$), and (c) the same spectrum on a log-log
scale. The noise line shape closely follows a Lorentzian, indicating
exponentially decaying hole spin correlations \cite{Li12}.
  Reprinted figure with permission from Y. Li et al.,
  Physical Review Letters 108, p. 186603, 2012 \cite{Li12}.
  Copyright \copyright (2012) by the American Physical Society.
  }
\label{fig3_6}
\end{figure}

When GaAs samples are grown with an admixture of indium atoms that
substitute gallium, indium does not disolve in the lattice uniformly
but rather creates InAs ``droplets'' of a few tens of nm
diameter and several nm height (shown in Fig.~\ref{fig3_6}(a)), which are called the self-assembled
quantum dots (QD). QDs provide a confining potential for a
localized single electron or hole state with an uncompensated spin.

In a relatively strong in-plane magnetic field, the spin noise power
spectrum  of hole-doped QDs was found to be in a good agreement with
theoretical expectations \cite{Crooker10}.  The spectrum consists of a shallow Gaussian
peak, which indicates the presence of the Larmor precession with
a strong inhomogeneous broadening due to different values of the
Lande $g$-factor of hole spins in different QDs. Moreover, the measured
anisotropy of the $g$-factor was in a good agreement with theoretical predictions
\cite{Crooker10}.  At zero magnetic field but relatively high temperatures, $T>7$K,
the experimental observations also agreed well with a theoretical expectation of
phonon-induced spin relaxation \cite{Li12}. It was found that at lower
temperatures, $T<7$K, the spectrum  no longer depended on $T$, which ruled
out the phonon origin of spin  relaxation and indicated
the transition to the regime where the spin relaxation was dominated by the coupling
to the nuclear spin bath.

An unexpected behavior was observed at $B=0$
and low temperature. It was found that the spectrum of hole-doped QDs
consists  of a single  narrow
Lorentzian peak (Fig.~\ref{fig3_6}(b)), indicating an exponential relaxation of hole spins
within $\sim 0.4$~$\mu$s \cite{Li12,ingaas-1}. To understand
why this behavior was unexpected, consider the  Hamiltonian that
describes the hyperfine coupling of the central spin  to a nuclear spin bath:
\begin{equation}
\hat{H}^{\rm hf} = \sum_{i=1}^N g^i_{||} \hat{S}_{z}
\hat{s}^i_{z}+\sum_{i=1}^N g^i_{\perp} (\hat{S}_{x}
\hat{s}^i_{x}+\hat{S}_{y}
\hat{s}^i_{y}),
\label{hole-ham1}
\end{equation}
where $\hat{S}_{\alpha}$, $\hat{s}^i_{\alpha}$  are the components of
the central and nuclear spin operators, respectively, $g^i_{||}$ and
$g_{\perp}^i$ are the out-of-plane and in-plane hyperfine
coupling constants describing the central spin and $i$th nuclear
spin interaction. Typically, the number of nuclear spins in an InGaAs quantum dot is $N\sim
10^5$. The hyperfine coupling is almost isotropic for electron-doped QDs
with $\la g^i \ra  \sim 1$ MHz. It is usually an order of magnitude smaller and
has a relatively strong out-of-plane anisotropy for
hole-doped QDs: $g_{\perp}^i/g_{||}^i \sim 0.2-0.5$.

Due to the large number of nuclear spins, one can justify the mean field
approximation, in which  the effect of the nuclear spin bath on the
central spin is described by an effective field, called the Overhauser
field $ {\bf B}_{\textnormal{O}}$ \cite{merkulov}:
\begin{equation}
\hat{H}_{\rm eff}^{\rm hf} = {\bf B}_{\textnormal{O}}\cdot \hat{\bf S}.
\label{over1}
\end{equation}
One can assume that this field points in a random direction, which is different in
different QDs. The components of the Overhauser field are
described by the Gaussian distribution, such that $\sqrt{\la B_{\textnormal{O},z}^2 \ra}  \sim g_{||} \sqrt{N}$ and $\sqrt{\la B_{\textnormal{O},x}^2 \ra },\sqrt{\la B_{\textnormal{O},y}^2 \ra } \sim g_{\perp} \sqrt{N}$.

\begin{figure}
\includegraphics[width=3in]{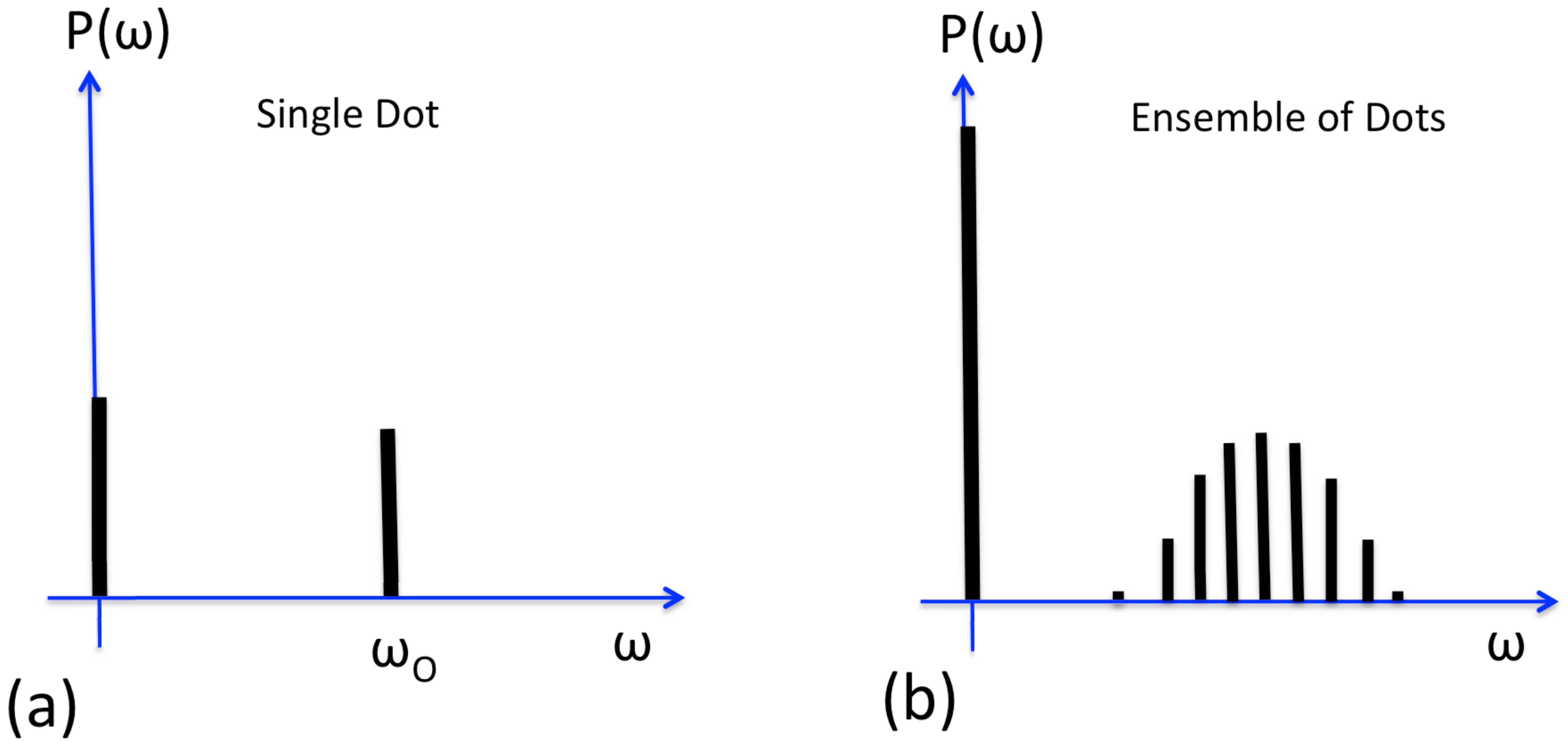}
\caption{ (a) The expected spin noise power spectrum  for a single quantum
  dot (as follows from Eq.~(\ref{dots-power})). (b) Inhomogeneous
  broadening of the finite frequency peak in an ensemble of quantum dots. }
\label{broad-dots}
\end{figure}
The size of the Overhauser field sets the frequency of precession of the
central spin around this field: $\omega_\textnormal{O}\approx g_{||} \sqrt{N} \sim
10$ MHz for hole-doped QDs. Let $\theta$ be the angle between
the Overhauser field and the measurement $z$-axis. A coherent spin polarization
precession around such a field is described by the correlator
\begin{equation}
\la S_z(t) S_z(0) \ra = \cos^2(\theta)+\sin^2(\theta) \cos(\omega_\textnormal{O} t),
\label{prec}
\end{equation}
where we normalized the correlator to 1 at $t=0$. While the spin polarization component
along the measurement axis remains constant, the other
part oscillates \cite{merkulov}. The Fourier transform
of this correlator produces two delta-functions (Fig.~\ref{broad-dots}(a)): one at zero frequency and another
at $\omega_\textnormal{O}$:
\begin{equation}
P(\omega)\equiv C_2(\omega) = \cos^2(\theta)\delta(\omega) +\sin^2(\theta) \delta(\omega-\omega_\textnormal{O}).
\label{dots-power}
\end{equation}

If  measurements are performed on an ensemble
of a large number of quantum dots, Eq.~(\ref{dots-power}) should be
averaged over the distribution of the Overhasuer field. Each quantum dot
then still contributes to the $\delta$-function at $\omega=0$ and, in addition,
produces a finite frequency contribution at its individual $\omega_\textnormal{O}$, so
that the high frequency peak is expected to be broadened and have the
Gaussian shape, as shown in Fig.~\ref{broad-dots}(b). The maximum of this peak is expected to be at $\omega \approx
g_{||} \sqrt{N} \sim 10$ MHz. There have been debates in theoretical
literature about the role of the slow dynamics of the Overhauser
field. This dynamics is responsible for the broadening of the zero
frequency peak. Numerical studies \cite{dobrovitski,Faribault2013,sinitsyn-12prl-2}  have shown that, within the model
(\ref{hole-ham1}), this dynamics is very slow so that only a small power-law broadening
of the zero frequency peak is expected, not exceeding the transverse hyperfine
coupling frequency scale $g_{\perp} <0.1$ MHz.

Apparently, the experimental observation \cite{Li12} of a single Lorentzian peak with a
width of $\sim 1$ MHz (Fig.~\ref{fig3_6}(b)) strongly contradicts the above
expectation. Not only the shape of the peak but also
its characteristic width could not be explained by the above mentioned theory. Moreover, this result can
not be related to an unusual inhomogeneous broadening.
In fact, it was possible to make an experiment in
which all but one quantum dots were  shielded from the probe beam so
that the Faraday rotation noise spectrum of a {\it single spin} was detected
\cite{single-spin}. At zero magnetic field, no deviations in measurement
results for a single quantum dot from the results for an ensemble of quantum dots
were found.


The experimental observations reported in \cite{Li12,ingaas-1} were
explained theoretically in Ref.~\cite{sinitsyn-12prl-2} by the presence
of quadrupole couplings, whose importance had been clearly exposed
earlier in Ref.~\cite{imamoglu-quad}. A more realistic Hamiltonian is written as
\begin{equation}
 \hat{H}= \hat{H}^{\rm hf}+ \sum_{i=1}^N \frac{\gamma_q^i}{2} (\hat{\bf s}^i \cdot {\bf n}^i)^2 ,
\label{Ham-q}
\end{equation}
where $\hat{H}^{\rm hf}$ is defined by Eq.~(\ref{over1}), $\gamma_q^i$ is
the strength of the quadrupole coupling of the $i$th nuclear spin, $\hat{\bf s}^i$ is the nuclear spin operator, and
${\bf n}^i$ is the unit vector along the direction of the local quadrupole coupling anisotropy.
The last term in Eq.~(\ref{Ham-q}) becomes nontrivial, i.e., different from a constant when $s^i>1/2$. In fact,
the most abundant isotopes of Ga and In are characterized by
$s=3/2$ and $s=9/2$, respectively.
The assumption \cite{sinitsyn-12prl-2} of a broad distribution of ${\bf n}^i$  inside
self-assembled quantum dots helped to explain the experimental observations.

An analytical solution to the system described by the Hamiltonian (\ref{Ham-q}) can
be found in the limit of a large quadrupole coupling, which is
relevant to the hole-doped dots. In this case, the nuclear spin dynamics
is dominated by the last term in (\ref{Ham-q}), which induces fluctuations of the
Overhauser field with a typical amplitude $\sim g_{||} \sqrt{N}$ and
correlation time $\sim \gamma_qs/2$.  The dynamics of the central spin subjected to
such a fluctuating Overhauser field was investigated in
\cite{sinitsyn-12prl-2}, where it was shown that in hole-doped InGaAs
quantum dots the central spin  relaxes exponentially. In the case of electron-doped dots, numerical simulations
\cite{sinitsyn-12prl-2,electron-dot} predict that the basic
physics leading to Fig.~\ref{broad-dots}(b) is correct except the
zero frequency peak is broadened by the quadrupole coupling effects.  This difference from the hole-doped quantum dots follows mainly from order of magnitude stronger hyperfine coupling, which reduces the relative role of the quadrupole coupling effects. Recent optical measurements of spin correlators in electron-doped dots \cite{alex-nat,electron-dot}  and spins confined near impurities \cite{impurity-noise-3} have confirmed this prediction. Moreover,
experiments with a single quantum dot revealed an additional oscillation in the spin correlator in the time-domain,
which was also explained by the quadrupole coupling effect \cite{alex-nat}.





\subsection{Magnetic Impurities, Thin Magnetic Films, and  Nanomagnets}

The spin noise power spectrum of localized electrons bound to
impurities has been observed in \cite{PhysRevB.87.045312}.
The noise from such bound electrons is similar to the noise from electrons confined
in quantum dots. This is explained by the same spin relaxation mechanism at low temperatures:
the electron spin interaction with local nuclear spins. Importantly,  impurities have
a reproducible confining potential. Consequently, the resulting spin
noise is less influenced by inhomogeneous broadening, which can be
very strong in quantum dots.

The spin noise power measurements of magnetic Mn ions in CdTe quantum
wells were reported in \cite{mn-noise}. In this system, the spin noise
power spectrum has a relatively complex form explained by multiple resonances of spin-5/2  Mn ions.

The authors of \cite{impurity-noise-3} have communicated the spin noise power spectrum
of electrons bound to Si impurities in GaAs.
The measured noise power spectrum contains two clear peaks - one centered at
zero and another one centered at the characteristic frequency of the Overhauser
field. The application of the external magnetic field removes the zero
frequency peak and produces a Gaussian broadened peak at the Larmor frequency, as expected.
The spin dephasing and relaxation times of electrons bound to Si impurities in GaAs are longer
compared to these in InGaAs quantum dots.


In \cite{balk2014critical}, the Kerr rotation spectroscopy was applied to
detect ferromagnetic fluctuations and random thermal motion of domain
walls in a thin film made of magnetic Co atoms.  The width of the film
varied so that  the magnetization fluctuations could be studied in different phases near the ferromagnetic phase transition. Due to the large size of the domain walls,  both temporal and
spatial noise characterization was performed.

In \cite{liu-echo}, the optical SNS was applied to study fluctuations of the magnetization in molecular nanomagnets. The latter are molecules with uncompensated spins of magnetic ions. Typically, spins of
nanomagnets experience magnetic crystal anisotropy, in which they have two or several degenerate ground states. This degeneracy is lifted by spin tunneling and interactions of nanomagnets with each
other and with nuclear spins. Fluctuations of the magnetization of nanomagnets can be used for precise parameter estimation of these technologically interesting nanostructures.


\subsection{Alternative SNS Techniques}

Non-optical SNS methods usually do not
probe the spin polarization directly but rather detect fluctuations of
the local magnetic field produced by flipping spins. The sensitivity
of such techniques to the relative position of probe and noise
sources complicates the interpretation of the signal. At the same time,
this can be used as an advantage in situations when a better spatial resolution is needed.

{\it Barkhausen noise.} Studies of fluctuations in magnetic
systems take roots in the work of Heinrich Barkhausen who proved already in 1919
that the magnetic hysteresis curve is not continuous,
but is made up of small random steps caused when the magnetic domains
move under an applied magnetic field \cite{barkhausen}.
This noise can be characterized by placing a coil of a conducting wire
near the sample. Motion of ferromagnetic domain walls produces changes
in the magnetization that induces noisy electrical signals in the
coil. Studies of Barkhausen noise have been used in practice as a
nondemolition tool to characterize distribution of elastic stresses and the micro-structure of magnetic samples \cite{barkhausen-rev}.

{\it SQUID-based spectroscopy.} Superconducting quantum
interference devices (SQUIDs) have been the method of choice for
studies of magnetic field fluctuations in superconductors
\cite{PhysRevB.57.10929}, spin glasses \cite{PhysRevLett.57.905,
  PhysRevLett.57.905,PhysRevB.40.7162}, ferroelectric liquid crystals \cite{PhysRevLett.79.1062}, and nanomagnets
\cite{PhysRevB.57.497}.  In this approach,
the sample is placed directly in the vicinity of a dc-SQUID chip with integrated
pickup loops and field coils that detects magnetic field fluctuations
that originate from
flipping magnetic moments. The frequency bandwidth of SQUID devices is
relatively small in comparison to optical SNS setups. It is most
suitable for materials made of ferromagnetic grains or large spins, such as nanomagnet arrays.
Strong disorder, dipole interactions, and anisotropy fields lead to a
broad distribution of spin relaxation rates in such samples so that
the measured noise power spectra usually were reported to have the
form of $\sim 1/f$ or $\sim 1/f^{\alpha}$ noise. Yet, the dependence
of such spectra on temperature and the magnetic field could be used to
extract important physics. A number of theoretical publications have
been devoted to this field, e.g., to  the origin of the
power laws and violation of the fluctuation-dissipation theorem in
spin glasses
 \cite{PhysRevLett.113.217002,PhysRevLett.104.247204,PhysRevLett.83.5038,PhysRevB.86.134414}.

{\it Cantilever-based spectroscopy.}
 Local fluctuations of spins can  also be probed with a nanomechanical
 resonator \cite{cantilever-science,cantilever-noise}. In this approach,
 fluctuations of the magnetization are detected via measuring their
 mechanical force that the time-dependent spin polarization produces on a small permanent
 magnet attached to the end of a sensitive silicon
 cantilever. This technique was successfully applied to ensembles
 of electronic localized spins  \cite{cantilever-science},  molecular
 nanomagnets \cite{cantilever-noise}, and nuclear spins \cite{cantilever-NMR}. Currently, the
 allowed frequency bandwidth is comparable to the optical techniques,
 while cantilever-based spectroscopy has a better spatial resolution
 and is not restricted to optically sensitive materials. Its
 applications are restricted, however, to spins near the surface of a
 sample. Interpretation of the data can be complicated by complex
 distribution of forces with which spins at different locations act on
 the probe mechanical resonator.

{\it STM-based spectroscopy.}
Measuring fluctuations of a spin polarized tunneling current from an STM tip can
be used to extract spectroscopic information on the temporal
susceptibility of a single magnetic atom. This technique has an
exceptional time and atomic-scale spatial resolution. However, its applications are restricted to
 surface spins. Moreover, technical problems such as the lack of a good
control of the magnitude of the tunneling current
remain to be resolved. For a detailed discussion of this method, we refer the reader to a recent
 review \cite{balatsky-stm-rev} and theoretical papers \cite{balatsky-noise, PhysRevB.70.033405,PhysRevB.73.184429,PhysRevLett.102.256802}.

\begin{figure}
\centerline{\includegraphics[width=0.75\columnwidth]{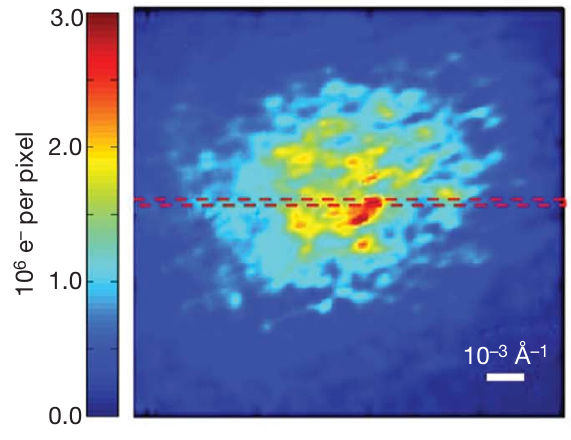}}
\caption{CCD image of the X-ray speckle observed for the chromium [200] lattice
Bragg reflection (for more details, see Ref. \cite{Shpyrko07a}).
Reprinted with permission from Macmillan
Publishers Ltd: Nature 447, pp. 68–-71, 2007, Copyright \copyright (2007) \cite{Shpyrko07a}.}
\label{fig3_10_X_ray}
\end{figure}

{\it X-ray photon correlation spectroscopy.}
This experimental technique
has been used for many different purposes including studies of
domain wall fluctuations in antiferromagnets at short wavelengths \cite{Shpyrko07a,Jacques14a}. In this application, a coherent beam
reflected from a sample surface (chromium (111) surface in Ref. \cite{Shpyrko07a}) creates an interference pattern known as speckle \cite{dainty1975laser,goodman1976some} (see Fig.~\ref{fig3_10_X_ray} for a speckle example).
The sensitivity of the speckle to the domain wall structure was used to obtain information regarding the microscopic dynamics of antiferromagnetic domains  \cite{Shpyrko07a}. In the speckle data analysis, the autocorrelation function $g_2(t)=\left< I(\tau)I(\tau+t) \right>_\tau /\left< I(\tau) \right>^2_\tau$ is found for a given pixel and used to extract relevant time scales. The theory of coherent light propagation in disordered media and speckle statistics was developed in \cite{andreev-speckles}.

The chromium antiferromagnetism is a complex phenomenon associated with spin- and charge-density waves of different periods. In Ref.~\cite{Shpyrko07a}, the spin-density wave dynamics was tested indirectly at a charge-density wave Bragg reflection angle. It's interesting that the reflection at the spin-density wave angle contains just a single stationary peak \cite{Jacques14a}. Therefore, we feel that there is a need of better understanding of
co-existence of the spin- and charge-density waves in chromium, and their coupling to light.

\subsection{Nuclear Spin Noise}
 The nuclear magnetic fields from mesoscopic volumes at equilibrium are extremely weak.
Nevertheless, it is possible to study
fluctuations of such fields \cite{nuclear-noise1,chandra2013spin}. For example,
SQUID probes were used to measure the nuclear spin noise as early as in 1985 \cite{nuclear-noise}.
An approach based on the STM tunneling spectroscopy was theoretically developed in \cite{PhysRevB.73.184429}.

A cantilever based
detector was used in \cite{cantilever-NMR} to resolve mesoscopic nuclear spin noise in real
time with submicron resolution.  This experiment not only allowed studies of nuclear spin diffusion but also demonstrated the capability
to manipulate the local nuclear spin distribution.

Considerable progress in the detection of nuclear spin noise has been
achieved recently using nanomagnetometry based on NV-centers in diamonds \cite{NV2,NV3,nV1}. Magnetic field
fluctuations, as small as 10nT, from a single nuclear spin have been resolved
with a nanometer precision \cite{NV4}. Electronic spins in
coupled GaAs quantum dots were also shown to work as a sensitive probe of the environmental field fluctuations \cite{yakoby-13prl}.

Finally, although optical beams do not couple to the nuclear spin
polarization directly, the noise power spectrum of nuclear
spins has been recently observed by optical SNS \cite{impurity-noise-3}. In
this experiment, the electronic spin noise in a silicon doped ultraclean GaAs
semiconductor was studied. Surprisingly, in addition to the expected
noise power peaks of donor-bound electrons, relatively weak peaks have
been observed in the low frequency (10-100kHz) range. Such peaks
responded to the
external magnetic field as expected from the nuclear spin noise of the  host isotopes: $^{75}$As, $^{69}$Ga and
$^{71}$Ga. This finding allowed the authors of
\cite{impurity-noise-3} to explore the nuclear spin noise at very low
magnetic field values and observe effects consistent with presence of quadrupole couplings.
Apparently, this feature in the optically probed noise power spectrum follows from fluctuations of local
Overhauser fields that nuclear spin fluctuations induce on observable electronic
spins \cite{PhysRevB.91.205301}, \cite{impurity-noise-3}(supplementrary).  It is possible, e.g., that a precessing nuclear spin polarization creates an effective ac-like Overhauser
field that can produce a noise power peak of electronic spins at
the precession frequency as a 2nd order effect in the field
amplitude \cite{crooker-noneq1}.


\section{Spin Correlators in Quantum Weak Measurement Theory}

In the case of a quantum evolution, the
measurements  influence the system's state. For example, if projective measurements
are performed at a high rate, the spin may not be able to precess due to
the quantum Zeno effect \cite{zeno1}. Moreover, quantum mechanical
spin correlators depend on products of operators at different moments of time.
Spin operators, generally, do not commute. Hence, their product can be
non-Hermitian and its trace with the equilibrium density matrix can be a complex
number. There are  different combinations of spin operator products
that, after being traced  with the density matrix, produce
real but different results. The important questions then emerge about which combination is the quantum
mechanically justified expression for the measured spin correlator in
a particular experiment and how to perform minimally invasive measurements.

\subsubsection{Weak Measurements.} Being based on the idea of weak-measurement \cite{sham,weak11},  SNS can perform measurements at a high rate without producing strong
disturbances. In optical SNS, the measurement beam does not
make ``hard" projective measurements of the total spin of a
mesoscopic spin system. Instead, the spins are allowed to entangle
weakly with the linearly polarized beam. One can describe such interaction by the effective Hamiltonian

\beq
\hat{H}_{\rm int} =- i\lambda_\textnormal{SO} \hat{S}_z \frac{\partial}{\partial \phi},
\
\label{wm1}
\eeq
where  $-i\partial /\partial \phi$ is the operator inducing rotation of the linear polarization axis of the probe beam, $\hat{S}_z$ is the quantum mechanical operator of the total spin polarization in the observation volume along z-axis, and $\lambda$ is the coupling strength. Consider a photon wave packet that is allowed to interact with our spin system during a time interval $\delta t$.
The wave packet is made of states with different polarization angles $\phi$:
  $$
  |\Psi_0 \ra= \int \textnormal{d}\phi \,  w(\phi) | \phi \ra,
  $$
  where  $w(\phi)$ is a Gaussian function peaked near $\phi =0$ and normalized so that $\int w^2(\phi) \textnormal{d}\phi =1$.
  Initially, the spin system is disentangled from this wave packet,
  i.e. the entire wave function is a product $ |\psi_s \ra   |\Psi_0 \ra
  $, where $ |\psi_s \ra $ is the state vector of a spin
  system. The interaction is described by the evolution operator $\hat{U}
  = \exp (-i\hat{H}_{\rm int} \delta t  )$. For example, if the initial density matrix of the spin system is given by a superposition of different projections of $\hat{S}_z$, i.e.
  \begin{equation}
  \hat {\rho}_s = \sum_{i,j} \rho_{ij} |s_z^i \ra \la s_z^j |, \label{initial_spin_rho}
  \end{equation}
then, after the interaction taking the time $\delta t$ of the passage of the wave packet through the sample, the state of the total density matrix  (the spin-photons state) is an entangled state:
\begin{eqnarray}
\nonumber
\hat{\rho} = \iint \textnormal{d} \phi_1 \textnormal{d} \phi_2 \, w(\phi_1) w(\phi_2) \\
\nonumber
 \quad\quad\quad\quad\quad\quad \times\sum_{i,j} \rho_{ij} |\phi_1 -\bar{\lambda} s_z^i \ra  |s_z^i  \ra  \la s_z^j |  \la \phi_2 -\bar{\lambda} s_z^j |,
 \end{eqnarray}
 where $\bar{\lambda}=\lambda \delta t$.

The detector then performs the projective measurement of the rotation angle.  The magnetization $S_z$  is inferred from the observed polarization rotation, $\phi_0$, as  $S_z=\phi_0/\bar{\lambda}$. Note that
such an $S_z$ is  not the exact value of the magnetization but rather it is its most likely estimate.
 The probability to obtain an estimate $S_z$ as a measurement result is
\beq
P(S_z)=  {\rm Tr} \left[ \hat{\rho} \delta(\hat{\phi} - \phi_0) \right]  = \sum_i \rho_{ii} w^2( \phi_0-\bar{\lambda} s_z^i).
\label{weak-p}
\eeq

In principle, any measurement disturbs the system. In the case of the weak measurement, such
disturbances are small. In order to evaluate the spin density matrix right after the measurement,
we project the density matrix $\hat \rho$ into the measured polarization state $\phi_0$ using the projection operator
$|\phi_0 \ra \la \phi_0|$. Up to a normalization constant, the result is
\beq
\hat {\rho}_s' \sim \sum_{i,j} \rho_{ij} w(\bar\lambda (S_z- s_z^i)) w(\bar{\lambda} (S_z- s_z^j))  |s_z^i \ra \la s_z^j |.
\label{newrho}
\eeq
Note the changes compared to the initial spin density matrix (\ref{initial_spin_rho}).

\subsubsection{Formulation Based on Kraus Operators.} Generally, one can reformulate (\ref{weak-p}) and (\ref{newrho}) by introducing the {\it Kraus operators}:
\beq
\hat{K}(S_z) = (2\bar{\lambda}/\pi)^{1/4} e^{-\bar{\lambda} (S_z-\hat{s}_z)^2},
\label{kraus1}
\eeq
so that the probability of measuring the outcome $S_z$ is given by  \cite{belzig-11njp}
\beq
P(S_z) = {\rm Tr} \left[ \hat{\rho}_1 (S_z)   \right], \quad \hat{\rho}_1(S_z) = \hat{K}(S_z) \hat{\rho}_s \hat{K}^\dagger (S_z).
\label{res1Kr}
\eeq
It is now easy to extend this analysis to find the probability of observing an arbitrary trajectory $S_z(t)$. First, we note that a single wave packet passage through a sample can be considered extremely fast in comparison with dynamics of the spin system. Because of this reason, the  spin Hamiltonian does not appear in Eqs. (\ref{res1Kr}). However, on a longer time scale, in the Heisenberg picture, the measurement operator evolves with time:
$$
\hat{S}_z(t) = e^{i\hat{H}t} \hat{S}_z e^{-i\hat{H}t},
$$
where $\hat{H}$ is the Hamiltonian of the spin system.

One can then introduce a Kraus operator in the Heizenberg picture that corresponds to the observed trajectory $S_z(t)$ as \cite{belzig-11njp}
$$
\hat{K} [S_z(t)] = C {\rm \hat{T}} e^{-\int \lambda (S_z(t) - \hat{S}_z(t))^2 \textnormal{d}t },
$$
where $C$ is a normalization factor and ${\rm \hat{T}}$ is time-ordering operator (later times on the left). Finally, the probability of a trajectory (the functional probability) is given by  \cite{belzig-11njp}
\beq
P[S_z(t)] = {\rm Tr} \left[\hat{K}^{\dagger}  \hat{K} \hat{\rho}_s \right].
\label{prob-w}
\eeq

By definition, $S_z(t)$ is a real-valued trajectory of the detector
output. Having obtained probabilities of such trajectories, one can
calculate arbitrary correlators by standard means. This task is
straightforward but somewhat lengthy to be explained here. We refer the reader to Ref.~\cite{belzig-11njp} for the detailed discussion and summarize here only some results.
It turns out that, within  this model, the measurement of correlators corresponds to finding traces of the following operator products with the initial density matrix:
\begin{eqnarray}
\nonumber && \la S_z(t_1) S_z(t_2) \ra = {\rm Tr} \left[  \breve{S}_z  \hat{U}(t_2,t_1)  \breve{S}_z \hat{U}(t_1,0) \hat{\rho}_s    \right], \\
\label{corr32} \\
\nonumber && \la S_z(t_1) \ldots S_z(t_n) \ra =  \\
&& \quad\quad\quad\quad\quad\quad\quad {\rm Tr} \left[  \breve{S}_z  \hat{U}(t_n,t_{n-1})  \ldots \breve{S}_z \hat{U}(t_1,0) \hat{\rho}_s    \right] , \nonumber
\end{eqnarray}
where
\beq
\breve{S}_z  \hat{X}  \equiv \frac{1}{2} \{ \hat{S}_z , \hat{X} \}
\label{breve1}
\eeq
 is normalized anticomutator of $\hat{S}_z$ with the matrix on the
 right of it, and $\hat{U}(t_a,t_{a-1})$ is the evolution operator of
 the density matrix from time $t_{a-1}$ to time $t_a$.

 According to \cite{belzig-11njp}, the effect of measurement on the system has the following two consequences: First, the measured Faraday rotation is not strictly proportional to the spin polarization but has an uncorrelated  background noise component:  $\phi(t) = \bar\lambda S_z (t) +
\zeta(t)$, with $\la \zeta(t) S_z(t) \ra=0$, where $S_z(t)$ is the true instantaneous spin polarization. In fact, such a
background noise can be measured separately and subtracted if one
considers cumulants of the spin variables. Second, due to the system-detector coupling,
the evolution of the system's density matrix has to be described by a {\it Lindblad}-type equation
\beq
\frac{\textnormal{d} \hat{\rho}_s}{\textnormal{d} t} =\hat{L} \hat{\rho}_s \equiv  -\frac{i}{\hbar}\left[\hat{H}, \hat{\rho}_s  \right] - \frac{\lambda}{2} [\hat{S}_z, [ \hat{S}_z, \hat{\rho}_s] ],
\label{corr34}
\eeq
so that the evolution operator is defined as $\hat{U} (t_a,t_{a-1}) = \hat{T} \exp \int_{t_{a-1}}^{t_a} \hat{L} \textnormal{d}t$.

An important consequence of Eq.~(\ref{corr34}) is that one can minimize the feedback effect of  a detector  by reducing the coupling constant $\lambda$. In the limit $\lambda \rightarrow 0$, the correlators (\ref{corr32}) become  a trace of the initial system density matrix with anticommutators of spin operators at different moments of time. One can calculate these correlators quantum mechanically without explicitly assuming the presence of a detector. There is a price to be paid for allowing the coupling to a detector to be small:
a decrease in coupling implies an increase in the background noise $\zeta (t)$ and, hence, a longer time to filter it out experimentally.

Equations~(\ref{corr32})-(\ref{corr34}) were derived within a specific
model of a detector. In the literature on  SNS, one can encounter
other detector models that correspond to a different choice of the
Kraus operator \cite{sham}. In the limit of weak coupling to the
detector, they produce the same prediction for the 2nd order
correlator. However, higher order spin correlators, as well as
the behavior of all correlators beyond the limit of the weak coupling to
detector, can depend on the choice of the detector model.

Finally, as it is discussed in \cite{belzig-12prl},  different
models of the weak measurement generally lead to different definitions
of the operator $\breve S_z$. For example, we may find that instead of (\ref{breve1}) one should rather use
\beq
\breve{S}_z \hat{X} = \frac{1}{2}  \{ \hat{S}_z, \hat{X} \} - i \beta [\hat{S}_z, \hat{X}],
\label{breve-def}
\eeq
with some real constant $\beta$. The latter becomes non-zero, e.g.,
when the detector interferes by absorbing energy.
Interestingly, in such a case one cannot guarantee the positivity of
the noise power spectrum \cite{belzig-12prl}. This fact may look
impossible considering the definition (\ref{noiseP1wk}), which is
the average of  squares of measured real numbers. This controversy is
removed by noticing that the measured signal is actually the physical
noise  plus the background noise. Only this
sum is constrained to have a positive noise power spectrum. However, when the
 power spectrum of the background noise is  measured separately and then subtracted, one can
discover that the physical spin noise power spectrum can be negative
 in a certain range of  frequencies. The case $\beta=0$, which we discussed in
this section, however, does not allow this  effect, as it follows from the following property.

\underline{The weak positivity property}: The physical part of the noise power spectrum of the correlator (\ref{corr32}) with $\breve{S}_z \hat{X} = \frac{1}{2}  \{ \hat{S}_z, \hat{X} \}$ is always positive definite.
Indeed, using the fact that $2 \la S_z^2(t) \ra \geq \la  \{ \hat{S}_z(t), \hat{S}_z(0) \} \ra$,  we find  that the expression $\la S_z(t) S_z(0) \ra$ (as it is defined in (\ref{corr32})) is maximized at $t=0$. The Fourier transform of such a real function of time is positive definite (according to  the Bochner's theorem \cite{reed}), which proves the statement.

In was shown that some properties of higher than 2nd order correlators are incompatible with classical physics even at $\beta=0$ in Eq.~(\ref{breve-def}). For example, the 3rd order correlator, measured at the thermodynamic equilibrium, may violate the 3rd order Onsager relations \cite{belzig-12prl}, and the 4th order correlator may break classical positivity constraints \cite{belzig-11njp}.
It should be kept in mind that experimental observations of such effects by SNS are yet to be
achieved. In our opinion, such observations are highly desirable to demonstrate
fundamental quantum mechanical phenomena at mesoscopic scale.

\section{Faraday Rotation}
\label{FR}

The optical methods used for  SNS do not always provide
a pure spin noise signal. Faraday rotation fluctuations can be sensitive to
other sources of physical fluctuations in a material, such as
fluctuations of the background charge in quantum dots and valley
polarization fluctuations in Dirac semiconductors. This section
explains some fundamental effects that justify the optical SNS.

\subsection{Faraday Effect}

In the optical SNS, a linearly polarized beam passes through a  volume of a material or a slab with atomic vapor. Polarization of the outgoing beam is rotated by an angle $\theta_F$. In magnetic materials, this angle is typically proportional to the magnetization of the illuminated region. On the other hand, in paramagnetic systems, this angle would be zero on average. Nevertheless, due to the thermally induced spin fluctuations, the total magnetization of the illuminated region fluctuates as $\delta S_z(t) \sim \sqrt{N}$ where $N$ is the number of spins in the observation volume. By analogy with the Faraday effect in ferromagnetic systems, one can expect that there is a linear dependence between the measured Faraday rotation signal and the instantaneous spin polarization:
\begin{equation}
\theta_F(t) = \alpha S_z(t),
\label{angle1}
\eeq
where $\alpha$ is a non-universal coefficient that depends on material and setup characteristics. The time correlators of the signal $\theta_F(t)$ can then be interpreted as correlators of  spin polarization times a setup specific constant.


 Let $k_{\pm} = \bar{k}\pm \Delta k/2$ be wave vectors in the material for left and right circularly polarized waves, respectively,
\begin{equation}
{\bf E}_{\pm} = E_0(\hat{\bf x} \pm i \hat{\bf y} )e^{-i(\omega t - k_{\pm} z)}.
\label{wave1}
\end{equation}
Note that directions of propagation and frequencies of the circular polarized components are the same in Eq.~(\ref{wave1}).
Suppose now that  the incident wave is linearly polarized along x-axis: ${\bf E }_{\rm in} ={\bf E}_{+}+{\bf E}_{-}$ and passes through a slab of a material of width $d$. The outgoing wave then is given by
\begin{eqnarray}
\nonumber  {\bf E}_{\rm out} &=& E_0e^{-i\omega t } \left[ \hat{\bf x}\left( e^{ik_{+} d} \right.\right. \\
&& \quad\quad\quad\quad \left.\left. +e^{ik_{-} d} \right) + i \hat{\bf y}  \left( e^{ik_{+} d} -e^{ik_{-} d} \right)  \right].
\label{wave11}
\end{eqnarray}
Leaving only terms linear in small  $\Delta k d \ll 1$, where $\Delta k \equiv (k_+ - k_{-}) $, we find:
\begin{equation}
{\bf E}_{\rm out} = E_0e^{-i\omega t +i\bar{k}d } (2\hat{\bf x} -   \Delta k d   \hat{\bf y} ),
\label{wave2}
\end{equation}
i.e. the out-going beam polarization is rotated by a small angle
\beq
\theta_F = \Delta k d/2 =( \omega d/2c) (n_+ - n_{-}),
\label{faraday1}
\eeq
where we used the fact that $k_{\pm}=\omega n_{\pm}/c$. Here $c$ is the speed of light, and $n_{\pm}$ are refraction coefficients for, respectively, clockwise ($+$) and counterclockwise ($-$) polarized beams.

For microscopic calculations, it is important to relate the difference $n_+ - n_{-}$ to the elements of  polarization tensor of the medium.
Consider, for simplicity,  the case of a uniaxial optical anisotropy, such that the measurement  $z$-axis coincides with the direction of the magnetization and the main optical axis of the system. The electromagnetic wave passing the sample induces the electric polarization
\begin{equation}
{\bf P}(\omega) = 4\pi \varkappa(\omega) {\bf E}(\omega).
\label{pol}
\end{equation}

The polarizability tensor $\varkappa$ generally has off-diagonal components, e.g., an electric field along the $x$-axis induces the  polarization along the $y$-axis. Disregarding dissipation, weak magnetic effects are described by such imaginary off-diagonal components in
\beq
\varkappa = \left(
\begin{array}{ccc}
\varkappa_{xx} & i \eta_{xy} & 0 \\
-i\eta_{xy} & \varkappa_{yy}  & 0 \\
0 & 0 & \varkappa_{zz}
\end{array}
\right),
\label{pol2}
\eeq
where, in the case of uniaxial anisotropy considered here, $\varkappa_{yy}=\varkappa_{xx}$.
More generally, the diagonal and off-diagonal components of the tensor (\ref{pol2})  have both real and imaginary parts. We refer to Ref.~\cite{FE1} for a more detailed discussion.
The tensor (\ref{pol2}) is diagonalized in the rotating basis, i.e.,
\beq
{\bf P}_{\pm} = 4 \pi (\varkappa_{xx} \mp \eta_{xy}) {\bf E}_{\pm},
\label{pol3}
\eeq
where $E_{\pm}$ are defined in (\ref{wave2}) and ${\bf P}_{\pm} \sim \left ( \hat{\bf x} \pm i \hat{\bf y} \right)$.
The electric susceptibility can be defined then separately for each circular polarization: $\varepsilon_{\pm} = 1+4\pi (\varkappa_{xx} \mp \eta_{xy})$.
Using $n_{\pm}=\sqrt{\mu_0 \varepsilon_{\pm}}$,  we find $n_{+}-n_{-} \approx -4\pi \eta_{xy}/\bar{n}$, where $\bar{n}=(n_++n_{-})/2$. Substituting the above into Eq. (\ref{faraday1}), we obtain

\beq
\theta_F \approx - 2\pi \eta_{xy}(\omega) \frac{\omega d}{c\bar{n}}.
\label{faraday2}
\eeq
The off-diagonal component $\eta_{xy}$ of the polarization tensor can be calculated microscopically by considering linear response of the charge polarization to the external electric field. There are differences in such calculations for conduction electrons and dielectric media.

\subsection{Faraday Rotation Fluctuations due to Conduction Electrons}

In the case of conduction electrons, one can associate the charge polarization with the current in the region: $\dot{P}_{\alpha}(t) = J_{\alpha}(t) =\int^t \sigma_{\alpha \beta }(t-t') E_{\beta}(t') \textnormal{d}t'$, where $\sigma_{\alpha \beta}$ are the elements of the conductivity tensor. Taking the Fourier transform we find $\eta_{xy}(\omega) = \sigma_{xy}(\omega)/\omega$ and, hence,
\beq
\theta_F= -2\pi \sigma _{xy}(\omega) \frac{d}{c\bar{n}}.
\label{faraday3}
\eeq
Equation~(\ref{faraday3}) is  an approximation valid in the linear order in magnetization and in the limit of weak beam absorption, which is usually the case for weak spin fluctuations near the thermodynamic equilibrium and when the beam is sufficiently detuned from resonant optical transitions. Absorption effects would lead to corrections to (\ref{faraday3}) that depend on $\sigma_{xx}$.

In paramagnetic materials, at zero external magnetic  field,
the off-diagonal conductivity is zero, i.e. $\sigma_{xy}=0$, which follows from the time reversal symmetry of the Hamiltonian.  On the other hand, in all conduction materials, conduction electrons experience the spin-orbit coupling.
This coupling leads  to the intrinsic AC spin Hall effect \cite{sinitsyn-04prl-2}, namely, electrons with spins $s_z=\pm 1/2$ deflect in transverse to the electric field direction, i.e. one can introduce spin-dependent conductivities $\sigma_{xy}^{\uparrow}$ and $\sigma_{xy}^{\downarrow}$ for up and down electronic spins, separately. The time-reversal invariance guarantees that $\sigma_{xy}^{\uparrow} = -\sigma_{xy}^{\downarrow}$ in a paramagnetic system on average. However, a local spin fluctuation creates an  imbalance of spins up and down. Consequently, the total charge Hall conductivity $\sigma_{xy} = \sigma_{xy}^{\uparrow}+\sigma_{xy}^{\downarrow}$ does not vanish, and for a weak spin fluctuation it is proportional to the instantaneous spin polarization: $\sigma_{xy} \sim S_z(t)$.

This mechanism is responsible for observation of Faraday rotation
fluctuations in conventional semiconductors, such as GaAs. However,
the spin orbit coupling is not always responsible for a nonzero Hall
conductivity. A notable example is the class of novel materials -
transition metal dichalcogenides (TMD)s \cite{sinitsyn-14prl1}, in which, in addition to
spins, conduction electrons possess the ``valley" discrete
degree of freedom.

Consider, e.g., the 2D semiconductor MoS$_2$ -- a prototypical group-VI dichalcogenide \cite{mos2-2}. A monolayer MoS$_2$ is a direct gap semiconductor with a band gap of approximately 1.8 eV. Its band structure is characterized by the conduction and valence-band edges located at the corners ($K$ and $K'$ points) of the 2D hexagonal Brillouin zone. A strong spin-orbit coupling due to the $d$ orbitals of the heavy metal atoms results in a significant spin-splitting of the valence band.
In the vicinity of $K$-points the Hamiltonian can be written as
\begin{equation}
\hat H=a t \left( \tau k_x \hat\sigma_x+k_y \hat\sigma_y\right)+\frac{\Delta}{2}\hat\sigma_z-\lambda \tau \ \frac{\hat\sigma_z-1}{2}\hat s_z,
\label{eq2_2_1}
\end{equation}
where $a$ is the lattice constant, $t$ is the effective hopping
integral, $\tau=\pm$ denotes the valley degrees of freedom $K$ and $K'$;
$\bf{\hat \sigma}$ are the Pauli matrices describing the sublattice degrees of freedom,
$\Delta$ is the bandgap, $2\lambda_\textnormal{SO}$ is the spin splitting at the valence band
caused by the spin-orbit coupling, and $\hat s_z$ is the Pauli matrix for spins.

Note that the spin is conserved by the Hamiltonian (\ref{eq2_2_1}). The energy dispersion of the Bloch bands for the same spin near each of the $K$-points
corresponds to a massive 2D Dirac band, as shown in Fig.~\ref{fig2_2_1}. Each of the electronic bands  has a nonzero Hall conductivity \cite{sinitsyn-06prl}.
However, the signs of such Hall conductivities  depend  on the valley index rather than spins. One can show that for the Hamiltonian (\ref{eq2_2_1}),  we have
\beq
\sigma_{xy}^{K\uparrow}  \approx \sigma_{xy}^{K\downarrow}, \,\,\, \sigma_{xy}^{K\uparrow} = - \sigma_{xy}^{K'\downarrow}, \,\,\, \sigma_{xy}^{K\downarrow} = - \sigma_{xy}^{K'\uparrow},
\label{cond1}
\eeq
where, e.g., $\sigma_{xy}^{K\uparrow} $ is the Hall conductivity of electrons in the $K$ valley having spins up  \cite{sinitsyn-14prl1}.

At the thermodynamic equilibrium, the number of electrons in $K$ and $K'$ valleys is, on average, the same so the total Hall conductivity of the material is, on average, zero.
However, intervalley scatterings and spin flips lead to the fluctuating imbalance of electrons in different bands,  and hence to the Faraday rotation.
The valley-dependent optical selection rules for interband transitions at $K$ and $K'$ points are shown schematically in Fig.~\ref{fig2_2_1}.
The frequency of the probe beam can be strongly detuned from the lower valence bands, while being almost in resonance between conduction electrons and the upper valence band.  Optical transitions are spin conserving, so   only electronic spins up in the $K$-valley and spins down in the $K'$ valley become optically sensitive for such a light frequency.
\begin{figure}[t]
\begin{center}
\includegraphics[width=0.7\linewidth]{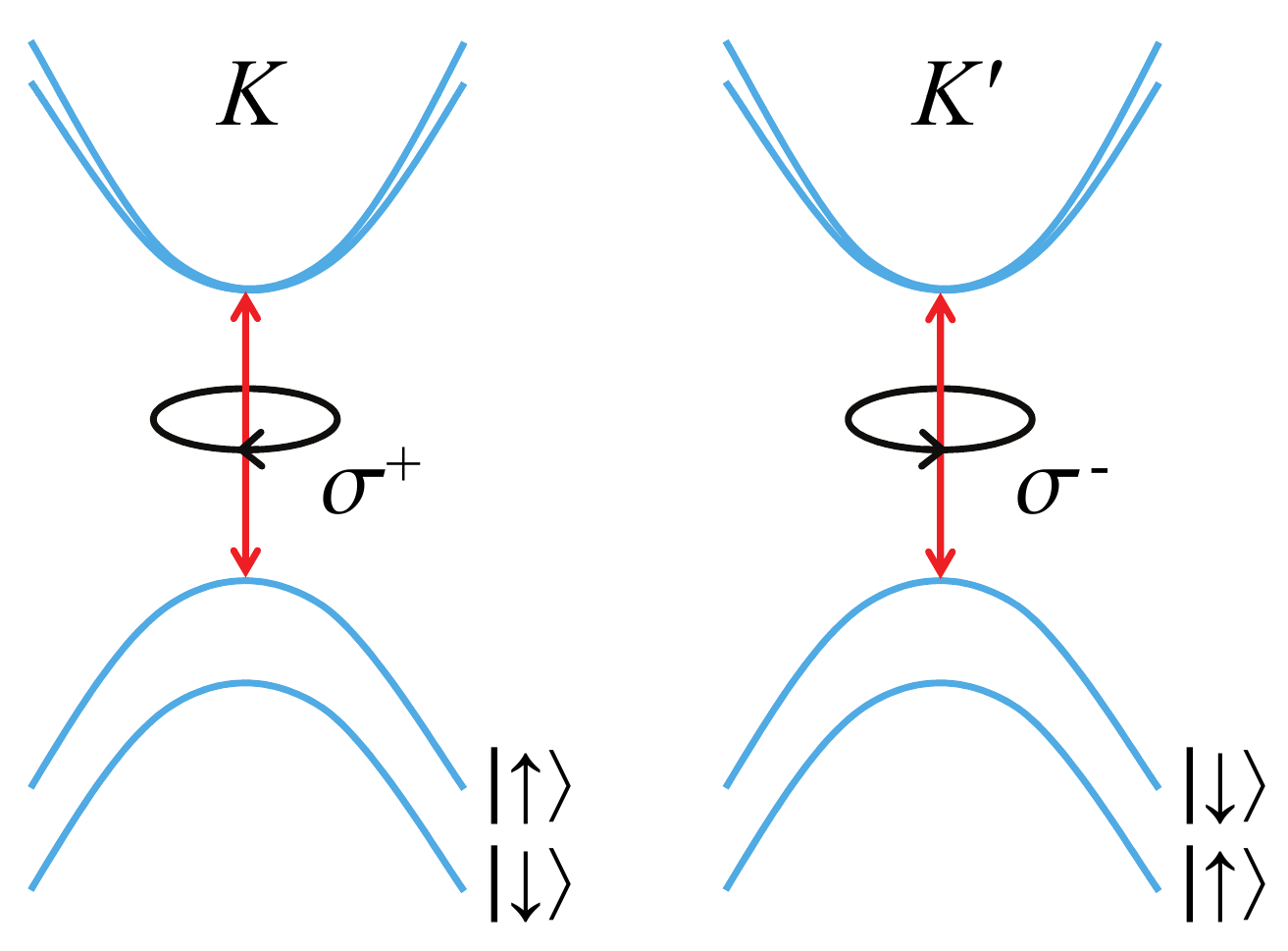}
\caption{Schematics of valley and spin optical transition selection rules for a circularly polarized light with a close to band-gap photon energy. \label{fig2_2_1}}
\end{center}
\end{figure}

As different valleys have different signs of the Hall conductivities, the corresponding Faraday rotation angle can be expressed as \cite{sinitsyn-14prl1,luyi-nat15}
\begin{equation}
\theta_F\sim N_{K\uparrow} - N_{K' \downarrow},
\label{eq2_2_4}
\end{equation}
where $N_{K \uparrow}$ is the excess number of electrons  in the observation region with spins up in the $K$ valley, and   $N_{K' \downarrow}$ is the excess number of electrons with spins down in the $K'$ valley.
It is now convenient to introduce the valley polarization, $\delta N_v= (N_{K \uparrow}+N_{K \downarrow}-N_{K' \uparrow}-N_{K' \downarrow})/2$, and the spin polarization, $\delta S_z=( N_{K \uparrow}-N_{K \downarrow}+N_{K' \uparrow}-N_{K' \downarrow})/2$, in terms of which the Faraday rotation angle is expressed as
\begin{equation}
\theta_F (t)\sim \delta N_{v} (t) + \delta S_z(t) .
\label{eq2_2_44}
\end{equation}

Since both $N_v$ and $S_z$ experience fluctuations, Eq.~(\ref{eq2_2_44}) shows explicitly that the Faraday rotation noise may not always be proportional to the pure spin noise, as it was suggested in Eq.~(\ref{angle1}). The optically measured noise power spectrum contains additional information here that corresponds to the dynamics of $N_v$. Fortunately, the contributions of spin and valley polarizations should be easy to distinguish because they differently respond to application of an external  magnetic field. Thus both types of fluctuations can potentially be explored by  optical SNS \cite{sinitsyn-14prl1}.

\subsection{Faraday Rotation from  Spins in Quantum Dots}
If a single electron is confined in a self-assembled semiconductor quantum dot, its wave function can be in an arbitrary superposition of two states:
\beq
|1/2 \ra =|E\rangle |\uparrow\rangle, \quad |-1/2 \ra =|E\rangle |\downarrow \rangle,
\label{e-band}
\eeq
 where $|E\rangle$ is the spatial component of the wave function.

For hole-doped quantum dots, in GaAs, the wave function is more complex. It can be represented as a superposition of states of the valence band with the total angular momentum equal to 3/2 \cite{Testelin}:
 \begin{eqnarray}
\label{hh-band}
|3/2,  3/2 \rangle &=& \frac{|X+iY\rangle}{\sqrt{2}} |\uparrow \rangle, \\
|3/2,  1/2 \rangle &=& \frac{|X+iY\rangle |\downarrow \rangle -2|Z\rangle |\uparrow \rangle }{\sqrt{6}}, \\
|3/2,  -1/2 \rangle &=& \frac{|X-iY\rangle |\uparrow \rangle +2|Z\rangle |\downarrow \rangle }{\sqrt{6}}, \\
|3/2,  -3/2 \rangle &=& \frac{|X-iY\rangle}{\sqrt{2}} |\downarrow \rangle,
\end{eqnarray}
where $|X\rangle$, $|Y\rangle$ and $|Z\rangle$ are orbital functions with symmetries of, respectively, $x$, $y$ and $z$.

The electronic states form a  Kramer's doublet and, hence, are degenerate.  In contrast, hole states in epitaxial InAs or InGaAs quantum dots  are split into the heavy hole (hh) and light hole (lh) doublets. So, one can speak about two characteristic splittings between hole and electronic states: $\Delta_{hh}$ and $\Delta_{lh}$ for, respectively, heavy and light hole states. The frequency of the measurement beam can be tuned close to the resonance with the hh-states so that lh-states can be considered optically inactive and irrelevant.

Heavy holes are predominantly made of states $|3/2,  \pm 3/2 \rangle $.  The time reversal invariance constrains  them to be in a superposition of the following two vectors:
\begin{eqnarray}
\nonumber \vert \psi_1 \rangle &=& \alpha (|3/2;  3/2 \rangle +\beta | 3/2;-1/2 \rangle +\gamma |3/2; 1/2 \rangle),\\
\label{psi12-1} \\
\nonumber \vert \psi_2 \rangle&=&\alpha( |3/2; -3/2 \rangle +\beta^* |3/2; 1/2 \rangle -\gamma^* |3/2; -1/2 \rangle),
\end{eqnarray}
where $\alpha=1/\sqrt{1+|\beta|^2+|\gamma|^2}$ ensures a proper normalization.  Nonzero values of parameters $\beta$ and $\gamma$ originate from  mechanical strains. The parameter $\beta$ is nonzero due to nonzero components   $\varepsilon_{xy}$ and $\varepsilon_{xx}-\varepsilon_{yy} $ of the strain tensor (Ref.~\cite{Testelin}), and the strain with $\varepsilon_{zx}-\varepsilon_{yz} \ne 0  $ corresponds to the  parameter $\gamma \ne 0$.  The relative sizes of these two types of strains are usually strongly different. In the bulk of 3-dimensional GaAs samples, $\varepsilon_{xx}-\varepsilon_{yy} $ is typically substantial, while  nonzero $ \varepsilon_{zx}-\varepsilon_{yz}$ can be induced in samples grown along an unusual crystal direction. Hence, the parameter $\gamma$ is typically assumed to be vanishingly small in the bulk of GaAs, while usually  $|\beta|^2 \sim 0.1$ (Ref.~\cite{Testelin}).

In the basis (\ref{psi12-1}),  the operator of the spin projection on the measurement axis has the following matrix form:
\begin{equation}
\hat{s}_{z}' = \frac{\alpha^2}{2} \left(
\begin{array}{cc}
1+\frac{|\gamma|^2-|\beta |^2}{3} & \frac{\gamma^* \beta^*}{3}\\
\\
\frac{\gamma \beta}{3} &- 1-\frac{|\gamma|^2-|\beta |^2}{3}
\end{array}
\right).
\label{sigmaz}
\end{equation}
This operator,  in the natural basis of states (\ref{psi12-1}), is not proportional to the Pauli $\hat{\sigma}_z$-matrix.
Hence,  the linear relation between the average spin and the Faraday rotation angle is not {\it a priori} obvious.

The optical beam field couples states of a single hole to the exciton states that consist of one electron and two holes with opposite spins.
We will assume that such an exciton state can be described similarly to the electronic state  (\ref{e-band}) with a different meaning of the spatial part of the wave function.
Since all states inside the quantum dot are localized, the matrix elements of the coordinate operators are well defined, e.g., 
\begin{eqnarray}
\label{matrix-el}
\langle E | \hat{x} |X\rangle &=&\langle E | \hat{y} |Y\rangle \equiv q \ne 0,\\
\langle E | \hat{x} |Y\rangle &=& \langle E | \hat{y} |X\rangle =\langle E | \hat{x} |Z\rangle=\langle E | \hat{y} |Z\rangle=0,
\end{eqnarray}
where $q$ is some constant parameter that characterizes the quantum dot.

Let us assume that the energy difference between the electron in the valence band and the exciton state is $\Delta_{hh}$.
Consider now the charge polarization induced by an ac-field ${\bf E}(t)={\bf y} E_y e^{i\omega t}$. The system is then  described by the Hamiltonian
\begin{equation}
\hat{H}=\hat{H}_0+ eE_ye^{i\omega t} \hat{y},
\label{ham}
\end{equation}
where $\hat{H}_0$ is the unperturbed Hamiltonian of the  quantum dot and $\hat{y}$ is the y-coordinate operator. The charge polarization of a quantum dot  in a state $|\Psi \rangle$, along the transverse to the electric field direction,
 is given by $P_x = e \la \Psi | \hat{x} | \Psi \ra$.  Using the linear perturbation theory, we find that the linear order in the electric field contribution to $P_x$ oscillates with the same frequency as the electric field:
\begin{equation}
P_x \approx \frac{-i e ^2E_y e^{i\omega t}}{\hbar}  \sum \limits_{s=\pm 1/2} \Im \left[ \frac{  \la\Psi | \hat{y} |s \ra  \la s | \hat{x}  | \Psi \rangle}{ \Delta_{hh}-\omega +i\Gamma } \right],
\label{faraday-dot1}
\end{equation}
where $s$ runs over localized exciton eigenstates (\ref{e-band}) of the Hamiltonian and $\Gamma$ is a phenomenological parameter describing broadening of the optical resonance. In derivation of (\ref{faraday-dot1}) we used the assumption that $\Delta_{hh} - \omega \ll \Delta_{hh}$, which has always been the case in the optical SNS applications.

\begin{figure}
\begin{center}
{\includegraphics[width=0.9\columnwidth]{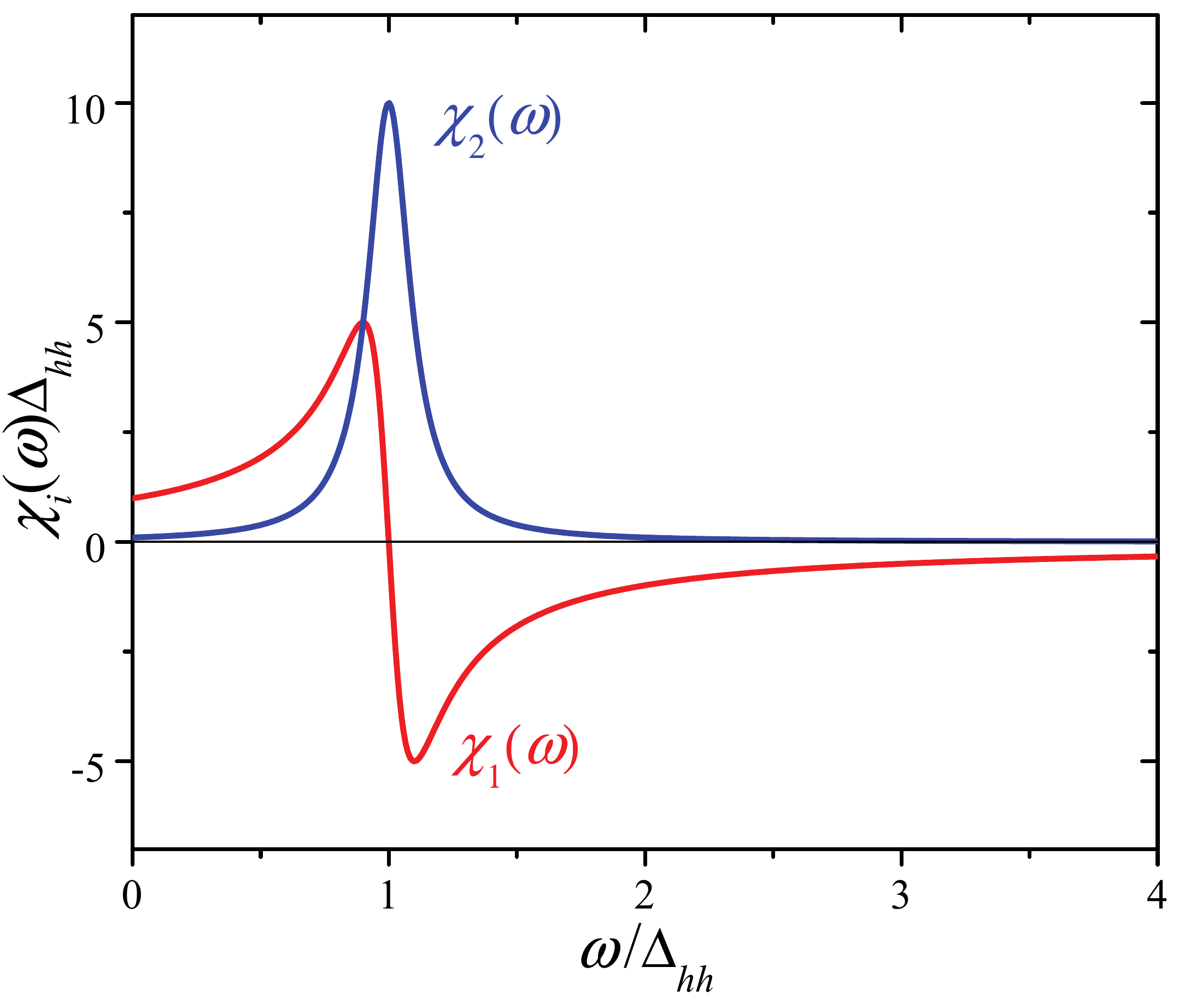}}
\end{center}
\caption{
Functions $\chi_1(\omega)$ and $\chi_2(\omega)$ (Eqs.~(\ref{chiw}), (\ref{chiww})) at $\Gamma/ \Delta_{hh}=0.1$. Far from the resonance ($|\omega-\Delta_{hh}| \gg \Gamma$), the real part $\chi_1(\omega)$ is much larger  in absolute value than the imaginary part $\chi_2(\omega)$.
  }
\label{fig4_1_chi}
\end{figure}
Using the relation (\ref{faraday2}) between the off-diagonal elements of the polarization tensor and the Faraday rotation angle, we find
\begin{equation}
\theta_{F} \sim \chi_1(\omega)  \sum \limits_{s=\pm 1/2} \Im   \left[  \la \Psi | \hat{x} |s \ra  \la s | \hat{y}  | \Psi \ra \right] ,
\label{faraday4}
\end{equation}
where 
\beq
\chi_1(\omega)= \frac{\Delta_{hh} -\omega}{ (\Delta_{hh}-\omega)^2+\Gamma^2},
\label{chiw}
\eeq
and where we disregarded dissipative effects that are proportional to a Lorentzian-like broadening at a single resonant transition:
\beq
\chi_2(\omega)= \frac{\Gamma }{ (\Delta_{hh}-\omega)^2+\Gamma^2},
\label{chiww}
\eeq
which decays as $\chi_2 \sim 1/(\Delta_{hh}-\omega)^2$ at large detuning, i.e. much faster than $\chi_1(\omega)$.

Consider now the wave function  in a superposition of pseudo-spin states (\ref{psi12-1}):
$$
|\Psi \ra = a \vert \psi_1 \rangle+ b\vert \psi_2 \rangle,
$$
with some coefficients $a$ and $b$.
One can verify now that
$$
\sum \limits_{s=\pm 1/2} \Im   \left[  \la \Psi | \hat{x} |s \ra  \la s | \hat{y}  | \Psi \ra \right]  = q^2  \la \Psi |\hat{s}'_z  | \Psi \ra,
$$
where parameter $q$ was defined in (\ref{matrix-el}).
Consequently, the contribution of the given quantum dot and its spin state to the Faraday rotation angle is given by
\begin{equation}
\theta_{F}  =Q \chi_1(\omega)   \la \Psi |\hat{s}'_z  | \Psi \ra ,
\label{faraday5}
\end{equation}
where $Q$ is a constant that depends on the dipole matrix elements between electronic and heavy hole states of the quantum dot.
Different dots contribute additively to the total observed Faraday rotation angle.
This calculation verifies that the Faraday rotation is proportional to the spin polarization of the dots.

The beam frequency dependence of this effect is described by the function $\chi_1(\omega)$ in (\ref{chiw}), which we show in Fig.~\ref{fig4_1_chi}. Generally, $\chi_1(\omega)$ has a more complex form due to strong Coulomb interaction in an electron-hole excitation, however, it is a rather general property that $ \chi_1(\omega)$ decays as $\sim 1/(\Delta_{hh}-\omega)$ at large detuning from the resonance \cite{sinitsyn-KK}.  This is in contrast with dissipative  absorption processes that are usually described by Gaussian or Lorentzian functions that decay much faster.
Hence, the fact that  the Faraday rotation decays slowly with detuning from the resonance allows the probe beam frequency choice at which absorption of energy from the beam  is tuned below a desired level.


The results of this subsection demonstrate that  SNS is justified by
the sensitivity of the Faraday rotation to the spin state in a quantum
dot.  Interestingly, a similar line of arguments was  used to design another
measurement technique \cite{yakoby-13prl,charge-noise} that, instead of spins, probes charge fluctuations in the
vicinity of a quantum dot. Such fluctuations occur relatively slowly
(below 1MHz frequencies) and usually do not interfere with
spin noise. This technique employs the fact that  electrostatic time-dependent potential produced by
such fluctuations modulates the size of the optical gap
$\Delta_{hh}$.




\subsection{Spin Noise as a Probe of a Homogeneous Line-Width of an
  Optical Transition}
Spin noise can be used not only to study spin interactions.
It was shown in \cite{two-color-1} that it can also be used as an alternative
probe of optical characteristics of electronic systems, which previously could be studied only by much more invasive nonlinear optical methods.

In the previous subsection, we discussed that the Faraday rotation of
a beam with a frequency $\omega$ has resonant character.
The interaction of a spin system with an optical beam is described by the response
function $\chi(\omega)=\chi_1(\omega)+i\chi_2(\omega)$. The imaginary part, $\chi_2(\omega)$, is responsible for an absorption
peak, similar to the Lorentzian in Eq.~(\ref{chiww}), and the Faraday
rotation is proportional to the real part $\chi_1(\omega)$, such as in Eq.~(\ref{chiw}).
 Imaginary
and real parts of $\chi(\omega)$ are not independent. They can be expressed in terms of each other via the Kramers-Kronig relations \cite{sinitsyn-KK}.

\begin{figure}
\centerline{\includegraphics[width=0.99\columnwidth]{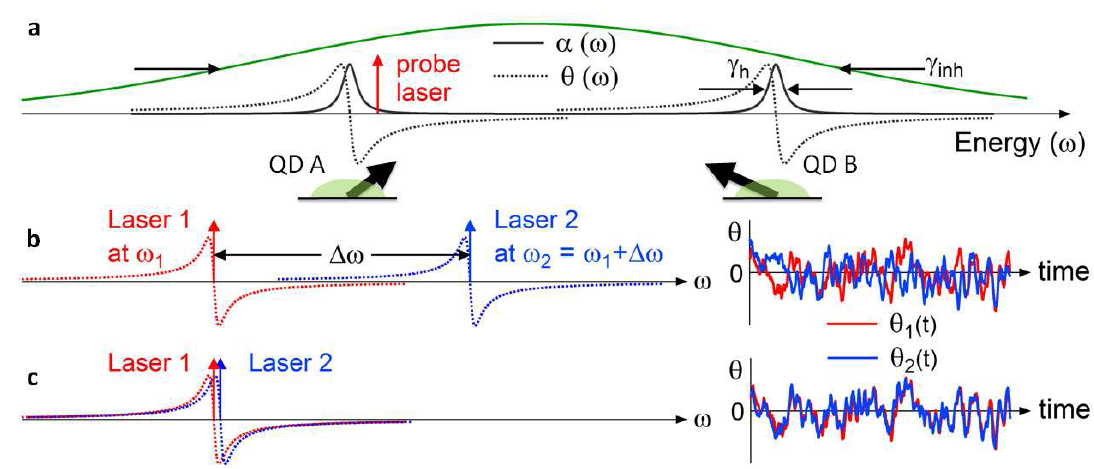}}
\caption{
Two-color spin noise spectroscopy of a quantum dot ensemble \cite{two-color-1}.
(a) Gaussian inhomogeneous broadening of the absorption resonance in an ensemble of quantum dots (green) compared with homogenous broadening of absorption and Faraday Rotation of  single dots (QD A and QD B) probed by beams with different frequencies (respectively, solid and dotted black curves). (b-c) Comparison of the Faraday rotation fluctuations in beams detuned from each other by a large frequency difference $\Delta \omega$ and two beams without detuning (see Ref. \cite{two-color-1} for more details). Reprinted with permission from Macmillan Publishers Ltd: Nature  Communications 5, 4949, 2014, Copyright \copyright (2014) \cite{two-color-1}.
  }
\label{fig3_9}
\end{figure}
A single quantum dot has a relatively sharp resonance, described by the
absorption curve $\chi_2(\omega)$ with a characteristic width
$\gamma_h$, however, due to the disorder in shapes of quantum dots, there is
a wide range of optical frequencies of the probe beam at which  spin
noise can be detected, as illustrated in Fig.~\ref{fig3_9}(a).  In
\cite{crooker-13prl}, such a  strong inhomogeneous
broadening of the resonant frequency was observed in an ensemble of InGaAs quantum dots.

One can ask a question whether it is possible to determine the
homogeneous broadening of a single dot when we can
probe only many quantum dots simultaneously within the
domain of linear optics. It was shown in \cite{two-color-1} that this
question can be resolved positively
if one can detect fluctuations of the optical
signal that are induced by spin fluctuations.

To illustrate the idea of Ref.~\cite{two-color-1}, let us formalize
the problem. Consider an abstract elementary system, such as a
molecule or a quantum dot, which presence can be detected by a probe
beam with the frequency $\omega$.
The system can be detected by  changes in some characteristic $X$, which can be a detector voltage signal  proportional to some characteristic of
the optical beam, e.g., the rotation of the beam polarization.
If interaction with the beam is weak, then $X$ is generally proportional  to the
intensity of the probe beam. We assume that the elementary system, e.g., a single quantum dot, couples noticeably to the beam at frequencies around
some resonant value $\omega_0$ so that we can write a linear relation:
\begin{equation}
X(\omega)=f(\omega_0-\omega) I(\omega),
\label{lin1}
\end{equation}
where the function $f(y)$ is peaked at $y=0$ and decays at a characteristic frequency $\gamma_h$. The latter is called the {\it homogeneous broadening}.

If there is only one quantum dot in the observation region, then
$\gamma_h$ can be determined simply by scanning the
response to different probe beam frequencies $\omega$. However, if we
 deal with a large ensemble of similar dots characterized by a distribution of $\omega_0$, then
the linear response changes to
\begin{equation}
X(\omega)=\int \textnormal{d}\omega_0 \, \rho(\omega_0) f(\omega_0-\omega) I(\omega),
\label{lin2}
\end{equation}
where $\rho(\omega_0)$ is the density, per unit frequency, of elementary systems with
a resonant frequency $\omega_0$. Quite often, it happens that the homogeneous broadening $\gamma_h$ is much smaller than the characteristic width of the {\it inhomogeneous} distribution $\rho(\omega_0)$. In such a case, one can approximately evaluate (\ref{lin2}) as
\begin{equation}
X(\omega)= g \rho(\omega) I(\omega),
\label{lin3}
\end{equation}
where $g=\int \textnormal{d}y \, f(y)$ is a constant that, alone, does not contain enough information to determine $\gamma_h$.
Equation (\ref{lin3}) shows that, for a large inhomogeneous broadening, the linear response characteristics cannot provide information about $\gamma_h$. Neither this information can be obtained if the ensemble of systems is probed by more than one beam at different frequencies, e.g., if we probe the system by two beams at frequencies $\omega_1$ and $\omega_2$ then

\begin{equation}
X= g [\rho(\omega_1) I(\omega_1)+\rho(\omega_2) I(\omega_2)],
\label{lin4}
\end{equation}
which again is not giving information about $\gamma_h$.

Now, imagine that $X$ is again the sum of responses from two beams at
frequencies $\omega_1$ and $\omega_2$ but
the function $f(\omega_0-\omega)$ is no longer time-independent but rather
experiences random fluctuations with time. For example, we know that in the case of
quantum dots  spin fluctuations induce fluctuations of the
Faraday rotation. In such a case, the probed characteristic $X$ will also be fluctuating. Let us
assume that such fluctuations are  independent for different quantum dots
in the ensemble.

Curiously, even when the linear laws (\ref{lin1}) and (\ref{lin4}) are
valid, by measuring  fluctuations of $X$ we can actually obtain the
information about $\gamma_h$. When the ensemble is probed by two
beams, the time averaged signal $\la X \ra$ is merely obtained from (\ref{lin4}) with $g\rightarrow \la g \ra$. However, the variance of $X$ appears to be nonlinear in probe beam intensities and
contains a  cross-correlation contribution:
\begin{eqnarray}
\nonumber {\rm var}(X) &\equiv&  \la [X(t)]^2\ra - \la X(t)\ra^2 \approx g_{11} \rho I^2(\omega_1) +\\
&+& g_{22} \rho I^2(\omega_2)  + g_{12}(\omega_1-\omega_2) \rho I(\omega_2)I(\omega_1),
\label{var0}
\end{eqnarray}
where dependence of coefficients $g_{11}$ and $g_{22}$ on frequencies of the probe beams, as well as the difference between $\rho(\omega_1)$ and $\rho(\omega_2)$,
can be disregarded if $\delta \omega \sim \gamma_h$, where $\delta
\omega\equiv \omega_1-\omega_2$. The
function $ g_{12}(y) $ generally decays at
$y\sim \gamma_h$, so it is sensitive to the difference between beam frequencies.

For example, the measured Faraday rotation  in \cite{two-color-1}
could be described by $f(\omega) = \alpha \chi_1(\omega_0-\omega)$, where $\chi_1(\omega)$ was the real part of
the response function and $\alpha$ was the fluctuating spin noise signal with some $\la \alpha \ra$ and ${\rm var}(\alpha)$.
Using the fact that the variance of the sum of signals from
independent quantum dots is the sum of variances from individual dots
(which is a good assumption for well separated quantum dots) one can find that
\begin{equation}
\label{var3-1}
g_{11} =g_{22} = {\rm var}(\alpha)  \int \textnormal{d}y \chi_1^2(y),
\end{equation}
\begin{equation}
g_{12} =2{\rm var}(\alpha)  \int \textnormal{d}y \, \chi_1(y)\chi_1(y- \delta \omega).
\label{var3}
\end{equation}
Eq.~(\ref{var3}) shows that $g_{12}$ is suppressed when $\delta \omega > \gamma_h$.
The above expression for $g_{12}$ is written in terms of the
real component of $\chi(\omega)$. Using the Kramers-Kronig relation
\begin{equation}
\chi_1(y) = \frac{1}{\pi}  \mathcal{P} \int \frac{\chi_2 (z)}{y-z} \, \textnormal{d}z,
\label{KK-1}
\end{equation}
and the identity \cite{pecselin-book}
\begin{equation}
\mathcal{P} \int \frac{\textnormal{d}y}{(y-z_1)(y-z_2)} = \pi^2\delta (z_1-z_2),
\label{KK-2}
\end{equation}
 we find that
$ \int \textnormal{d}y\, \chi_1(y)\chi_1(y- \delta \omega) =  \int
 \textnormal{d}y\, \chi_2(y)\chi_2(y- \delta \omega)$, so
\begin{equation}
g_{12}  \sim  \int \textnormal{d}y\, \chi_2(y)\chi_2(y- \delta \omega),
\label{KK-3}
\end{equation}
i.e. even though the experiment measures the Faraday
rotation, the correlator $g_{12}$ can be expressed via the dissipative
part $\chi_2(\omega)$ of the optical response function.

Consider the Lorentzian broadening
\begin{equation}
\chi_2(\omega) = \frac{\gamma_h}{(\omega-\omega_0)^2+\gamma_h^2}.
\label{lor-ex}
\end{equation}
Substituting (\ref{lor-ex}) into (\ref{KK-3}) one can find that $g_{12}$ is of the Lorentzian form  \cite{two-color-1}
\begin{equation}
g_{12}(\delta \omega) \sim \frac{2\pi \gamma_h}{(\delta \omega)^2+4\gamma_h^2}.
\label{lor-ex2}
\end{equation}

As another example, consider that individual quantum dots have a Gaussian absorption due to, e.g., substantial fluctuations of the resonance frequency $\omega_0$  caused by charge  fluctuations in the vicinity of the dots. Then,
\begin{equation}
\chi_2(\omega) = \frac{\gamma_h}{\pi} e^{-\gamma_h^2 (\omega-\omega_0)^2},
\label{gauss-ex}
\end{equation}
and
\begin{equation}
g_{12}(\delta \omega) \sim \frac{ \gamma_h}{\sqrt{2\pi}}
e^{-\frac{\gamma_h^2 (\delta \omega)^2}{2}},
\label{gauss-ex2}
\end{equation}
i.e. we find the Gaussian form of $g_{12}$.
In the experiment  \cite{two-color-1}, a Lorentzian shape of $g_{12}(\delta
\omega)$ was observed (Fig.~\ref{fig3_9a}), which was also used to
determine the homogeneous broadening $\gamma_h$ in hole-doped InGaAs quantum dots.

\begin{figure}
\centerline{\includegraphics[width=0.99\columnwidth]{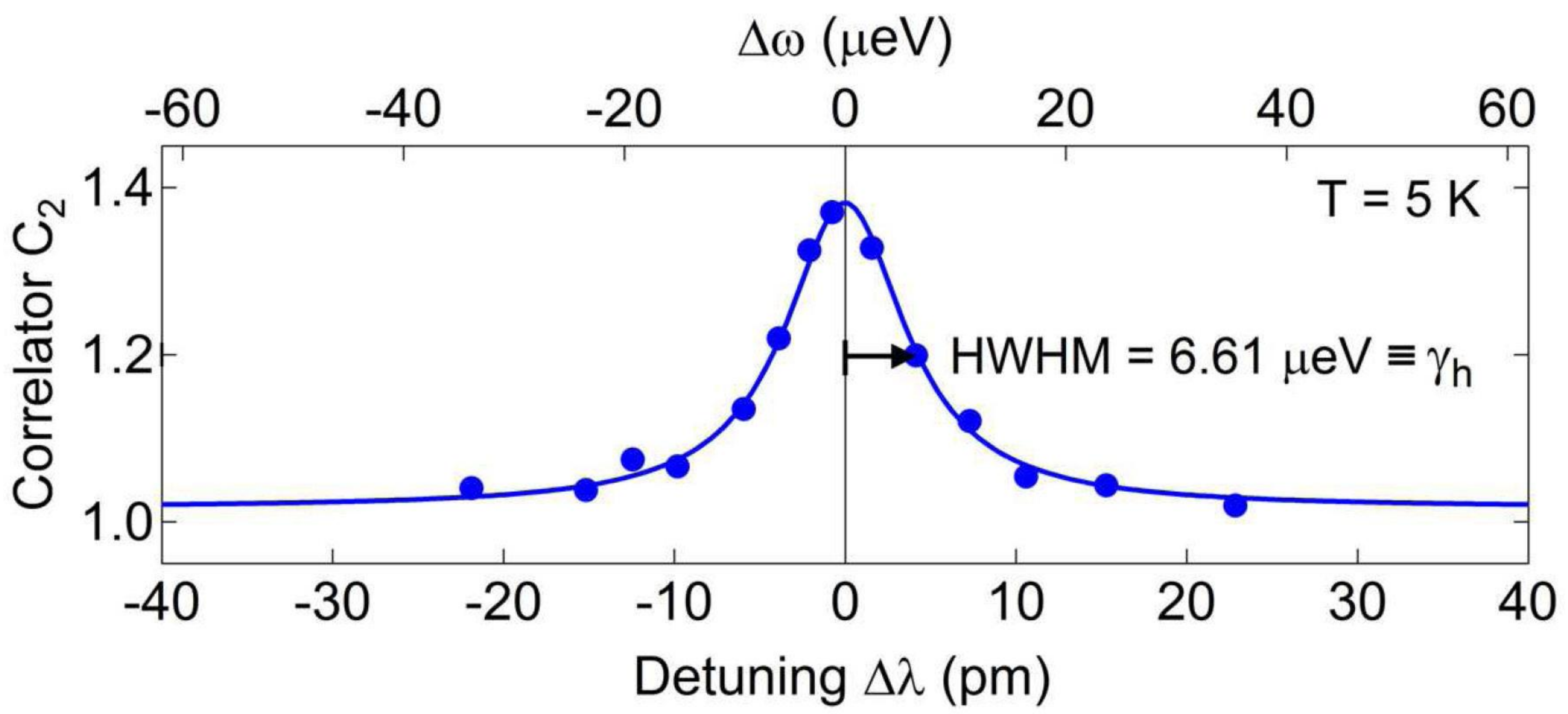}}
\caption{Two-color spin noise correlator as a
function of the detuning between the probe lasers \cite{two-color-1}. The
solid line is a Lorentzian fit.
Reprinted with permission from Macmillan Publishers Ltd: Nature  Communications 5, 4949, 2014, Copyright \copyright (2014) \cite{two-color-1}.
  }
\label{fig3_9a}
\end{figure} 
\section{Thermodynamic Constraints}

The statistics of spontaneous fluctuations at the thermodynamic equilibrium should be in agreement with the laws of thermodynamics. One of such laws is the existence of the thermodynamic equilibrium, which is described by the Boltzmann distribution of microstate probabilities at a given
temperature. In this section, we discuss some notable consequences
of such constraints on spin noise correlators that can be observed experimentally.

\subsection{Sum Rules for Integrated Noise Power}

Consider the total area $A_{\alpha \beta}$ under a curve representing the noise power spectrum for any two spin variables:
\beq
 A_{\alpha \beta} \equiv \int\limits_{-\infty}^{\infty} P_{\alpha \beta} (\omega) \, \textnormal{d}\omega,
\label{area1}
\eeq
where
\beq
 P_{\alpha \beta} (\omega) =\int\limits_{-\infty}^{\infty} e^{i\omega t} \la S_{\alpha}(t) S_{\beta}(0) \ra  \, \textnormal{d}t.
\label{area2}
\eeq
Here, the indices $\alpha,\beta=x,y,z$ can be either the same or different. Moreover, the spin variables in Eq.~(\ref{area2}) can even correspond to spin components of different electrons or atoms. Substituting Eq.~(\ref{area2}) into Eq.~(\ref{area1}), the integral over $\om$  results in the time delta function removed by the integration over $t$. The result is
\begin{equation}
A_{\alpha \beta} = 2\pi \la S_{\alpha}(0)S_{\beta}(0)  \ra.
\label{area3}
\end{equation}

Thus, we have found that the area under the noise power spectrum does not depend
on the dynamics of spin fluctuations in the sense that it depends only on the equal time
correlator of spin polarization components. This correlator is a
thermodynamic characteristic. Its value at the thermodynamic
equilibrium can be found from the knowledge of the  free energy of
the system as a function of the static external magnetic field.

SNS experiments are often performed in
relatively weak magnetic fields (below 0.1T). At typical temperatures of spintronics
experiments (above 3K), the corresponding Zeeman splitting
energy is then about two orders of magnitude smaller than $k_BT$. Therefore, while the spin dynamics can be very sensitive to such fields,
the probabilities of microstates do not change
substantially, and, consequently, such external fields do not affect the equal time correlators,  such as the one given by Eq.~(\ref{area3}).
This observation leads to the following approximate but valuable sum
rule:

\underline{Area Conservation Rule}: {\it The area under the noise power spectrum curve
remains  unchanged after the application of an external magnetic field or any other static perturbation
if the corresponding coupling energy, such as Zeeman level splitting, is much smaller than $k_BT$. }

This rule is quite handy,
e.g., it helps making a quick check of a result of lengthy theoretical calculations or to
interpret experimental data. We note again that the application of a weak magnetic field can
change the spectrum $P_{\alpha \beta} (\omega)$ considerably, e.g. by
shifting the peaks of the noise power to  new Larmor frequencies.
At the same time, such changes do not significantly affect the area under the spectral curve.

In hot atomic gases, the  temperature energy scale is much
larger than energies of  characteristic spin interactions. In such
situations, all relevant spin microstates can be considered as equally
probable and statistically independent. Moreover, for any state with
a spin polarization $S_{\alpha}$, there is an equally probable state with
spin polarization $-S_{\alpha}$. This leads to the following
rule:

\underline{No-Go Rule}: {\it At the thermodynamic equilibrium and in
  the limit of  large temperature, the integrated
  noise power spectrum of cross-correlators} ($\alpha \ne \beta$ {\it in Eq.} (\ref{area2})), {\it is zero. }

This rule restricts our
ability to use the total area of the noise power spectrum for studying
cross-correlations when the temperature exceeds characteristic energy
scales of spin dynamics. In other words, in cross-correlators,
the useful information  is contained
only in the functional form of $P_{\alpha \beta}(\om)$.
The corrections to the area conservation and no-go rules are of the order of $gB/k_BT$, where $B$ is the magnetic field strength and $g$ is the corresponding $g$-factor. In the case of warm alkali vapors, where $T\sim 400$K but Zeeman and hyperfine energies are $\lesssim 1$K in the temperature scale, the no-go sum rule holds with a high accuracy.

\subsection{Fluctuation-Dissipation Theorem}
The fluctuation-dissipation theorem is another important result that
is frequently used in theoretical calculations. It relates the
noise power spectrum  to linear response characteristics
\cite{pecselin-book, kogan-book}. In what follows, this connection is
explicitly demonstrated.

The quantum-mechanically justified expression for the noise power spectrum is given by
\beq
P(\omega) = \int_{-\infty}^{\infty} \textnormal{d}\tau e^{i\omega \tau} {\rm Tr} \left(\hat{\rho} \frac{1}{2}\{\hat{S}_z (\tau),\hat{S}_z(0) \}  \right),
\label{noiseP1}
\eeq
where $\{...,...\}$ is the anti-commutator, and $\hat{\rho}$ is the density matrix at the thermodynamic equilibrium.
The probability  $w_m$ of a  microstate $m$ can be written as
\beq
w_m=e^{\frac{F-E_m}{k_BT}}.
\label{wm1XX}
\eeq
Here, $F$ is the free energy and $E_m$ is the energy of the microstate. One can now rewrite Eq.~(\ref{noiseP1}) as
\begin{eqnarray}
\nonumber
P(\omega) = \frac{1}{2} \int_{-\infty}^{\infty} \textnormal{d}\tau \\
\quad \sum \limits_{mn} w_m \left( S_{mn}(\tau) S_{nm}(0) +S_{mn}(0) S_{nm} (\tau) \right) e^{i\omega \tau},
\label{noiseP111}
\end{eqnarray}
where, in terms of eigenstates of the unperturbed Hamiltonian,
$$
S_{mn}(\tau) \equiv \la u_m| \hat{S}_z (\tau) |u_n \ra.
$$
Note that since $ \hat{S}_z (\tau)  = e^{i\hat{H}_0 \tau/\hbar}  \hat{S}_z (0) e^{-i\hat{H}_0 \tau/\hbar} $,
$$
S_{mn}(\tau) = e^{-i\omega_{nm} \tau} S_{mn}\quad \textnormal{with} \quad \omega_{nm} = \frac{E_n-E_m}{\hbar}.
$$
Here we also highlight that in this subsection we do not use the convention $\hbar=1$ adopted in the most other parts of the review. This is related to the fact that the quantum and classical versions of the fluctuation-dissipation theorem look differently. The classical version is obtained from the quantum one in the limit $\hbar \omega \ll k_BT$, in which the Plank constant cancels out, as it will be
clear from the final result.

Substituting the above relations into Eq.~(\ref{noiseP111}), and performing the integration over $\tau$ we find
$$
P(\omega) =  \pi \sum \limits_{mn} w_m  |S_{mn}|^2  \left(\delta (w-\omega_{nm}) +\delta(\omega+\omega_{nm}) \right) =
$$
$$
=  \pi \sum \limits_{mn} (w_m+w_n)  |S_{mn}|^2 \delta(\omega- \omega_{nm}).
$$
Finally, using $w_n = e^{-\hbar \omega_{nm}/(k_BT)} w_m$, the following expression for the noise power spectrum can be obtained
\beq
P(\omega) =  \pi   \sum \limits_{mn} w_m \left(1+e^{-\frac{\hbar \omega_{nm}}{k_BT}}  \right) |S_{mn}|^2 \delta(\omega- \omega_{nm}),
\label{noiseP2}
\eeq
which, taking into account the time delta function, can be rewritten in the equivalent form
\beq
P(\omega) =  \pi \left(1+e^{-\frac{\hbar \omega}{k_BT}}  \right)  \sum \limits_{mn} w_m  |S_{mn}|^2 \delta(\omega- \omega_{nm}).
\label{noiseP2}
\eeq

Consider now the following expression:
\beq
C(\omega) \equiv \frac{1}{2\hbar} \int_{-\infty}^{\infty} \textnormal{d}\tau e^{i\omega \tau} \la [ \hat{S}_z(\tau), \hat{S}_z(0) ] \ra,
\label{corrC1}
\eeq
where the square brackets are the standard commutator of operators. Going through the same steps, we find

\bea
\nonumber C(\omega) = \frac{1}{2\hbar} \int_{-\infty}^{\infty} \textnormal{d}\tau e^{i\omega \tau}   \sum \limits_{mn} w_m  \left(e^{-i\omega_{nm}\tau} - e^{i\omega_{nm}\tau} \right) = \\
=\frac{\pi}{\hbar} \left( 1- e^{-\frac{\hbar \omega}{k_BT}} \right)  \sum \limits_{mn} w_m |S_{mn}|^2 \delta (\omega - \omega_{nm}).
\label{noiseC2}
\eea
Comparing (\ref{noiseP2}) and (\ref{noiseC2}), we obtain
\beq
P(\omega) =  \hbar \; {\rm coth} \left( \frac{\hbar \omega}{2 k_BT} \right) C(\omega).
\label{noisePC1}
\eeq

Let us now provide the physical interpretation of $C(\omega)$.
Consider the linear response of the spin system to a time-dependent magnetic field $h_z(t)$ applied along the $z$-axis.
The Hamiltonian term describing this interaction is $\hat{H}_h(t) = -h_z(t) \hat {S}_z $. Up to the linear order in $h(t)$, we have
\bea
\nonumber && \langle \hat{S}_z(t) \rangle = \\
\nonumber && \quad\quad =\langle e^{(i/\hbar) \int_{-\infty}^t \hat{H}_h(t') \textnormal{d}t'}  \hat{S}_z(t) e^{ -(i/\hbar) \int_{-\infty}^t \hat{H}_h(t') \textnormal{d}t'} \rangle \approx \\
 &&  \quad\quad \approx \frac{i}{\hbar} \int_{-\infty}^{t} \textnormal{d}t' h_z(t')  \langle [\hat{S}_z(t),\hat{S}_z(t')] \rangle.
 \label{noiseC3}
\eea

 Let us now introduce the {\it linear response function} $A_{zz} (t)$, such that
$$
 A_{zz}(t)=0\,\,\,\ {\rm for} \,\,\, t<0,
$$
and
\beq
 \langle \hat{S}_z(t) \rangle = \int_{-\infty}^{t} \textnormal{d}t' A_{zz}(t-t') h_z(t').
\label{response1}
\eeq
Comparing (\ref{noiseC3}) and (\ref{response1}) we find
\beq
A_{zz}(\tau) =\frac{i}{\hbar} \theta(\tau)   \langle [\hat{S}_z(\tau),\hat{S}_z(0)] \rangle.
\label{response2}
\eeq
Hence,
\beq
\int_0^{\infty} \textnormal{d} \tau e^{i\omega \tau}  \langle [\hat{S}_z(\tau),\hat{S}_z(0)] \rangle  = -i\hbar A_{zz}(\omega).
\label{aw1}
\eeq
Note also that
\bea
\nonumber  \int_{-\infty}^{0} \textnormal{d} \tau e^{i\omega \tau} \langle [\hat{S}_z(\tau),\hat{S}_z(0)] \rangle = \\
\nonumber  \quad\quad\quad =  - \int_{0}^{\infty} \textnormal{d} \tau e^{-i\omega \tau} \langle [\hat{S}_z(\tau),\hat{S}_z(0)] \rangle = \\
\quad\quad\quad = i\hbar A_{zz}(-\omega)=-i\hbar A_{zz}^*(\omega),
\label{aw2}
\eea
where we used the fact that, due to the translation in time invariance of the equilibrium correlator, $ \langle [\hat{S}_z(\tau),\hat{S}_z(0)] \rangle = -\langle [\hat{S}_z(-\tau),\hat{S}_z(0)] \rangle$, and that $A_{zz}(t)$ is real valued.
Substituting (\ref{aw1}) and (\ref{aw2}) into (\ref{corrC1}) and then (\ref{noisePC1}), we find:

\bea
\nonumber P(\omega) &=& -\frac{i}{2}  \hbar \left(A_{zz}(\omega) - A^*_{zz}(\omega) \right) {\rm coth} \left( \frac{\hbar \omega}{2 k_BT} \right)= \\
&=& \hbar \Im \left[ A_{zz}(\omega) \right] {\rm coth} \left( \frac{\hbar \omega}{2 k_BT} \right).
\label{FDT1}
\eea
Equation~(\ref{FDT1}) is the famous fluctuation-dissipation theorem stating that the equilibrium noise power spectrum is related to the imaginary part of the response function.

\subsection{Higher Order Fluctuation Relations}

Higher order spin correlators also satisfy their own fluctuation relations that connect these correlators to nonlinear
response characteristics and lower order cumulants at
nonequilibrium conditions. Such relations encountered, for example, in the theory of spin glass dynamics \cite{spin-glass-hoc}.

Nonlinear and nonequilibrium thermodynamics is currently a highly
active field of research because profound symmetries were identified at the level of the full counting
statistics that became known as {\it Fluctuation Relations}. These are
exact formulas that hold true even in systems that are driven
arbitrarily far away from the thermodynamic equilibrium
\cite{bochkov-81,jarzynski-97prl,sinitsyn-11pre,sinitsyn-11jstat,sinitsyn-11jpa,feldman-15}. In particular,
 fluctuation relations for spin currents were considered in \cite{feldman-15}. The experimental verification
of fluctuation relations  can provide information about system parameters that cannot be easily extracted by standard approaches \cite{sinitsyn-11prb}.

Numerous relations between  higher order cumulants
and  nonlinear response characteristics have been reviewed in
\cite{stratonovich-book1,stratonovich-b2}. Specifically, the simplest of the known fluctuation
relations, beyond the standard fluctuation-dissipation theorem, relate
the response of the spin correlator to a time-dependent perturbation
with the nonlinear response of the average spin polarization.  Since
SNS beyond the thermodynamic equilibrium and linear response has already
been successfully demonstrated in \cite{crooker-noneq1}, it is likely that
some of the predictions of nonlinear thermodynamics will be
tested in the near future. Below, we discuss
two examples that are valid in the domain of classical
overdampped stochastic dynamics.

\subsubsection{Higher Order Onsager Relations.}
Onsager relations are equalities between different cross-correlators of variables at the thermodynamic equilibrium \cite{stratonovich-book1}.

Consider a mesoscopic classical interacting spin system with Markovian stochastic
evolution among $N$ discrete states. Let ${\bf p}(t)$ be the vector of probabilities of all
possible microstates of the spin system. Each classical microstate $|i\ra$ is
characterized by the energy $E_i$ and eigenvalue of the spin
polarization operator
$\hat{s}_{\alpha}$, where $\alpha =x,y$ or $z$, or other spin indexes in the  case of a multicomponent spin system.
Note that projections of the classical spin commute with each other.

The Markovian evolution is described by the master operator $\hat{H}$ according to
\beq
\frac{\textnormal{d}}{\textnormal{d}t} {\bf p} = \hat{H} {\bf p}.
\label{markov1}
\eeq
Section~4 of Ref.~\cite{stratonovich-b2} explains the derivation of the master equation from microscopic Hamiltonian equations of motion.
At the thermodynamic equilibrium, kinetic rates satisfy the detailed
balance condition that can be included by writing the elements of the  Liouville operator
matrix in the Arrhenius parametrization \cite{sinitsyn-08prl,jarzynski-08prl}: $H_{ij} =
ke^{\beta (E_j-W_{ij})}$, $H_{jj}=-\sum_{i \ne j}  H_{ij}$, where $W_{ij}=W_{ji}$ are real parameters describing the ``barriers" between the states, $k$ is a constant real
parameter, and $\beta = 1/(k_BT)$. This parametrization guarantees that
the evolution converges to the Boltzmann  distribution $\left[ {\bf
  P}_{eq}\right]_i =Ce^{-\beta E_i}$, where $C=\left( \sum_{i=1}^N e^{-\beta
    E_i} \right)^{-1}$ is the normalization constant. Obviously, the operator $\hat{H}$ satisfies
the condition
\beq
\hat{H}^{\dagger} =e^{\beta \hat{E}} \hat{H} e^{-\beta \hat{E}},
\label{transv1}
\eeq
where $\hat{E}={\rm diag} \{ E_1,\ldots, E_N \}$.

The spin correlator $C_{\alpha_n\ldots \alpha_1} \equiv \la s_{\alpha_n}(t_n) \ldots s_{\alpha_1}(0) \ra $ is then given by
\beq
C_{\alpha_n\ldots \alpha_1}  =\la \hat{\bf 1} |
\hat{s}_{\alpha_n} e^{\hat{H}(t_n-t_{n-1})}  \ldots \hat{s}_{\alpha_2} e^{\hat{H}t_1}\hat{s}_{\alpha_1} |{\bf P}_{\rm eq}\ra,
\label{ons1}
\eeq
where $\la \hat{\bf 1} |=(1,\ldots,1)$ is the vector with all unit entries. Since
the correlator (\ref{ons1}) is a real  function, it coincides with its
complex  conjugated expression:
\beq
C_{\alpha_n\ldots \alpha_1}=\la {\bf P}_{\rm eq} |
\hat{s}_{\alpha_1}e^{\hat{H}^{\dagger} t_1} \hat{s}_{\alpha_2} \ldots
e^{\hat{H}^{\dagger} (t_n-t_{n-1})}  s_{\alpha_n} | \hat{\bf 1}\ra,
\label{ons2}
\eeq
Using Eq.~(\ref{transv1}) and the relation $| \hat{\bf 1}\ra = (e^{\beta \hat{E}}/C) |{\bf
  P}_{\rm eq}\ra $, Eq.~(\ref{ons2}) leads to a higher order Onsager
relation \cite{stratonovich-book1,spin-glass-hoc}:
\begin{eqnarray}
\nonumber \la s_{\alpha_n}(t_n) \ldots s_{\alpha_1}(0) \ra= \\
 \quad \la
s_{\alpha_1}(t_n) s_{\alpha_2}(t_n-t_2) \ldots
s_{\alpha_{n-1}}(t_n-t_{n-1}) s_{\alpha_n}(0) \ra.
\label{ons3}
\end{eqnarray}
Here, $n=2$ corresponds to the standard Onsager reciprocity relation $\la
s_{\alpha_2}(t) s_{\alpha_1}(0) \ra = \la
s_{\alpha_1}(t) s_{\alpha_2}(0) \ra$, while $n=3$ corresponds to a higher order relation
\beq
\la s_{\alpha_3}(t_2) s_{\alpha_2}(t_1) s_{\alpha_1}(0) \ra =
 \la s_{\alpha_1}(t_2) s_{\alpha_2}(t_2-t_1) s_{\alpha_3}(0)  \ra.
\label{ons4}
\eeq

\subsubsection{Bochkov-Kuzovlev Type of Fluctuation Relations.}
Other classical higher order relations are most easily derived from
the Bochkov-Kuzovlev formulas \cite{bochkov-81} (see also chapter 1 in
\cite{stratonovich-b2} for a textbook introduction).
Let $U({\bf s})$ and $S({\bf s})$ be the energy and entropy of a spin system
with a known spin fluctuation size ${\bf s}$, assuming that all other degrees of freedom quickly equilibrate at a given temperature, and let us define a spin-dependent conditional free energy: $F({\bf s})=U({\bf s})-k_BTS({\bf s})$.
The application of a constant external magnetic
field ${\vh}$ leads to a change of the free energy $F({\bf s})$:
$F({\bf s})= f({\bf s})-{\vh}\cdot {\bf s}$, where $ f({\bf s})$ is the independent of $\vh$ part, and where the Bohr magneton and
$g$-factor are included into the definition of ${\vh}$.
A system at a constant $\vh$ always relaxes to the thermodynamic equilibrium with the probability distribution (see, e.g., Sec. 2.2.7 of Ref.~\cite{stratonovich-book1} for details):
\beq
p_{\rm eq} ({\bf s}) =\frac{e^{-\beta F({\bf s}) } }{Z},
\label{eqp1}
\eeq
where $Z=Z(\vh)$ is the partition function.
In order to guarantee such a relaxation, kinetic rates must satisfy the detailed balance constraints, so that probabilities of processes that connect microstates with spin polarizations ${\bf s}$ and ${\bf s}'$ are connected by
\beq
\frac{p_{{\bf s}\rightarrow {\bf s}'}} {p_{{\bf s}'\rightarrow {\bf s}}}=e^{\beta(F({\bf s})-F({\bf s}'))}.
\label{eqp2}
\eeq

Consider now the case when the system is  initially at the thermodynamic equilibrium with $\vh=0$ but then the field is suddenly switched on to a constant value. We are interested in the ratio of probabilities of observing
 transitions from the state ${\bf s}$ to ${\bf s}'$ and backwards. This ratio is influenced by the initial state probabilities:
\beq
\frac{p_{\rm eq} ({\bf s}) p_{{\bf s}\rightarrow {\bf s}'}} {p_{\rm eq} ({\bf s}') p_{{\bf s}'\rightarrow {\bf s}}}=e^{\beta\vh \cdot ({\bf s}'-{\bf s})}.
\label{eqp3}
\eeq
Note that equilibrium distribution prefactors in (\ref{eqp3}) completely cancelled the dependence of the rhs on $f({\bf s})$.

Bochkov and Kuzovlev showed that Eq.~(\ref{eqp3}) can be extended to an arbitrary driving protocol ${\bf h}(t)$. Let the external field be zero up to time $-T_m/2$, and assume that it  changes
with time during the measurment interval $t\in (-T_m/2,T_m/2)$, returning to zero
value at the end: ${\vh}(T_m/2) ={\vh}(-T_m/2)=0$.
One of the Bochkov-Kuzovlev
formulas relates the probability $P({\bf s}(t) | {\vh}(t))$  of observation of a trajectory of the
spin system ${\bf s}(t)$ under the driving protocol $ {\vh}(t)$ with
the probability  $P({\bf s}^R(t) |{\vh}^R(t))$ of observation of the
reversed in time trajectory ${\bf s}^R(t)={\bf s}(-t)$
 under the action of the reversed in time driving protocol ${\vh}^R(t)={\vh}(-t)$ [Eq.~(2.11) in Ref.~\cite{bochkov-81}]:
\beq
\frac{P({\bf s}(-t) |{\vh}(-t))}{P({\bf s}(t) | {\vh}(t))} = e^{-\beta W},
\label{bc1}
\eeq
where
\beq
W=\int\limits_{-T_m/2}^{T_m/2} \textnormal{d}t \, {\vh}(t) \cdot \dot{\bf s}(t)
\label{work1}
\eeq
is the work done by the time-dependent field on the spin system.
To derive (\ref{bc1}) we should split the time interval in infinitesimal steps $(t_i, t_{i+1})$, $i=0,\ldots,N$ with $N \gg 1$. At each step, the field $\vh (t_i)$ can be considered constant. Hence, Eq.~(\ref{eqp2}) can be applied at given $\vh (t_i)$ during this interval.
Using this constraint and the definition $P({\bf s}(t) | {\vh}(t))=p_{\rm eq} ({\bf s}_0) \prod_i p_{{\bf s}_i\rightarrow {\bf s}_{i+1}}$, and  $P({\bf s}(-t) | {\vh}(-t))=p_{\rm eq} ({\bf s}_N) \prod_i p_{{\bf s}_{i+1}\rightarrow {\bf s}_{i}}$, one can verify the validity of Eq.~(\ref{bc1}).

The next observation made by Bochkov and Kuzovlev was that the expression for the probability functionals (\ref{bc1}) can be used to generate numerous relations between higher and lower order correlators at the thermodynamic equilibrium. For example, following \cite{spin-glass-hoc}, we move the denominator in (\ref{bc1}) to the right-hand side,
multiply both sides of equation by $s_j(t_1)$ and sum over all
possible trajectories.  Then, in the sum over all trajectories, each  ${\bf s}(t)$ encounters with the time-reversed one, ${\bf s}(-t)$. Hence,
$$
 \sum \limits_{{\bf s}(t)} P({\bf s}(-t) |{\vh}(-t)) s_j(t_1) =\sum \limits_{{\bf s}(t)} P({\bf s}(t) |{\vh}(-t)) s_j(-t_1).
$$ 
 This finally gives us
 \beq
\la s_j(-t_1) \ra_{{\vh}^R} = \la s_j(t_1) e^{-W} \ra_{{\vh}},
\label{genf1}
\eeq
where the index ${\vh}$ means that the average is taken under the action
of the protocol ${\vh}(t)$. By equating the linear in ${\vh}(t)$
terms in (\ref{genf1}) we find

\beq
\beta \frac{\partial}{\partial t_2} \la s_j(t_1) s_k(t_2) \ra =
\frac{\delta \la s_j(t_1) \ra }{\delta h_k (t_2)} -\frac{\delta \la s_j(-t_1) \ra }{\delta h_k (-t_2)},
\label{fr1-1}
\eeq
where the averages are already over the equilibrium
probability distribution of microstates of the system. The causality requires $\delta
  \la s_j(-t_1) \ra / \delta h_k (-t_2)=0$ at $t_1>t_2$. Therefore,

\beq
 \beta \frac{\partial}{\partial t_2} \la s_j(t_1) s_k(t_2) \ra =
\frac{\delta \la s_j(t_1) \ra }{\delta h_k (t_2)}, \quad t_1>t_2.
\label{fr1-2}
\eeq
In the frequency domain, Eq~(\ref{fr1-2})  reproduces the fluctuation-dissipation relation
(\ref{FDT1})  in the ``classical" limit $ \omega \ll k_BT$. To show this, consider $k=j=z$ and identify $\frac{\delta \la s_j(t_1) \ra }{\delta h_k (t_2)}$ with $A_{zz}(t_1-t_2)$ in (\ref{FDT1}), and note that due to the time translation invariance,
$ \frac{\partial}{\partial t_2} \la s_z(t_1) s_z(t_2) \ra = -\frac{\partial}{\partial t_1} \la s_z(t_1) s_z(t_2) \ra $.
Afterwards we multiply both sides of Eq.~(\ref{fr1-2}) by $2\sin(\omega \tau)$, with $\tau=t_1-t_2$,  and integrate over $\tau$ from zero to infinity. We then obtain an equation
$$
-2\int_0^{\infty} \textnormal{d}\tau \, \sin(\omega \tau) \partial_{\tau} \la s_z(\tau) s_z(0) \ra =  \int \textnormal{d}\tau \, A_{zz} (\tau) \sin(\omega \tau).
$$
On the left-hand side of this equation, we perform integration by parts and use the Wiener-Khinchine theorem in Eq.~(\ref{c2-1}). For the right-hand side, we note that  $\int \textnormal{d}\tau \, A_{zz} (\tau) \sin(\omega \tau) =2\Im [A_{zz}(\omega)]$. The final result is
 \beq
P(\omega) = \frac{2k_BT}{\omega}  \Im [A_{zz}(\omega)],
\label{fr1-22}
\eeq
which coincides with (\ref{FDT1}) in the limit $ \omega \ll k_BT$.

Similarly, we can derive relations between higher order correlators. Multiplying both sides of (\ref{bc1}) by $s_j(t_1)
s_l(t_2)$ and repeating the same steps for $-t_2>t_1>0>t_2$, we find a higher order
fluctuation relation:
\begin{eqnarray}
\nonumber \beta \frac{\partial}{\partial t_2}\la s_j(-t_2) s_k(t_1) s_l(t_2) \ra &=&\frac{\delta \la s_j(-t_2) s_l(t_2) \ra }{\delta h_k (t_1)}-\\
&-&\frac{\delta \la s_l(-t_2) s_j(t_2) \ra }{\delta h_k (-t_1)}.
\label{order3}
\end{eqnarray}
It relates the 3rd order correlator at the equilibrium to a linear
response of the 2nd order correlators.

Quantum theory predicts that higher order relations,
like (\ref{ons4}) and (\ref{order3}), are not generally satisfied
\cite{belzig-12prl}. At the same time, some other extensions of
higher order fluctuation relations to the quantum domain are known
\cite{stratonovich-book1}. Their implications for spin fluctuations remain to be understood.

As a side note, we point that the modern research on higher order fluctuation relations has shifted from
relations between correlators to studies of global properties of statistics of work and information flow \cite{jarzynski-97prl,sinitsyn-11jpa,spin-td}. Therefore, the article  \cite{bochkov-81} by Bochkov and Kuzovlev is often associated with a special consequence of Eq.~(\ref{bc1}), which is obtained by moving $P({\bf s}(t) | {\vh}(t))$ from the denominator
to the  right-hand side of this equation and averaging over all possible trajectories:
\beq
\la e^{-\beta W}\ra_{\bf h} =1.
\label{bc5}
\eeq
For example, Eq.~(\ref{bc5}) is often mentioned as the first but only one of many other similar relations, the most widely known of which is the Jarzynski equality \cite{jarzynski-97prl}
\beq
\la e^{-\beta W_J}\ra_{\bf h} =e^{-\beta \Delta F},
\label{jarz5}
\eeq
where $\Delta F$ is the change of the equilibrium free energy between the initial and final values of the time-dependent parameter ${\bf h}(t)$, and $W_J$ is the work defined differently from $W$, namely, $W_J=\int {\bf s}(t) \cdot (\textnormal{d}{\bf h}/\textnormal{d}t)\,\textnormal{d}t$, in our notation.

  Jarzynski equality (\ref{jarz5}) generally carries a different meaning than Eq.~(\ref{bc5}), and it has appeared quite influential, e.g., for applications to nonequilibrium free-energy sampling.
It seems, however, that  Bochkov-Kuzovlev relations like (\ref{bc1}), (\ref{genf1}), and (\ref{bc5}), with the definition of work (\ref{work1}), carry special importance because they provide the most straightforward  way to derive  higher order relations between various correlators at the equilibrium, such as Eq.~(\ref{order3}).




\section{Methods I: Spin Noise Phenomenology}

The phenomenological approach and relaxation time approximation are justified
by essentially the same arguments. In both cases, the spin dynamics is described by equations with several phenomenological parameters.
The microscopic derivation of these parameters is beyond the scope of the phenomenological approach. 
Its basic power is related to the fact that the phenomenological approach  can be formulated without paying a close attention to specific system details. This makes its predictions particularly useful as a ``first guess"  in many situations.
We note, however, that by no means the  phenomenological approach is universally applicable. For example,  it cannot  be generally applied to explain $1/\omega^{\alpha}$ type of spectra that encounter in disordered strongly interacting spin systems.

\subsection{Spin Fluctuation Dynamics}

\subsubsection{Bloch Equation.}
Let's consider  a spontaneous spin fluctuation $\vS$ emerged at $t=0$ in the observation volume.
The Bloch equation describes  the dynamics of this fluctuation, namely, its precession in an external magnetic field ${\bf B}$ and relaxation:
\beq
\frac{\textnormal{d}S_{\alpha}}{\textnormal{d}t}=  g \varepsilon_{\alpha \beta \gamma}   B_{\beta} S_{\gamma} - \frac{S_{\alpha}}{\tau_s^{\alpha}},
\label{sdyn1}
\eeq
where $\alpha, \beta = x,y,z$, $g$ is the electron $g$-factor (the Bohr magneton is included in its definition), and we also set $\hbar=1$. In semiconductors,  the $g$-factor is usually a tensor due the atomic lattice anisotropy. For the sake of simplicity, we disregard this detail here. We  assume that the measurement beam is directed along the $z$-axis and the magnetic field vector points in the $yz$-plane, making an angle $\theta_B$ with the  $z$-axis (see Fig.~\ref{fig6_1_scheme}).

The parameters $\tau_s^{\alpha}$ are the relaxation times along different axes.
In many semiconductors, the relaxation time is anisotropic, i.e. $\tau_s^x\ne \tau_s^y \ne \tau_s^z$.  However, in frequently discussed SNS applications (the hot atomic vapors and bulk GaAs), the relaxation time can be considered as almost isotropic: $\tau_s\equiv \tau_x=\tau_y=\tau_z$. Instead of relaxation times, we will also often use {\it the relaxation rates} defined as the inverses of relaxation times, e.g.,
\beq
\Gamma_s= 1/\tau_s.
\label{relrates1}
\eeq

\begin{figure}
\centerline{\includegraphics[width=0.5\columnwidth]{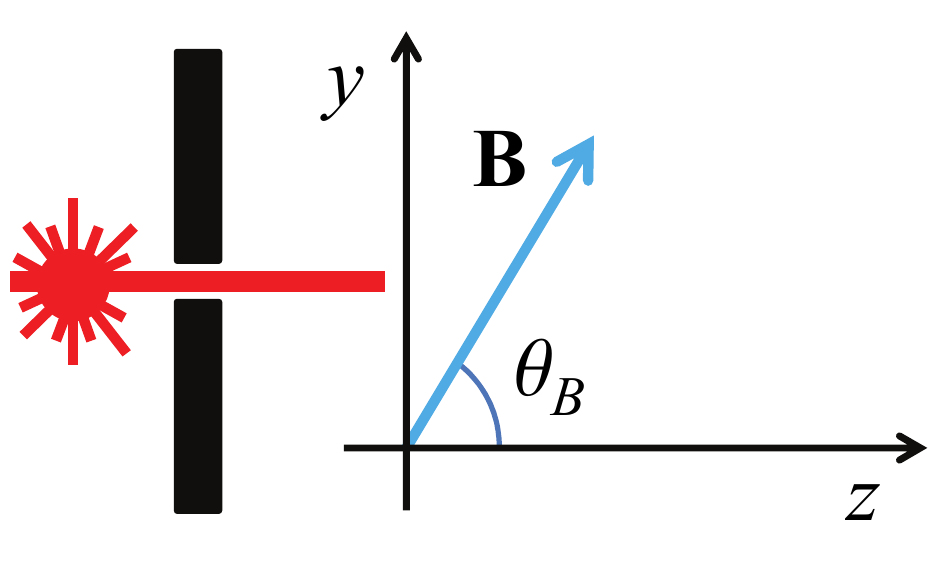}}
\caption{ Schematics of the measurement geometry.}
\label{fig6_1_scheme}
\end{figure}

The solution of Eq.~(\ref{sdyn1}) for an isotropic relaxation rate shows that a spontaneous spin fluctuation  ${\bf S} = \hat{z} S_{z0}$ initially emerged along the  measurement axis precesses  around the magnetic field and  decays according to
\begin{eqnarray}
\nonumber
S_z(t) = S_{z0} \cos^2(\theta_B) e^{-t/\tau_s} + \\
\quad\quad\quad\quad\quad\quad\quad\quad +S_{z0} \sin^2(\theta_B) e^{-t/\tau_s} \cos(\omega_Lt),
\end{eqnarray}
where $ \omega_L = g|B|$ is the Larmor frequency.

The Bloch equation (\ref{sdyn1}) can also be solved for the case of an oscillating magnetic field. In SNS setup, the multiphoton absorption processes due to the oscillating field can be observed in spin noise \cite{Braun07a}.

\subsubsection{Noise Power Spectrum in a Tilted Magnetic Field.}
The noise power spectrum is obtained as the Fourier transform of the spin-spin correlator:
\begin{eqnarray}
\label{ssc11}
 \nonumber P(\omega) =2\int\limits_0^{\infty}  \textnormal{d}t\, \cos {\omega t} \la S_z(t) S_z(0) \ra =\\
 = \la S_{z0}^2 \ra \left(  \frac{2\cos^2(\theta_B)/\tau_s }{ \omega^2 +\frac{1}{\tau_s^2}}+\sum_{s=\pm}  \frac{\sin^2(\theta_B)/\tau_s }{ (\omega-s \omega_L)^2 +\frac{1}{\tau_s^2}}  \right).
\label{ssc12}
\end{eqnarray}
Here, $s=\pm 1$.
Due to the symmetry of the spectrum under  $\omega \rightarrow -\omega$ transformation, only the positive values of $\omega$ are informative. Figure~\ref{fig6_2_P} shows that, in a tilted magnetic field, the noise power spectrum typically consists of two Lorentzian peaks  centered at zero and Larmor frequencies (see the blue curve in Fig.~\ref{fig6_2_P}). In the limiting cases of $\theta=0$ $(\pi/2)$ only the zero (Larmor) frequency peak is present. These limiting cases are presented by the black (zero frequency peak) and red (Larmor frequency peak) curves in Fig.~\ref{fig6_2_P}.

\begin{figure}
\centerline{\includegraphics[width=0.9\columnwidth]{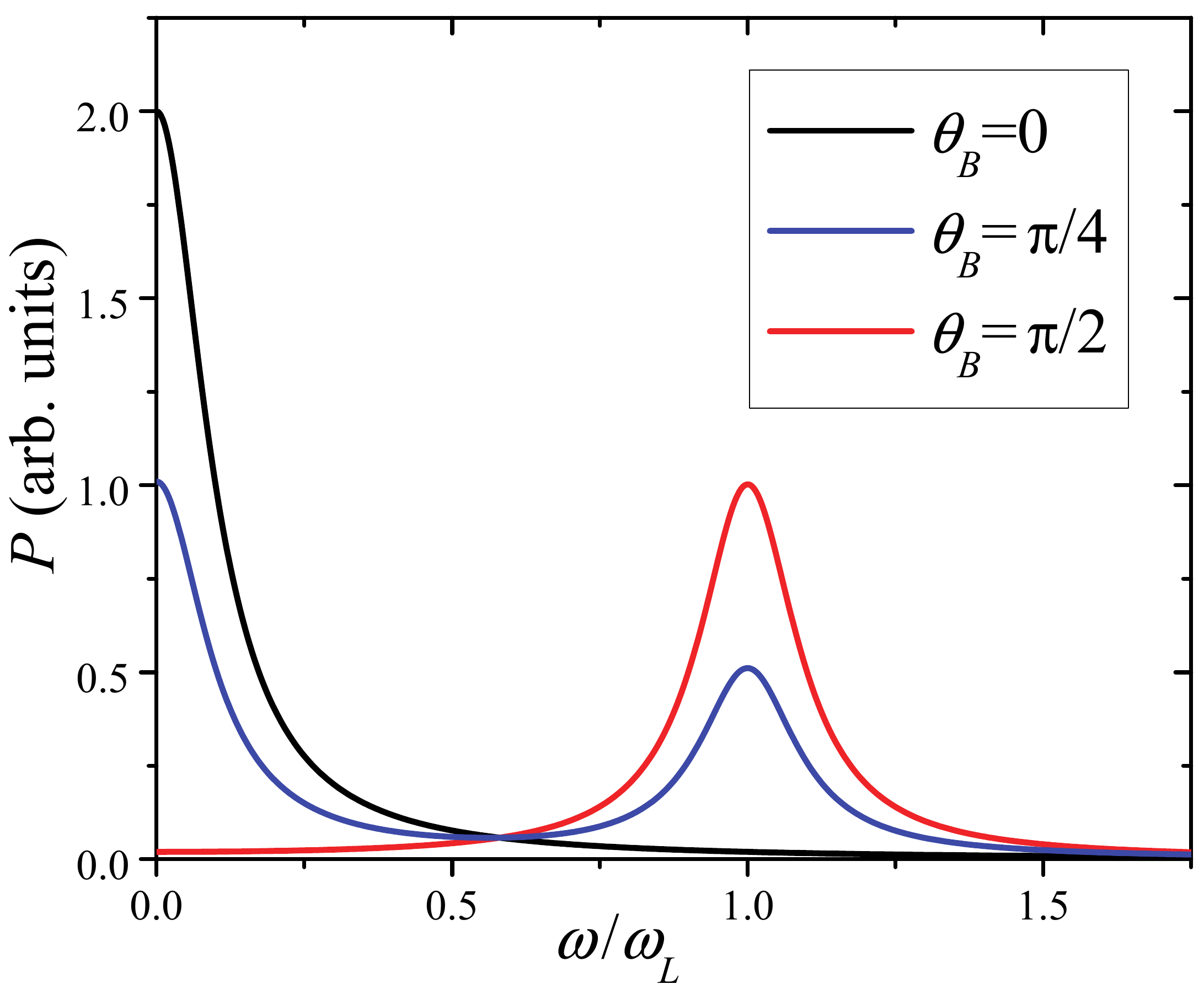}}
\caption{
The noise power spectrum $P(\omega)$ (Eq.~(\ref{ssc12})) shown for several values of the angle  $\theta_B$ between the external field and the measurement axis. Spin relaxation is isotropic with $\omega_L\tau_s=10$.}
\label{fig6_2_P}
\end{figure}

The relative areas of peaks are controlled by the angle $\theta_B$ that the external field makes with the measurement axis.
An important property is independence of the total area under the spectrum on the direction and magnitude of the external field. Moreover, the total area under the spectrum is  a thermodynamic characteristic independent of kinetic rates. Indeed, due to the property
\beq
\int\limits_{-\infty}^{\infty}  \textnormal{d}\omega \, \frac{\gamma }{ \omega^2+\gamma^2} = \pi,
\label{int-lor}
\eeq
the total area of the noise power spectrum is given by
\beq
\int\limits_{-\infty}^{\infty} P(\omega)\, \textnormal{d}\omega = 2\pi \la S_{z0}^2 \ra ,
\label{power-area1}
\eeq
i.e. it depends only on statistical properties of the equal time correlator.  The latter can be estimated from the knowledge of the equilibrium spin distribution.

\subsubsection{Characteristic Sizes of Spin Fluctuations.}
The variance of a fluctuation size, $\la S_{z0}^2 \ra$, enters  the expression for the noise power spectrum (\ref{ssc12}) as a parameter. In many situations, it can be easily estimated without resorting to complex microscopic techniques. Consider, for example,  spin noise of noninteracting conduction electrons that form a Fermi sea. The Fermi-Dirac distribution
\beq
f_D(\epsilon)=\frac{1}{1+e^{-\epsilon/k_BT}}
\label{fdirac1}
\eeq
describes the occupancy of energy levels by electrons in the system. Each level is doubly degenerate due to the two possible spin orientations of electron. There is no contribution to $\la S_{z0}^2 \ra$ from either empty levels or levels populated by two electrons. Spin fluctuations originate only from the energy levels occupied by a single electron. Such electrons can be found in the up- or down-spin state with the same probability.
The probability that a given energy level  $\epsilon$ is populated by a single electron is given by $f_{\uparrow} (1-f_{\downarrow}) + f_{\downarrow} (1-f_{\uparrow}) = 2 f_D(\epsilon) (1-f_D(\epsilon))$, where $f_{\uparrow/\downarrow}$ are the probabilities that coincide with (\ref{fdirac1}) at the thermodynamic equilibrium.

Spin fluctuations from independent singly occupied energy levels (orbitals) contribute additively to the variance of the total spin fluctuation. The average of the spin projection squared of a single electron is $\la \hat{s}_z^2 \ra = 1/4$. Taking into account that, at sufficiently low temperatures, the main contribution to spin noise comes from the conduction electrons near the Fermi surface, we find
\beq
\la S_{z0}^2 \ra = \frac{D}{4} \int \textnormal{d}\epsilon \, 2f_D(\epsilon) (1-f_D(\epsilon)) = \frac{1}{2} Dk_B T,
\label{td1}
\eeq
where $D$ is the density of states per spin direction. This example shows that in a sufficiently cold Fermi liquid ($k_BT \ll \epsilon_F$), the Pauli exclusion principle leads to a linear temperature dependence of the area under the spin noise power spectrum curve.

Another frequently encountered situation is when  spin noise is created by $N \gg 1 $ truly noninteracting electrons. This situation is found, e.g. in an insulating state when  spin noise is produced by localized electrons that are well separated and do not interact with each other. In this case, each localized spin-degenerate state is always populated by a single electron, so that
\beq
\la S_{z0}^2 \ra = \frac{N}{4},
\label{td2}
\eeq
i.e. each of $N$ electrons provides the same contribution ($1/4$) to the variance. Moreover, a similar situation takes place at high temperatures $k_BT \gg \epsilon_F$, when one can disregard the Pauli exclusion principle and assume that all electrons are uncorrelated.
Figure~\ref{noise-T} shows that the experimentally obtained integrated noise power depends on temperature linearly, in agreement with (\ref{td1}), but has a finite  offset value at $T=0$ in agreement with (\ref{td2}).

\subsubsection{Relaxation Time Anisotropy.}
A significant application area for SNS is the field of semiconductors, including novel 2D semiconducting materials \cite{sinitsyn-14prl1}. In many of such systems, due to an intrinsic anisotropy,  the spin relaxation rate depends on the direction of  spin polarization. For example, the in-plane spin relaxation in MoS$_2$ was estimated to be about an order of magnitude faster than the out-of-plane spin relaxation \cite{luyi-nat15}. Such anisotropy leads to the deviation of the  behavior of the spin noise power spectrum from the one described by  Eq.~(\ref{ssc12}).

Let's consider only the most important case of a purely in-plane magnetic field.
In the presence of anisotropy in relaxation rates, $\Gamma_s^{\perp} \ne \Gamma_s^z$, where $\Gamma_s^{\perp}$ is the in-plane relaxation rate,
the dynamics of spin polarization can be described by the equation
 \beq
\frac{\textnormal{d}{\bf S}}{\textnormal{d}t} =g B_y \hat{y} \times {\bf S} - \Gamma_s^{\perp} \left(S_x  \hat{x}  + S_y \hat{y}  \right) - \Gamma_s^z S_z \hat{z}.
\label{sp3}
\eeq
The solution of Eq.~(\ref{sp3}) with the initial condition ${\bf S}=S_0 \hat{z}$ can be written explicitly. Depending on the strength of the magnetic field and anisotropy, the spin polarization exhibits either a monotonous decay or oscillating behavior.

{\bf I.}  First,  assume the situation with $\Gamma_s^{\perp} > \Gamma_s^z$ and a relatively weak magnetic field $gB_y<(\Gamma_s^{\perp}-\Gamma_s^z)/2$. The spin polarization then relaxes monotonously:
\begin{eqnarray}
\nonumber S_{z}(t) = S_{z0} e^{-\frac{(\Gamma_s^z+\Gamma_s^{\perp})t}{2}} \times \\
\quad\quad\quad
\times \left( \cosh (\Omega t) +\frac{ (\Gamma_s^{\perp}-\Gamma_s^z) \sinh(\Omega t )}{2 \Omega}  \right),
\label{st1}
\end{eqnarray}
where
$$
\Omega = \frac{\sqrt{(\Gamma_s^{\perp}-\Gamma_s^z)^2-4(gB_y)^2}}{2}.
$$

The noise power spectrum consists of a single peak centered at zero frequency (Fig.~\ref{anisotr-fig}, black curve). The shape of this peak is described by the sum of two Lorentzians with different amplitudes and half-widths:
\beq
P(\omega) =  \la S_{z0}^2 \ra \left( \frac{A_+ \gamma_+}{\omega^2+\gamma_+^2}+ \frac{A_{-}  \gamma_{-}}{\omega^2+\gamma_{-}^2} \right),
\label{noisep11}
\eeq
where
$$
A_{\pm} = 1\pm \frac{ \Gamma_s^{z} - \Gamma_s^\perp }{2\Omega}, \quad \gamma_{\pm}=\bar{\Gamma}_s \pm \Omega, \quad \bar{\Gamma}_s=\frac{\Gamma_s^z+\Gamma_s^{\perp}}{2}.
$$

\begin{figure}
\centerline{\includegraphics[width=0.9\columnwidth]{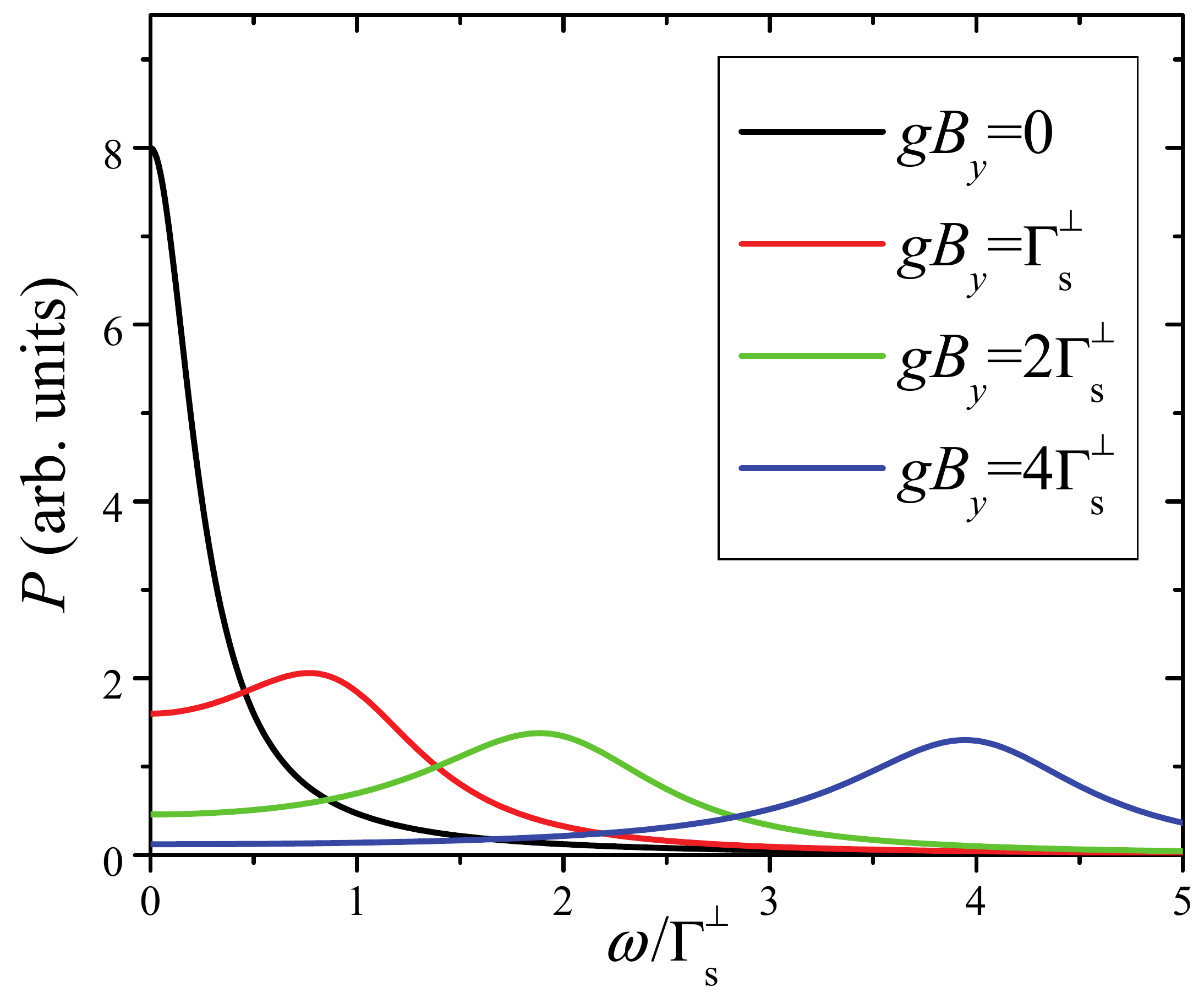}}
\caption{
The noise power spectrum $P(\omega)$ (Eqs.~(\ref{noisep11}), (\ref{noisep12})) shown for several values of the external in-plane magnetic field at $\Gamma_s^z=0.25\Gamma_s^{\perp}$. The peak at zero magnetic field is noticeably narrower than the peak at  large field values. }
\label{anisotr-fig}
\end{figure}

{\bf II.} When $gB_y>(\Gamma_s^{\perp}-\Gamma_s^z)/2$ and  $\Gamma_{s}^{\perp}>\Gamma_s^z$, the spin polarization shows an oscillatory behavior:
\begin{eqnarray}
\nonumber S_{z}(t) = S_{z0} e^{-\frac{(\Gamma_s^z+\Gamma_s^{\perp})t}{2}} \times \\
\quad\quad\quad  \times \left( \cos (\Omega t) +\frac{ (\Gamma_s^{\perp}-\Gamma_s^z) \sin(\Omega t )}{2 \Omega}  \right),
\label{st2}
\end{eqnarray}
where
$$
\Omega = \frac{\sqrt{4(gB_y)^2-(\Gamma_s^{\perp}-\Gamma_s^z)^2}}{2}.
$$
The noise power spectrum consists of two nearly Lorenzian peaks centered at $\omega=\pm \Omega$  (Fig.~\ref{anisotr-fig}, blue, green, and red curves), whose half-widths are determined by the average of  two relaxation rates:

\beq
P(\omega) = \sum \limits_{s=\pm} \frac{\la S_{z0}^2 \ra \left(\Gamma_s^{\perp}+s\beta \omega \right)}{(\omega+s\Omega)^2+\bar{\Gamma}_s^2}, \quad \beta=\frac{\Gamma_s^{\perp}-\Gamma_s^z}{2\Omega} .
\label{noisep12}
\eeq

{\bf III.} Finally, consider the regime of $\Gamma_s^{\perp}<\Gamma_s^z$, which takes place, for example, in the Rashba 2D electron system. This system is analyzed in more detail in Sec.~8. One can show that this regime is also characterized by the same equations as presented in the paragraphs {\bf I.-II.} above. However, now the transition from the monotonous relaxation to  oscillatory behavior occurs at $gB_y=(\Gamma_s^z-\Gamma_s^{\perp})/2$.




\subsection{Langevin Equation and Fluctuation-Dissipation theorem}

The approach based on the Bloch equation  is not self-consistent. It uses the size of a typical spin fluctuation as an input parameter, and does not include the dynamics leading to the appearance of spin fluctuations.
Such an approach is not easy to generalize, e.g., to include nonlinear interactions or calculate higher order correlators. A more rigorous approach is based on the Langevin equation that includes a noise term as a source of spin fluctuations:


\beq
{\dot{\vS}} = -{\bf \hat{R}} {\bf S} + \vxi,
\label{lang1}
\eeq
where  ${\bf \hat{R}}$ is called  the {\it relaxation matrix} (since its elements contain information about characteristic relaxation rates), and $\vxi$ is the noise term. The elements of ${\bf \hat{R}}$ can be read directly from the Bloch equation (\ref{sdyn1}):
\beq
R_{\alpha \beta} = -g \varepsilon_{\alpha \gamma \beta} B_{\gamma} +\frac{\delta_{\alpha\beta}}{\tau_s^{\alpha}}.
\label{R-matrix1}
\eeq

In mesoscopic systems, unless higher order spin correlators are needed, the noise term can be usually well approximated by a Gaussian delta-correlated noise:
\beq
\la \xi_{\alpha}(t) \xi_{\beta} (t') \ra = G_{\alpha \beta} \delta(t-t'),
\label{corr-mt1}
\eeq
where $\alpha, \beta =x,y,z$ and $G_{\alpha \beta}$ are elements of the {\it correlation matrix} ${\bf \hat{G}}$. This approximation is justified by the fact that  when the number of observed spins is large, $N\gg1$, one can choose a time scale  $\delta t$ such that it is much smaller than the ensemble average spin relaxation time but sufficiently large for many random spin flips to happen. Since the time interval $\delta t$ is much smaller than the spin relaxation time, the number of  spins that experience spin flips during  $\delta t$ is  much smaller than the total number of spins $N$. Hence, it is highly unlikely for any given spin to flip more than once in two consecutive time intervals of the size $\delta t$ or influence the dynamics of other spins. In turn, this means that spin fluctuations in nearby time intervals can be considered as statistically independent. Therefore, on a much longer time scale of spin relaxation, one can assume that the spin fluctuations are produced by a delta-correlated white noise.

At the thermodynamic equilibrium, one can derive the correlation matrix ${\bf \hat{G}}$ from the knowledge of the relaxation matrix ${\bf \hat{R}}$ and the condition that the equilibrium distribution of electronic spins is described by the Fermi-Dirac statistics. The arguments run as follows:

First, one can prove the following formula (see also Eq.~(4.4.51) in \cite{gardiner}), which is valid for any steady stochastic process described by the equation of the type (\ref{lang1}):
\beq
{\bf \hat{R}} \hat{\vsigma} +  \hat{\vsigma} {\bf \hat{R}^{\dagger}} ={\bf \hat{G}},
\label{gard1}
\eeq
where $\hat{\vsigma}$ is the matrix of equal time spin correlators, with components
\beq
\sigma_{\alpha \beta} = \la S_{\alpha} (t) S_{\beta}(t) \ra.
\label{sig-mt}
\eeq

\underline{Proof}:
A formal solution of Eq.~(\ref{lang1}) can be written as
\beq
\vS(t) = \int\limits_{-\infty}^{t} \textnormal{d}t_1 \, e^{-{\bf \hat{R}}(t-t_1)} \vxi(t_1).
\label{sol11}
\eeq
Substituting (\ref{sol11}) and (\ref{sig-mt}) into the left-hand side of Eq.~(\ref{gard1}), and then applying (\ref{corr-mt1}) we find:
\begin{eqnarray}
\nonumber {\bf \hat{R}} \hat{\vsigma} + \hat{\vsigma} {\bf \hat{R}^{ \dagger}} = \int\limits_{-\infty}^t \textnormal{d}t'  \, \left[ {\bf \hat{R}}  e^{-{\bf \hat{R}}(t-t')} {\bf \hat{G} }  e^{-{\bf \hat{R}^{\dagger}} (t-t')} + \right. \\
 \nonumber
\quad\quad\quad\quad\quad\quad\quad\quad  \left. + e^{-{\bf \hat{R}}(t-t')} {\bf  \hat{G}}  e^{-{\bf \hat{R}^{\dagger}} (t-t')}{\bf \hat{R}^{\dagger}} \right] = \\
\nonumber
\quad\quad\quad = \int\limits_{-\infty}^{t} \textnormal{d}t'\,
\frac{\textnormal{d}}{\textnormal{d}t'} \left[ e^{-{\bf \hat{R}}(t-t')} {\bf \hat{G}}  e^{-{\bf \hat{R}^{\dagger} } (t-t')} \right] = {\bf \hat{G}}.
\end{eqnarray}
 This proves the formula (\ref{gard1}).

Usually, Eq.~(\ref{gard1}) cannot be used to derive the correlation matrix because the matrix $\hat{\vsigma}$, by itself, has to  be found self-consistently by solving the Langevin equation. However, for a system at the thermodynamic equilibrium, the equal time correlator is directly related to the equilibrium Boltzmann-Gibbs distribution and can be derived whenever the free energy function of the system is known.
 Thus, for the equal time correlator of the Fermi gas,  one can verify that the magnetic field dependent terms in the correlation matrix (\ref{R-matrix1})  do not contribute to the matrix ${\bf \hat{G}}$.
Equation~(\ref{td1}) then gives
\beq
G_{\alpha \beta} =\delta_{\alpha \beta} \frac{Dk_BT}{\tau_s},
\label{G-matrix1}
\eeq
and for the system of $N$ independent spins, Eq.~(\ref{td2}) gives
\beq
G_{\alpha \beta} = \frac{N}{2\tau_s}\delta_{\alpha \beta}.
\label{G-matrix2}
\eeq

The experimentally measurable characteristic is the Fourier transform of the fluctuating spin polarization. For a large measurement time interval $T$, it is given by
\beq
S_{\alpha}(\omega) =\lim  \limits_{T\rightarrow \infty} \frac{1}{\sqrt{T}} \int\limits_{-T/2}^{T/2} e^{i\omega t } S_{\alpha}(t) \, \textnormal{d}t,
\label{freq1}
\eeq
from which the spin correlators in the frequency domain are obtained as
\beq
P_{\alpha \beta} = \la S_{\alpha}(\omega) S_{\beta}(-\omega) \ra.
\label{npower2}
\eeq
Taking the Fourier transform of Eq.~(\ref{lang1}) we find
\begin{equation}
 \vxi(\omega) = (i\omega \hat{\bf 1} + \hat{\bf R}) {\bf S} (\omega),
\end{equation}
and noting that $\la \xi_{\alpha}(\omega) \xi_{\beta}(-\omega) \ra = G_{\alpha \beta}$, the  spin correlators can be written as
\beq
P_{\alpha \beta} (\omega)= \frac{1}{i\omega \hat{\bf 1} + \hat{\bf R}} \hat{\bf G}  \frac{1}{-i\omega \hat{\bf 1} + \hat{\bf R}^{\dagger}}.
\label{npower3}
\eeq

The noise power spectrum is an element of ${\bf \hat{P}}$: $P(\omega) \equiv P_{zz}(\omega)$.
For example, for the Fermi gas electrons in a strong in-plane magnetic field  along the $y$-axis and anisotropic spin relaxation, we find using this approach:
\begin{eqnarray}
 P(\omega)&=& \sum_{s=\pm} \frac{k_BTD/(2\tau_s)}{(\omega-s\omega_L)^2+(1/\tau_s)^2}, \\
\nonumber \frac{1}{\tau_s} &=& \frac{1}{2} \left( \frac{1}{\tau_s^z} +\frac{1}{\tau_s^x}\right).
\label{power-fen}
\end{eqnarray}


\subsection{Cross-Correlation SNS and Multicomponent Spin Systems}

\subsubsection{Off-Diagonal Correlators.}
Consider the setup shown in Fig.~\ref{cross-beams} consisting of two measurement beams directed along different axes and probed separately by individual detectors. Signals from these detectors will be correlated because in the overlap region both beams  interact with the same electrons/atoms.  Let Detector 1 measure the spin polarization along the $z$-axis and convert it into the Fourier transform $S_z(\omega)$, as defined by Eq.~(\ref{freq1}). Respectively, Detector 2 obtains $S_x(\omega)$. Multiplying the two and averaging over repeated measurements during equal time intervals, one can obtain the cross-correlation spectrum, $P_{zx}(\omega) = \la S_{z}(\omega) S_{x}(-\omega) \ra$. As this expression is generally complex-valued, it is convenient to consider its real and imaginary parts separately:
$$
P_{zx}^{Re} (\omega)\equiv {\rm Re} \la S_{z}(\omega) S_{x}(-\omega) \ra,
$$
$$
P_{zx}^{\Im} (\omega) \equiv \Im \la S_{z}(\omega) S_{x}(-\omega) \ra.
$$

\begin{figure}
\centerline{\includegraphics[width=0.75\columnwidth]{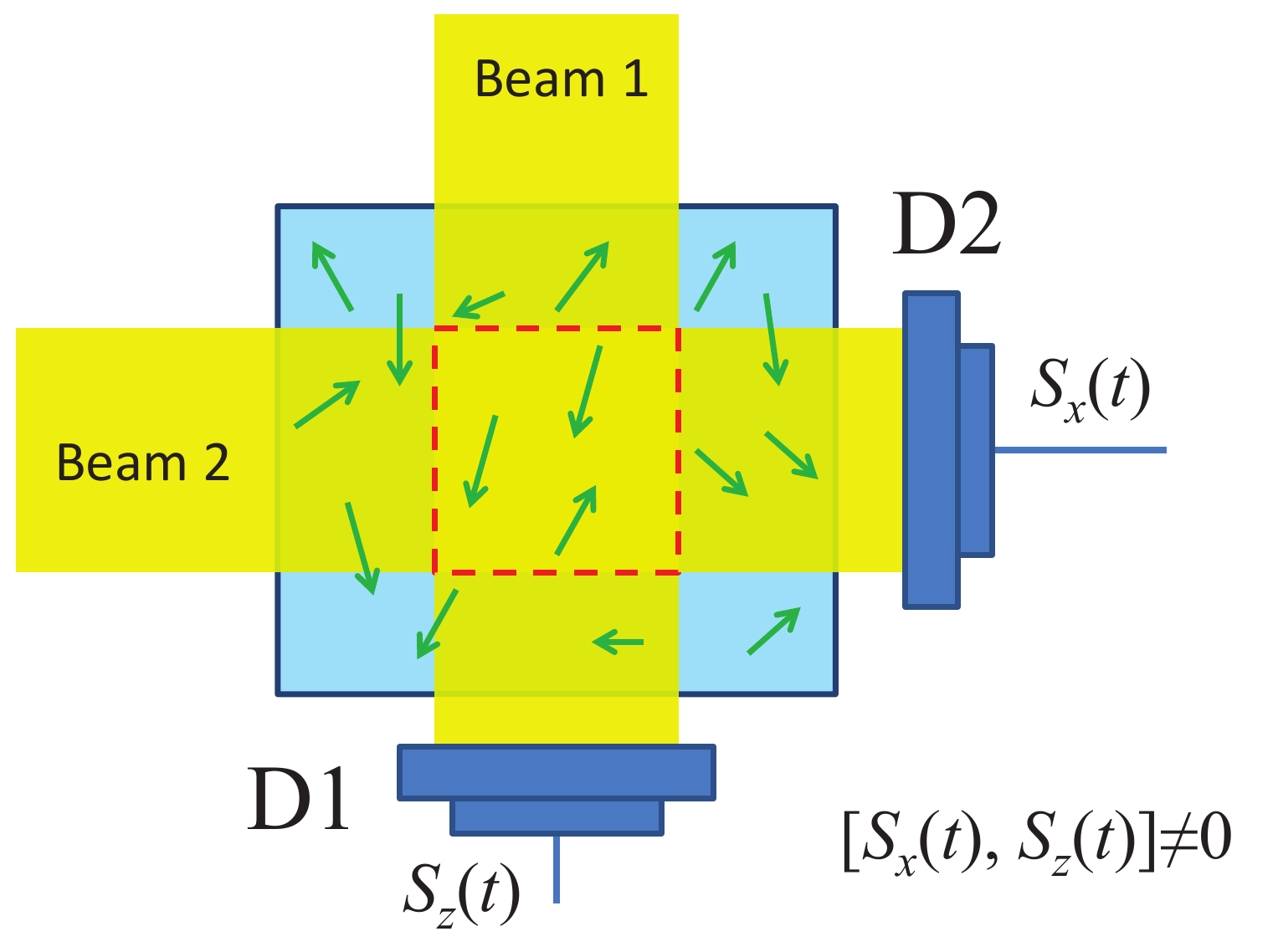}}
\caption{Two probe beams to study correlations between different spin polarization components. Here, D1 and D2 are detectors
measuring Faraday rotation angles. The crossing region is denoted by the red dashed line.}
\label{cross-beams}
\end{figure}

\begin{figure}
\centerline{\includegraphics[width=0.9\columnwidth]{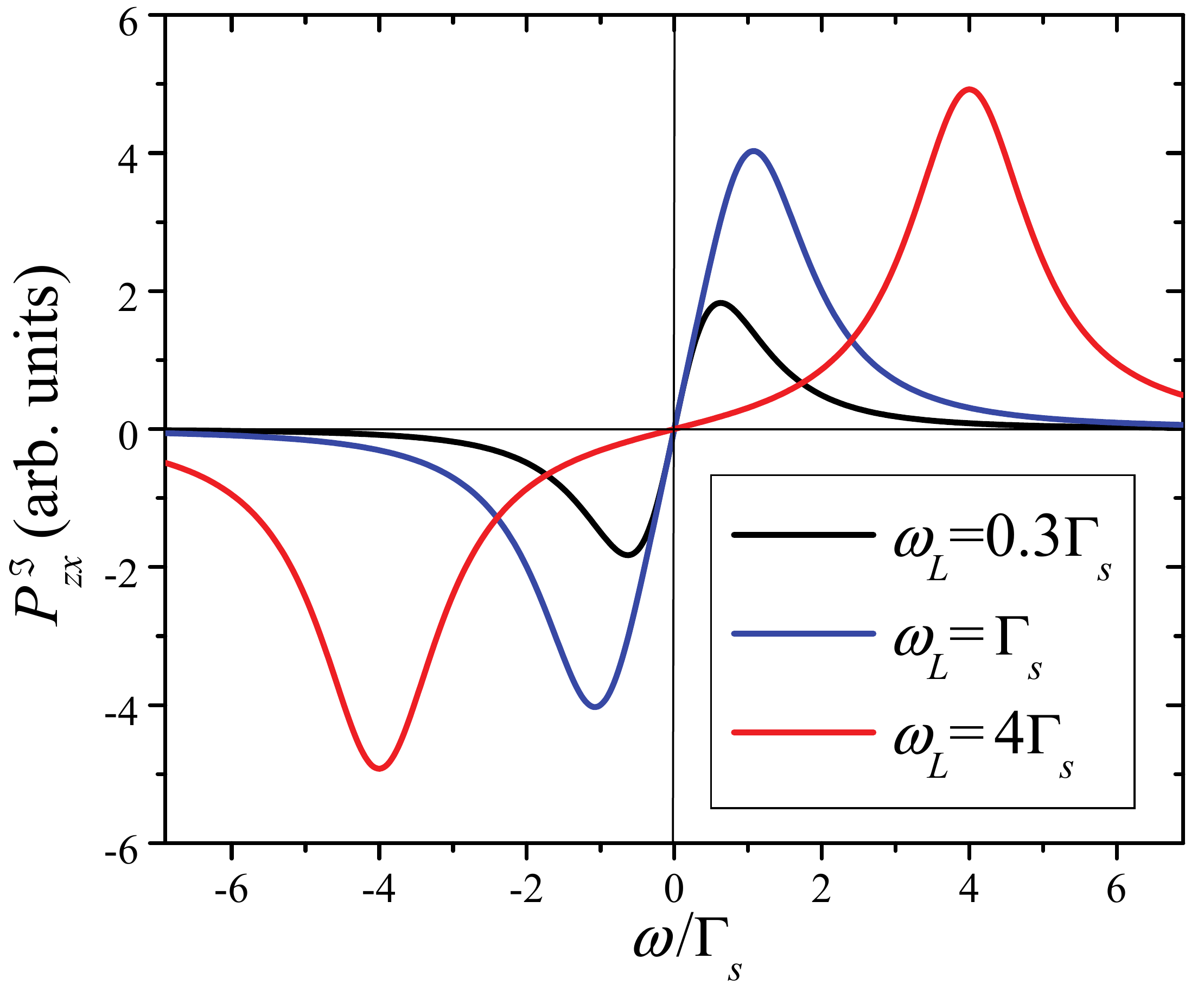}}
\caption{
The cross-correlation spectrum $P(\omega)$ (Eq.~(\ref{imcor})) shown for several values of the external in-plane magnetic field $gB_y \equiv \omega_L$.}
\label{imaginary-fig}
\end{figure}

For example, consider an atomic vapor of $N$ atoms in the observation region subjected to a weak magnetic field along the $y$-axis at a high temperature. Then, the spin relaxation rate is almost isotropic, so that the relaxation and correlation matrices read:
$$
R_{\alpha \beta} = -gB_y \varepsilon_{\alpha y \beta} +\Gamma_s \delta_{\alpha \beta}, \quad G_{\alpha \beta} = \frac{N\Gamma_s \delta_{\alpha \beta}}{2}.
$$
Using the above expressions, one can find  $P_{zx}^{Re} (\omega)=0$ and
\beq
P_{zx}^{\Im} (\omega) = \frac{N \omega_L \omega}{2(\omega^2+\omega_L^2+\Gamma_s^2)}  \sum_{s=\pm}  \frac{\Gamma_s}{(\omega-s\omega_L)^2+\Gamma_s^2},
\label{imcor}
\eeq
where $\omega_L=gB_y$.  At zero magnetic field ($B_y=0$) this correlator vanishes, while at large fields ($\omega_L \gg \Gamma_s$) its peak at $\omega>0$ has a Lorenzian shape: $P_{zx}^{\Im} (\omega)  \sim N\Gamma_s/[4[(\omega-\omega_L)^2+\Gamma_s^2]]$. Behavior of $P_{zx}^{\Im}(\omega)$ at several different values of the external magnetic field is shown in Fig.~\ref{imaginary-fig}.

\subsubsection{2D Dirac Materials.}
There are emerging applications of SNS requiring an extension of SNS theory (e.g., Eq.~(\ref{lang1})) to include some additional discrete degrees of freedom. For example, in the family of new 2D materials called the transition metal decalcogenides, the spectrum of electrons contains two Dirac valleys, $K$ and $K'$ (Fig.~\ref{fig2_2_1}). The Dirac valleys are coupled as electrons can scatter between the valleys. Moreover, the spin dynamics is valley-dependent because of the spin-orbit coupling, which can be considered as an effective out-of-plane magnetic field leading to a Zeeman splitting with opposite signs of interaction in different valleys. In such a case, it is convenient to think that this system consists of spins of two different kinds, marked by additional index $\tau=1(-1)$ for $K$ $(K')$. Spins from different valleys can interact with each other and even transform into each other by short-range impurity scattering.

The generalization of the Bloch equation to such a situation was discussed in \cite{sinitsyn-14prl1,luyi-nat15}:
\beq
\frac{\textnormal{d} \vS^{\tau} }{\textnormal{d} t}+\tau \vS^{\tau}\times
\vOm_{\mathrm{SO}}+\vS^{\tau}\times \vOm_{\mathrm{L}}
= -\Gamma_s \vS^{\tau}+\frac{\vS^{-\tau}-\vS^{\tau}}{2T_v},
 \label{dirac1}
\eeq
where $\vOm_{\mathrm{SO}} =\Omega_{\mathrm{SO}}(k_F)\bm{z}$ is the effective out-of-plane  magnetic
 field induced by the spin-orbit coupling, $\Gamma_s$ is the spin relaxation rate, $T_v$ the valley relaxation time, $\vOm_{\mathrm{L}} =
g\bf{B}$, and, as an index in $\vS^{\tau}$, $\tau=K,K'$ and $-K=K'$.

The elements  of the relaxation matrix $\hat{\bf R}$ now can be indexed by a complex index $\alpha \tau$, $\alpha =x,y,z$, $\tau=\pm$.
One can read the elements of this matrix directly from Eq.~(\ref{dirac1}). It is then straightforward to upgrade Eq.~(\ref{dirac1}) to a Langevin equation by adding a Gaussian noise term,  whose correlation matrix components can be obtained by application of the fluctuation-dissipation theorem:
\beq
G_{\alpha \tau, \beta \tau'} =2Dk_BT \left[   \frac{(\delta_{\tau \tau'} - \delta_{-\tau \tau'})
\delta_{\alpha \beta}}{2T_v} +
\Gamma_s\delta_{\tau \tau'} \delta_{\alpha \beta} \right].
\label{dirac-corm}
\eeq
The detailed discussion of the Langevin equation based on (\ref{dirac1}) can be found in \cite{sinitsyn-14prl1}. Here we note also that, in application to  MoS$_2$, there is a strong experimental evidence that the
valley relaxation is much faster than the spin relaxation \cite{luyi-nat15}. In such a case, the fast valley degrees of freedom can be integrated out and an effective description can be obtained in terms of the standard Bloch equation (\ref{sdyn1}) with a strong anisotropy of relaxation rates: $\Gamma_s^z \gg \Gamma_s^{\perp}$ that we have already discussed. The arguments go as follows.

Consider the total spin polarization $ {\bf S} = {\bf S}^{K} +{\bf S}^{K'} $, which changes relatively slowly with time, and the quickly relaxing combination,  $ {\bf S}^- = {\bf S}^{K} -{\bf S}^{K'} $.
In terms of these variables, Eq.~(\ref{dirac1}) reads:
\begin{eqnarray}
\frac{\textnormal{d}{\bf S}}{\textnormal{d}t} & = & g B_x \hat{x} \times {\bf S} + \Omega_{\rm SO} \hat{z} \times {\bf S}^-  -\Gamma_s {\bf S}, \\
\label{phen21}
\nonumber \\
\frac{\textnormal{d}{\bf S}^{-}}{\textnormal{d}t} & = & g B_x \hat{x} \times {\bf S}^{-} + \Omega_{\rm SO} \hat{z} \times {\bf S} - \frac{  {\bf S}^-}{T_v}-\Gamma^{s} {\bf S}^{-}.
\label{phen22}
\end{eqnarray}
Several approximations follow  \cite{luyi-nat15}: Due to the fast valley relaxation, it is safe to disregard the left-hand side in Eq.~(\ref{phen22}) and the last term on the right hand side. Moreover, estimates show that $ \Omega_{\rm SO}$ is much larger than the typical external magnetic field, which can be also disregarded in (\ref{phen22}). Hence,
\beq
{\bf S}^- = T_v \Omega_{\rm SO} \hat{z} \times {\bf S}.
\label{sm1}
\eeq
Let us now introduce a new relaxation rate
\beq
\Gamma_v = T_v \Omega_{\rm SO}^2.
\label{Tv}
\eeq
Substituting (\ref{sm1}) into (\ref{phen21}), we find the equation for the total spin polarization:
\beq
\frac{\textnormal{d}{\bf S}}{\textnormal{d}t} =g B_x \hat{x} \times {\bf S} - (\Gamma_v+\Gamma_s)  \left( \hat{x} S_x + \hat{y} S_y \right) - \hat{z} \Gamma_s S_z,
\label{sp2}
\eeq
which has the form of the Bloch equation with anisotropic relaxation rates.

\subsubsection{Mixtures of Hot Atomic Vapors.}

Consider a mixture of two interacting hot atomic gases, A and B.
After some approximations, the spin kinetics of this mixture is described by a set of coupled Bloch equations for spin polarizations of both atomic species \cite{two-color-2,Dellis13}:
\bea
\nonumber \f{\textnormal{d}{\bf S}^{A}}{\textnormal{d}t}&=&g_A{\bf S}^A\times {\bf B}-\gamma_A{\bf S}^{A}-\f{\gamma_{AB}}{2}(N_B{\bf S}^A-N_A{\bf S}^B),\\
\label{av-rel}
\\
\nonumber \f{\textnormal{d}{\bf S}^{B}}{\textnormal{d}t}&=&g_B{\bf S}^B\times {\bf B}-\gamma_B{\bf S}^{B}-\f{\gamma_{AB}}{2}(N_A{\bf S}^B-N_B{\bf S}^A),
\eea
where $A\equiv $ Cs and $B\equiv $ Rb, and $N_A$, $N_B$ are numbers of, respectively, Cs and Rb atoms in the observation region.  The kinetic rate $\gamma_{AB}$ describes the rate of random spin exchange interactions at scatterings of atoms of different kind. Note also that $g$-factors and individual relaxation rates for different atoms are generally different.
The relaxation matrix elements can be read from Eq.~(\ref{av-rel}):
\begin{eqnarray}
\nonumber
R^{\alpha \beta }_{ij} &=& \delta_{ij} \left[ \delta_{\alpha \beta} \gamma_{\alpha} + \frac{\gamma_{AB}(\delta_{\alpha \beta}N_{\bar{\beta}}  -\delta_{\alpha\bar{\beta}} N_{\alpha} )}{2} \right]- \\
&& \;\;\;\;\;\;\;\;\;\;\;\;\;\;\;\;\;\;\;\;\;\;\;\;\;\;\;\;\;\;\;\;\;\;\;\;\;\;\; - \delta_{\alpha \beta} g_\alpha B_x  \varepsilon_{xij},
\label{rel-mt}
\end{eqnarray}
where we defined a bar-operation as $\bar{A} = B$ and $\bar{B}=A$.

In recent experiments performed with hot atomic vapors \cite{two-color-2,Dellis13}, the temperature was several orders of magnitude higher than the characteristic energy scale for spin dynamics. This condition fixes the form of the correlation matrix elements:
 \beq
 G^{\alpha \beta }_{ij}= \frac{\delta_{ij}}{2} \left[ \delta_{\alpha \beta} \gamma_{\alpha} N_\alpha  +\frac{\gamma_{AB}N_A N_B}{2} ( \delta_{\alpha \beta} - \delta_{\alpha \bar{\beta}}) \right].
\label{corr-mt}
\eeq
\begin{figure}
\centerline{\includegraphics[width=0.9\columnwidth]{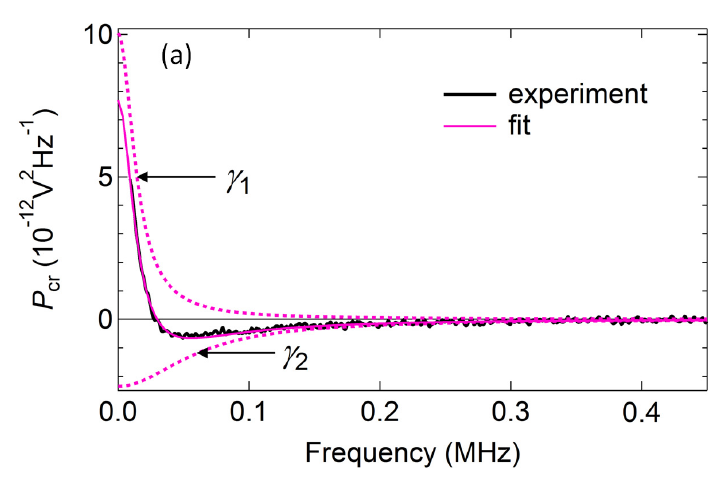}}
\caption{Cross-correlator of a mixture of Cs and Rb atomic vapors at zero magnetic field. Reprinted with permission from Macmillan Publishers Ltd: Scientific Reports 5, article number: 9573, 2015, Copyright \copyright (2015) \cite{two-color-2}.}
\label{cross-corr-fig}
\end{figure}
In Ref. \cite{two-color-2}, the  spin cross-correlator that corresponds to the Langevin equation with matrices (\ref{rel-mt}) and (\ref{corr-mt}) was investigated theoretically and compared with an experiment on the mixture of Cs and Rb. Experimental and theoretical results appeared to be in excellent agreement with each other. For example, Eq.~(\ref{npower3}) predicts that at zero magnetic field the cross correlator has a  simple form
\begin{equation}
P_{\rm  cr} (\om)=
 Q \Big(\f{\gamma_1}{\om^2+\gamma_1^2}-\f{\gamma_2}{\om^2+\gamma_2^2}\Big),
 \label{CSNPS4}
\end{equation}
which is the difference of two equal-area Lorentzians with widths $\gamma_{1}$ and $\gamma_{2}$. Here $Q= N_AN_B/(N_A+N_B)$, and the parameter $\gamma_1$ has the meaning of the effective total spin relaxation rate, while the difference $\gamma_2-\gamma_1$ is the spin exchange rate. Note that even though the exchange interactions conserve the total spin polarization, the spin exchange rate can be extracted by means of the two-color SNS from measurements of the cross-correlator. Figure~\ref{cross-corr-fig} shows perfect agreement of the theoretical prediction (Eq.~(\ref{CSNPS4})) with experimental results.

\subsubsection{Spatial Correlations.}

\begin{figure}
\centerline{\includegraphics[width=0.9\columnwidth]{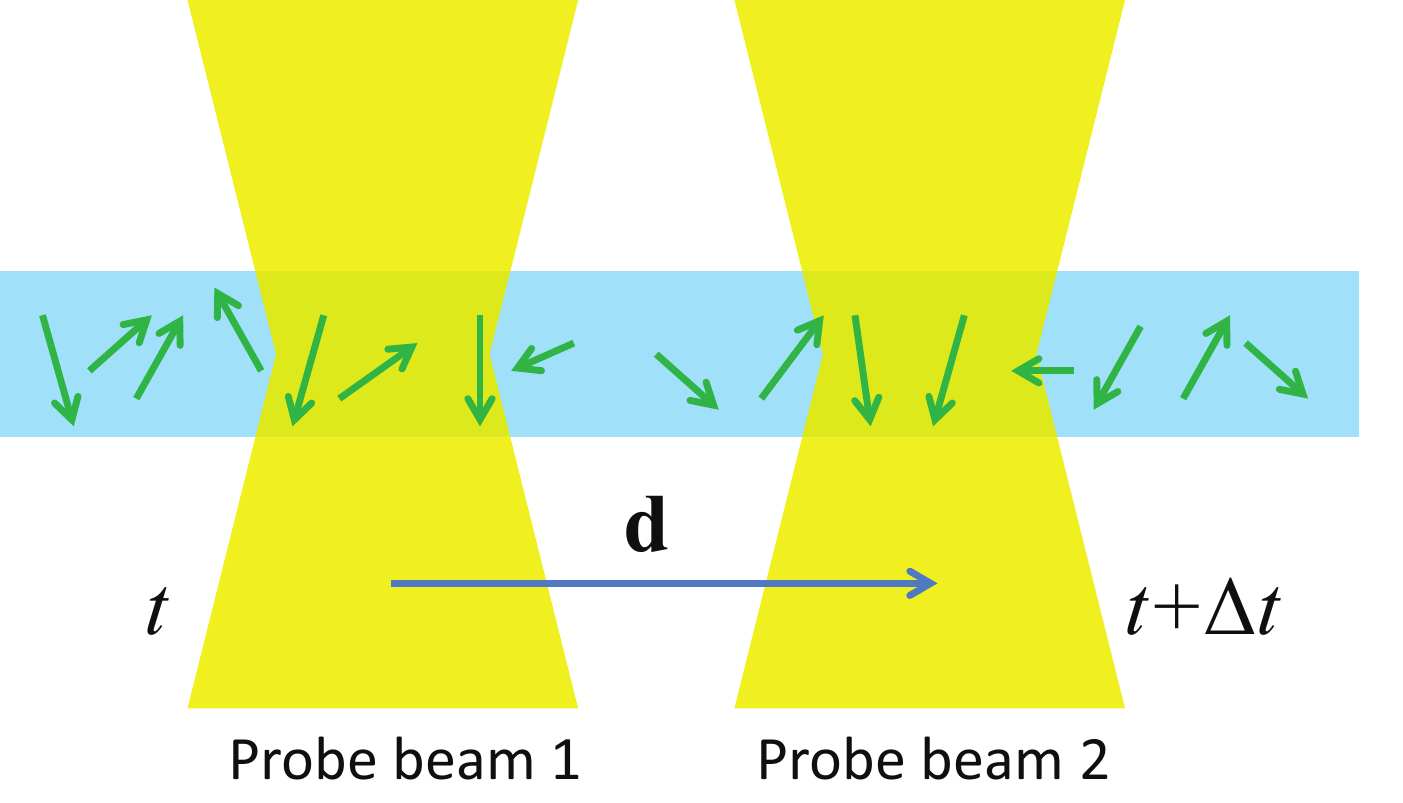}}
\caption{Two laser beams separated by a distance $\mathbf{d}$ measure correlations between time-shifted spin fluctuations at different space
locations. Y. V. Pershin et al., Applied
Physics Letters 102, p. 202405, 2013 \cite{sinitsyn-13apl}. Copyright (2013), American Institute of Physics.}
\label{two_beams}
\end{figure}

Up to this point, we considered correlation and cross-correlation functions of spin fluctuations coming from the same spatial location.
However, (cross-)correlation functions in both time and space provide more information regarding a system compared
to correlations in time alone. These correlators are particularly important for problems involving spin transport/diffusion
when we want to know how a spin fluctuation propagates through a semiconductor material or a nanowire. Optical SNS allows investigation
of the spin transport at equilibrium, namely, without any external excitation or bias voltage.

Ref. \cite{sinitsyn-13apl} presents a theory of two-beam SNS in semiconductor
wires with Bychkov-Rashba spin-orbit interaction taking into account several possible spin
relaxation channels and finite size of laser beams. A possible experiment geometry is shown in Fig.~\ref{two_beams}.
This theory predicts a peak shift with respect to
the Larmor frequency to higher or lower frequencies depending on the strength of spin orbit
interaction and distance between the beams.

\section{Methods II: Spin Noise of Conduction Electrons}


The basic framework for theoretical studies of spin noise
of itinerant electrons was developed in Ref. \cite{green1}. Here, we illustrate some of the theoretical methods
applicable to conduction electrons. Our illustration is based on the model of the Rashba 2D electrons described by the Hamiltonian
\beq
\hat{H} = \frac{\hbar^2 k^2}{2m_e} +\lambda_\textnormal{SO} (k_y \hat{\sigma}_x - k_x \hat{\sigma}_y) + V({\bf r}),
\label{rashbaH}
\eeq
where ${\bf k} $ is the electron momentum, $m_e$ is the electron mass,  $\lambda_\textnormal{SO}$ is the Rashba spin orbit coupling
and $ V({\bf r})$ is a static disorder potential. In simplified model calculations, the disorder potential is often
assumed to be Gaussian $\delta$-correlated, i.e. $\la V({\bf r}) \ra =0$ and $\la V({\bf r}) V({\bf r'})\ra =V_0^2 \delta ({\bf r}- {\bf r'})$,
where $V_{0}$ is a constant. Due to the Rashba spin-orbit interaction, the electron spins experience an effective momentum-dependent in-plane magnetic field. The elastic impurity scattering leads to fast fluctuations of this field, which in turn are responsible for the relaxation and fluctuations of spins of electrons.

\subsection{Drift-Diffusion Approach}

The spin drift-diffusion (DD) approach stays in between more rigorous calculations based on quantum Boltzmann equation
(or the Kubo formula) and the phenomenological theory. Instead of postulating the form of the Langevin equation with
phenomenological parameters, DD equations are based on the microscopic Hamiltonian (\ref{rashbaH}). These equations
are derived using a semiclassical description of the electron spatial motion
and quantum-mechanical description of the dynamics of electron spins.

DD approach is most useful to reveal the microscopic physics leading to spin noise and relaxation and to develop an intuition
for systems with more complex interactions. However, it has to be used with extra caution because it lacks the mathematical
rigor of more advanced calculation techniques.


\subsubsection{Spin Diffusion Equations}

Following Ref. \cite{arXiv_1007_0853v1}, we present a simple derivation
of spin diffusion equations for the 2D non-degenerate electron gas.
For the sake of simplicity, we assume the electrical neutrality and the absence of external
electromagnetic field.

Within DD approach, 2D electrons are
characterized by the momentum relaxation time $\tau$ and  the mean
free path $\ell$, so that the average velocity of electrons is
$v=\ell/\tau$. From elementary gas-kinetic considerations
\cite{Reif65a} we can write an equation for the change of electron
spin polarization $\Delta\mathbf{S}(x,y,t)$ in a region of
dimensions $2\ell \times 2\ell$ with the center at $(x,y)$ during
the time interval $\tau$:
\begin{eqnarray}
(2\ell)^2\Delta\mathbf{S}(x,y,t)= \frac{1}{4}v\tau
(2\ell)\left\{\mathbf{S}^\prime(x-2\ell,y,t)\right.
\label{S_flux}~~~~~\nonumber\\ +\mathbf{S}^\prime(x+2\ell,y,t)+
\mathbf{S}^\prime(x,y-2\ell,t)+\mathbf{S}^\prime(x,y+2\ell,t)~~~ \nonumber \\
\label{DDeqinit}
 \left.-4\mathbf{S}(x,y,t)\right\}.
\end{eqnarray}

In the right hand side of Eq.~(\ref{S_flux}), the first four terms
are the spin polarization fluxes into the region from four sides with
length $2\ell$, and the last term is the flux out of this region.
The prime symbols  in Eq.~(\ref{S_flux}) denote a change of spin polarization
because of SO interaction-induced spin precession by the angle $2\Omega \tau=4\lambda_\textnormal{SO} m_e\ell/\hbar=2\eta\ell$, where
$\eta=\Omega \tau / \ell$ is the spin precession angle per unit length.
For example,
\begin{eqnarray}
\mathbf{S}^\prime(x-2\ell,y,t)=\cos(2\eta\ell)\mathbf{S}(x-2\ell,y,t)-
\sin(2\eta\ell)\mathbf{ y}~~\nonumber\\
\times\mathbf{S}(x-2\ell,y,t)+ 2\sin^2(\eta\ell)\mathbf{
y}\cdot\mathbf{S}(x-2\ell,y,t) \mathbf{ y},~~~ \label{S_prime}
\end{eqnarray}
where $\bf{y}$ is the unit vector along $y-$axis.

In order to obtain spin diffusion equations, we
substitute the expressions for $\mathbf{S}^\prime$ into Eq.~(\ref{DDeqinit}), and expand
 trigonometrical functions up to quadric terms  with respect to small $2\eta \ell$
 and $\mathbf{S}^\prime$ terms up to  quadric terms with respect to $2\ell$. The resulting system of diffusion equations
for spin polarization has a form
\begin{eqnarray}
\frac{\partial S_x}{\partial t}=D\Delta S_x+C\frac{\partial
S_z}{\partial x}-2\gamma S_x \label{SxEq}, \\ \frac{\partial
S_y}{\partial t}=D\Delta S_y+C\frac{\partial S_z}{\partial
y}-2\gamma S_y \label{SyEq}, \\
 \frac{\partial S_z}{\partial t}=D\Delta S_z-C\left(\frac{\partial S_x}{\partial
x}+\frac{\partial S_y}{\partial y}\right)-4\gamma S_z
\label{SzEq},
\end{eqnarray}
where
\begin{equation}
C=2\eta D,~~\gamma=\frac{1}{2}\eta^2 D, \label{constants}
\end{equation}
and
\begin{equation}
D=\frac{\ell^2}{2\tau}. \label{Dh}
\end{equation}
Here $D$ is the coefficient of diffusion, $C$ describes spin
rotations, and $\gamma$ is the coefficient describing spin
relaxation. The same spin diffusion equations
(\ref{SxEq}-\ref{constants}) can be obtained for the model of 2D
localized electrons on a lattice in the hopping regime with the only difference that
$D=\ell^2/(4\tau)$, where $\tau$ is now the characteristic hopping time and
$\ell$ is the distance between the lattice sites. We note that Eqs. (\ref{SxEq}-\ref{constants})
can be easily extended to account for a drift term.

\subsubsection{Average Spin Relaxation.}

Next, we consider the relaxation of homogeneous spin polarization in the DD limit.
In this case, different components of spin polarization are uncoupled (see Eqs. (\ref{SxEq})-(\ref{SzEq}))
and the dynamics of their relaxation is characterized by the following spin relaxation times:
\begin{eqnarray}
\frac{1}{\tau_s^z}&=&4\gamma=\left( \frac{2\lambda_\textnormal{SO}m_ev}{\hbar}\right)^2\tau , \label{relTz}\\
\tau_s^x&=&\tau^y_s=2\tau^s_z. \label{relTxy}
\end{eqnarray}
According to Eqs. (\ref{relTz})-(\ref{relTxy}), the spin relaxation time for $S_z$ is two times shorter than the spin relaxation times for $S_x$ and $S_y$. The reason is that the effective in-plane
spin-orbit field (causing the spin relaxation) is always perpendicular to $S_z$.

\subsubsection{Spin Fluctuations.}

In order to find various correlation functions, one can assume \cite{sinitsyn-13apl} that at an arbitrary selected moment of time $t=0$ the vector of spin polarization density is given by a vector of continuous random variables $\mathbf{S}(\mathbf{r},0)=\boldsymbol{\xi}(\mathbf{r})$ such that $\langle  \xi_i(\mathbf{r}) \rangle =0$ and $\langle \xi_i(\mathbf{r}) \xi_j(\mathbf{r'}) \rangle=\lambda \delta(\mathbf{r}-\mathbf{r'})\delta_{ij}$, where $\langle .. \rangle$ denotes averaging over different realizations, $i,j=x,y,z$, and $\lambda$ is a parameter describing the strength of spin fluctuations. Using statistical considerations~\cite{Reif65a}, one can show that $\lambda=n/4$, where $n$ is the 2D electron density.

For a given realization of initial spin polarization density, the spin polarization at $t>0$ can be written using a Green's function $G_{ij}$ of the spin diffusion equations
\begin{equation}
S_i(\mathbf{r},t)=\int\limits_A G_{ij}(\mathbf{r},t;\mathbf{r'},0)S_j(\mathbf{r'},0)\textnormal{d}\mathbf{r'}, \label{eq0}
\end{equation}
where $A$ is the sample area, and  $\textnormal{d}\mathbf{r'}=\textnormal{d}x'\textnormal{d}y'$. Any new noise created in the system at $t>0$ is uncorrelated with the initial noise and thus is not included into Eq.~(\ref{eq0}).
Next, we note that a probe beam used in typical SNS setup averages the space distribution of the Faraday rotation angle
$\theta (\mathbf{r},t)=\kappa S_z(\mathbf{r},t)$ according to
\begin{equation}
\bar{\theta}(t) =\frac{1}{P_0} \int\limits_A I_m(\mathbf{r})\theta (\mathbf{r},t)\textnormal{d}\mathbf{r}= \frac{\kappa}{P_0}\int\limits_A I_m(\mathbf{r})S_z (\mathbf{r},t)\textnormal{d}\mathbf{r}. \label{av_theta}
\end{equation}
Here, $\kappa$ is a constant that couples $z$-component of spin polarization density with a local value of Faraday rotation angle, $P_0$ is the integrated laser beam intensity (power), and $I_m(\mathbf{r})$ is the space distribution of beam intensity.

In SNS experiments, the most typical determined correlation function is
\begin{equation}
R(t)=\langle \bar{\theta}(0) \bar{\theta}(t)\rangle. \label{R}
\end{equation}
Using Eqs.~(\ref{eq0})-(\ref{R}), and the expression $\langle \xi_i(\mathbf{r}) \xi_j(\mathbf{r'}) \rangle=\lambda \delta(\mathbf{r}-\mathbf{r'})\delta_{ij}$ (introduced above Eq.~(\ref{eq0})), we find the general equation determining the second order spin noise correlation function:
\begin{equation}
R( t)=\frac{\lambda\kappa^2}{P_0^2}\int\limits_A \int\limits_A I(\mathbf{r}) I(\mathbf{r'}) G_{zz}(\mathbf{r},t;\mathbf{r'}, 0) \textnormal{d}\mathbf{r}\textnormal{d}\mathbf{r'}. \label{R_final}
\end{equation}
The Fourier transform of $R(t)$ with respect to $t$ is the noise power spectrum
\begin{equation}
S(\omega)=2\int\limits_0^\infty R(t)\cos(\omega t)\textnormal{d}t. \label{S(f)}
\end{equation}
We emphasize that although Eq.~(\ref{R_final}) contains only $zz$ component of the Green function, the latter (as a solution of a system of spin diffusion equations) incorporates both transverse and longitudinal dynamics of spin polarization.

As a practical example of this approach, let us consider spin noise in quantum wires subjected to an in-plane magnetic field. Assuming a Gaussian distribution of the incident laser beam intensities along the $x$-direction, namely, $I(x)\propto \exp(-x^2/(2R_0^2))$, where $R_0$ is the beam radius, and using the Green's function of one-dimensional spin diffusion equation,
\begin{eqnarray}
G_{zz}(x,t;x',0)= \nonumber \\
\quad\quad\quad \frac{1}{\sqrt{4\pi D t}}e^{-\frac{(x-x')^2}{4Dt}}\cos\left(\eta (x-x')-\omega_Lt \right), \label{green_func}
\end{eqnarray}
where $\omega_L$ is the Larmor frequency, we find (with a help of Eq.~(\ref{R_final})) the noise correlation function
\begin{equation}
R(t) \propto \frac{\cos \left(\omega_L t\right)}{\sqrt{Dt+R_0^2}} e^{-\frac{R_0^2\eta^2Dt}{Dt+R_0^2}}. \label{res1}
\end{equation}

Assuming that the short times provide the main contribution to the Fourier transform of Eq.~(\ref{res1}) \cite{sinitsyn-13apl}, one can obtain
\begin{equation}
 S(\omega )\propto
 \frac{\tau_{DP}}{1+\tau_{DP}^2(\omega_L-\omega )^2}, \label{lorentz}
 \end{equation}
where $\tau_{DP}=(\eta^2D)^{-1}$ is the D'yakonov-Perel' spin relaxation time~\cite{Dyakonov72a,Dyakonov86a}. According to Eq.~(\ref{lorentz}), the spin noise spectrum shows a Lorentzian peak centered at $\omega_L $. The peak width is determined by the spin relaxation time.

In addition to the regular Rashba (Eq.~(\ref{rashbaH})) and/or  Dresselhaus \cite{Dresselhaus55a} spin-orbit coupling, any semiconductor structure has a random contribution to the strengths of spin-orbit interactions due to various types of disorders (such as interface fluctuations, random doping, etc.). A theory of SNS in quantum wires with randomness in the  spin-orbit coupling was developed in \cite{sod3}. This work analyzes various transport regimes and demonstrates that the spin relaxation can be very slow, and the resulting noise power spectrum can increase algebraically as the frequency goes to zero \cite{sod3}.

\subsection{Kubo Formula}
The most rigorous way to calculate the spin noise power spectrum at the thermodynamic
equilibrium is based on the Kubo formula. The application of this approach to SNS
was demonstrated in Ref. \cite{kubo-diffusion}, which
explores the effects of spatial diffusion of electrons. Here we provide a simplified example
that rederives results of some phenomenological calculations from previous sections.

First we note that, via the fluctuation-dissipation theorem, the 2nd order
symmetrized spin correlator can be expressed through the linear response function as
\beq
\chi_{zz}(t-t')= i\theta(t-t') \la [\hat{s}_z(t) , \hat{s}_z (t') ] \ra.
\label{chi1}
\eeq
 The diagrammatic technique to calculate the linear response function in terms of a power series expansion in the disorder potential  can be justified after switching to the imaginary time
 (Matsubara) representation \cite{diagram-book}. Importantly, one does not have to perform all calculations at the finite temperature before  returning to the real time. Instead,  one can consider first the formal expression for (\ref{chi1}) in the imaginary time, then take the low temperature limit and transform the expression back to the real time. The result is the Kubo-type expression for the susceptibility  \cite{burkov-kubo}:
\beq
\chi_{zz}(\Omega) = -i\Omega R(\epsilon_F-i\delta, \epsilon_F+\Omega+i\delta), 
\label{chi2}
\eeq
where
\begin{eqnarray}
\nonumber R(\epsilon_F-i\delta, \epsilon_F+\Omega+i\delta) = \\
\quad\quad
=\frac{1}{4}{\rm Tr} \left[  \hat{\sigma}_z  \la \hat{G} (\epsilon_F-i\delta )  \hat{\sigma}_z \hat{G} (\epsilon_F+\Omega+i\delta) \ra_{dis} \right].
\label{pp1}
\end{eqnarray}
Here, $\hat{G} (\epsilon_F)$ is the single electron Green function taken at the Fermi energy. We note that in (\ref{pp1}) and henceforth, the hats mark objects that are matrices in spin states, and $\la \ldots \ra_{dis}$ denotes averaging over disorder.
The trace in (\ref{pp1}) corresponds to the summation over all occupied single electron states, including the summation over the spin indexes. We also note that Eq.~(\ref{chi2}) appears different from the analogous expression, e.g., in \cite{burkov-kubo}, since we define the Fourier transform $\chi_{zz}(\Omega)$ of $\chi_{zz}(t)$ without a $2\pi$ denominator and the factor $1/4$ accounts for transition from spin-1/2 operators $\hat{s}_z$ to Pauli matrices $\hat{\sigma}_z$.

The eigenstates of the disorder free part of the Rashba Hamiltonian (\ref{rashbaH})
\beq
|u^{\pm}_{\bf k} \ra =\frac{1}{\sqrt{2}}\left( \begin{array}{l}
1 \\
\pm i e^{i\phi}
\end{array} \right)
\label{u1}
\eeq
correspond to the eigenvalues
\beq
\epsilon^{\pm}_{\bf k}=\frac{k^2}{2m} \mp \lambda_\textnormal{SO} k,
\label{e1}
\eeq
 where $k_x+ik_y \equiv ke^{i\phi}$. 
The disorder free Green function is given by
\beq
\hat{G}_0(\omega \pm i\delta, {\bf k} ) = \frac{|u_{\bf k}^{+} \ra \la u_{\bf k}^{+}| }{\omega-\epsilon_{\bf k}^{+}  \pm i\delta} +\frac{|u_{\bf k}^{-}  \ra \la u_{\bf k}^{-} |}{\omega-\epsilon_{\bf k}^{-}  \pm i\delta}.
\label{g1XX}
\eeq

The averaging over disorder in Eq.~(\ref{pp1}) can be performed in two steps. First, one should obtain the average of a single Green function.
In the self-consistent Born approximation,  this leads to the appearance of a finite self-energy.
According to \cite{sinitsyn-07prb-2}, for a Gaussian weak white-noise type of disorder, this effect for the Rashba system reduces to a simple renormalization of the parameter $\delta$
$$
\delta \rightarrow \Gamma,
$$
where
\beq
\Gamma = n_iV_0^2\frac{\nu_{+} + \nu_{-}}{4}.
\label{g1}
\eeq
Here $n_i$ is the impurity concentration, $V_0^2$ is the average square of the impurity potential $V_i({\bm r}) =V_0\delta ({\bm r} -{\bf r}_i)$,
and
$$
\nu_{\pm} = k_F^{\pm}\left| \frac{ \partial \epsilon_{\bf k}^{\pm}}{\partial k} \right|^{-1}
$$
are the Fermi surface densities of states of the two bands of the Rashba Hamiltonian.

The second effect of disorder averaging, in the self-consistent Born approximation, is the renormalization of the vertex in between two Green functions:
\beq
\hat{\sigma}_z  \rightarrow \hat{\Theta}_z \equiv a \hat{\sigma}_z,
 \label{s1}
\eeq
where
\beq
\hat{\Theta}_z= \hat{\sigma}_z + n_i V^2 \iint  \frac{\textnormal{d}^2 {\bm k}}{(2\pi)^2}  \hat{G}_0 (\epsilon_F-i\Gamma )  \hat{\Theta}_z \hat{G}_0 (\epsilon_F+\Omega+i\Gamma).
\label{s2}
\eeq
To calculate this effect, it is useful  to note that
$$
\la u_{\bf k}^{\pm}| \hat{\sigma}_z | u_{\bf k}^{ \pm} \ra =0,
\quad
\la u_{\bf k}^{\pm} |\hat{\sigma}_z | u_{\bf k}^{\mp} \ra =1.
$$
Next, in Eq.~(\ref{s2}) we switch to polar coordinates in the momentum space. Up to the off-diagonal terms that integrate to zero, the integration over the polar angle simplifies some expressions, e.g.,
$$
 | u_{\bf k}^{ \pm} \ra \la u_{\bf k}^{\mp}| \rightarrow \frac{ \hat{\sigma}_z}{2}.
$$
In the intermediate calculations, defining $\Delta \equiv \epsilon_{\bf k}^- - \epsilon_{\bf k}^+ \approx 2\lambda_\textnormal{SO} k_F$, we use the identity
$$
\frac{1}{(\epsilon_F-\epsilon_{\bf k}^+ -i\Gamma)(\epsilon_F-\epsilon_{\bf k}^- +\Omega \ +i\Gamma)} \approx \frac{1}{\Omega-\Delta +2i\Gamma} \times
$$
$$
\times \left(\frac{1}{\epsilon_F-\epsilon_{\bf k}^+ -i\Gamma} -\frac{1}{\epsilon_F-\epsilon_{\bf k}^- +\Omega \ +i\Gamma}  \right).
$$
The expression in parentheses is simplified using the fact that only its imaginary parts are substantial in the physical limit
\beq
 \epsilon_F \gg \Gamma \gg \lambda_\textnormal{SO} k_F \gg \Omega \sim \frac{(\lambda_\textnormal{SO} k_F)^2}{\Gamma}.
 \label{limit1}
 \eeq
There are many other simplifications in this case, for example,
$$
\nu_+ + \nu_{-} \approx 2m_e, \quad \nu_+ - \nu_{-} \approx -m_e (\lambda_\textnormal{SO} k_F)/\epsilon_F.
$$
Hence we can replace, e.g.,
$$
\frac{1}{\epsilon_F-\epsilon_{\bf k}^+ -i\Gamma} \approx i\pi \delta (\epsilon_F-\epsilon_{\bf k}^{+}),
$$
and then use
$$
i\pi  \iint \frac{\textnormal{d}^2 {\bm k}}{(2\pi)^2} \delta  (\epsilon_F-\epsilon_{\bf k}^{+}) = i\frac{\nu^+}{2}.
$$

After these manipulations, we obtain an equation that determines the parameter $a$:
\beq
a= 1 +\frac{2ia\Gamma}{\nu_++\nu_{-}} \left(\frac{\nu_+}{\Omega-\Delta +2i\Gamma}  +\frac{\nu_{-}}{\Omega+\Delta +2i\Gamma} \right).
\label{a1}
\eeq
Using Eq.~(\ref{g1}) and all the relations (\ref{limit1}), we find
\beq
a\approx \frac{2i\Gamma}{\Omega+2i(\lambda_\textnormal{SO} k_F)^2/\Gamma}.
\label{a2}
\eeq
Note that at this point the spin relaxation time appears in the calculations:
\beq
1/\tau_s \equiv 2(\lambda_\textnormal{SO} k_F)^2/\Gamma = (2\lambda_\textnormal{SO} k_F)^2 \tau_{\rm tr},
\label{sr1}
\eeq
where $\tau_{\rm tr}$ is the transport lifetime of conduction electrons \cite{sinitsyn-07prb-2}, defined as $\tau_{\rm tr}=1/(2\Gamma)$.

Substituting Eqs. (\ref{s1}), (\ref{a1}), into Eq.~(\ref{pp1}) we get
\beq
\chi_{zz}(\Omega) =\frac{i\Omega}{4} {\rm tr}  \iint  \frac{\textnormal{d}^2 {\bm k}}{(2\pi)^2}  \hat{G}_0 (\epsilon_F-i\Gamma )  \hat{\Theta}_z \hat{G}_0 (\epsilon_F+\Omega+i\Gamma),
\label{cc1}
\eeq
where ``${\rm tr}$" means here merely the trace over the spin indexes. In the limits (\ref{limit1}) we obtain
$$
\chi_{zz}(\Omega) =  \frac{i\Omega a}{4} \frac{\nu_+ + \nu_{-} }{2\Gamma} = -\frac{\Omega}{4} \frac{\nu_+ + \nu_{-} }{\Omega +i/\tau_s}.
$$

Finally, we can apply the fluctuation-dissipation theorem (\ref{FDT1})  to calculate the noise power spectrum.
Before making further calculations we emphasize that the majority of experiments are performed at relatively high temperatures, $k_BT \gg \hbar \Omega$. In this limit,  $ {\rm coth} \left( \frac{\hbar \Omega}{2 k_BT} \right) \approx 2k_BT/(\hbar \Omega)$ and
\beq
 P(\Omega) =\frac{ k_B T (\nu_{+} +\nu_{-}) }{2} \frac{1/\tau_s}{\Omega^2+(1/\tau_s)^2},
\label{pp44}
\eeq
Since $\left( \nu_{+} +\nu_{-}\right)/2 $ corresponds to the density of states per spin parameter $D$ in Eq.~(\ref{td1}),
the result (\ref{pp44}) coincides with Eq.~(\ref{ssc12}) at $\omega_L=0$, which was derived previously with the phenomenological approach.

\subsection{Quantum Boltzmann Equation}
An alternative calculation tool for conduction electrons is the upgraded quantum Boltzmann equation
approach  introduced in \cite{sinitsyn-13prl}, which was applied to a 2D electron system with
Rashba and Dresselhaus spin orbit coupling.

\begin{figure}
\centerline{\includegraphics[width=0.5\columnwidth]{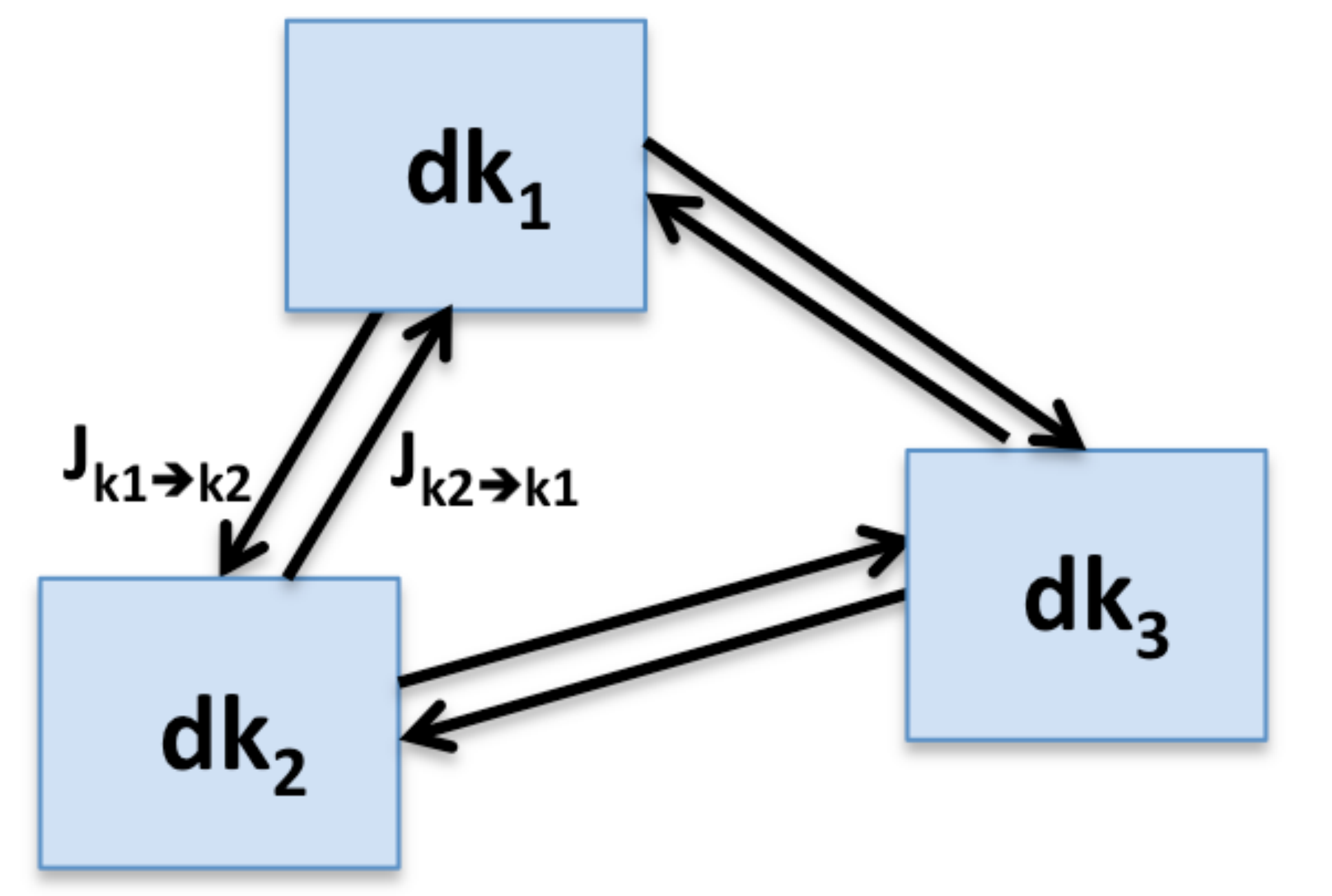}}
\caption{Spin fluctuations originate from  the shot noise of spin currents
  between mesoscopic phase space volumes.
  }
\label{phase-space-fig}
\end{figure}

The quantum Boltzmann equation is similar to the standard kinetic equation. It describes
the evolution of the single particle density matrix $\hat{\rho}$, which includes both charge
and spin degrees of freedom, changing due to ``hydrodynamic" evolution in external fields and random scatterings:
\begin{equation}
\partial_t \hat{\rho}-i[\hat{\rho},\hat{H}]=\hat{I}_{col} + \hat{\zeta} (\hat{\rho},\hat{H}),
\label{qbe}
\end{equation}
where $\hat{H}$  is the part of the Hamiltonian without random interactions, and $I_{col}$ is the collision integral describing for scatterings.
All but the last term in (\ref{qbe}) can be derived by previously developed diagrammatic
and semiclassical methods  \cite{sinitsyn-07prb-1,sinitsyn-07prb-2,sinitsyn-09jcmp,keldysh5,sn1}.
The last term in Eq.~(\ref{qbe}), $ \hat{\zeta} (\hat{\rho},\hat{H})$,   is the main addition that describes the local spin-charge {\it fluctuations} due to random scattering events of all kinds. This term becomes important only when small ($\sim
\sqrt{N}$) fluctuations are considered, which is the case of our interest now.
Solving Eq.~(\ref{qbe}), one can  derive various correlators, such as $\langle
\rho_{\alpha}(t,r') \rho_{\beta}(0,r) \rangle$, e.t.c, where $\rho_{\alpha}(t,r') $ are variables parametrizing the density matrix $\hat{\rho}$; index $\alpha$ can run, e.g., over x,y,z coefficients that define spin-1/2 density matrix in the Pauli matrix basis. For this purpose, the correlation properties of
$\zeta_{\alpha}$, such as $\langle \zeta_{\alpha}(t,r') \zeta_{\beta}(0,r) \rangle$, should be priorly obtained quantum-mechanically with scattering theory methods.

An approach to derive the statistical properties of the noise term was suggested in  \cite{sinitsyn-13prl}.
It starts from the observation that scattering processes in the momentum space happen much faster than the local spin relaxation in different phase space volumes that comprise mesoscopic numbers of electrons.
Therefore, on the time scale of the transport lifetime,  one can assume that different phase space volumes are described by approximately constant values of the spin density, while such volumes exchange
quickly fluctuating spin currents with each other, as shown in Fig.~\ref{phase-space-fig}. The spin current from  a phase space volume with a characteristic momentum ${\bf k}_1$ to the volume with a characteristic momentum ${\bf k}_2$ is denoted by ${\bf J}_{{\bf k}_1 \rightarrow {\bf k}_2}$, where different elements of the spin current vector  are $J^{\alpha}_{{\bf k}_1 \rightarrow {\bf k}_2}$, and $\alpha =x,y,z$ are spin projection components.

The problem of finding properties of the noise term in (\ref{qbe}) thus reduces to the problem of finding statistical properties of spin currents between large ``reservoirs" of electrons with different single particle density matrices
$\hat{\rho}_{{\bf k}_i}$ and $\hat{\rho}_{{\bf k}_j}$. This type of problems has been studied within the theory of electronic counting statistics \cite{nagaev-cascade}. This theory predicts that if one chooses a time scale $\delta t$ such that the number of scattered electrons during $\delta t$ is large but $\delta t$ is still much smaller than the spin relaxation time, then cumulants of the transferred spin $\delta {\bf s}_{{\bf k}_i \rightarrow {\bf k}_j} \equiv \int_{0}^{\delta t} \textnormal{d}t \, {\bf J}_{{\bf k}_i \rightarrow {\bf k}_j}$ become linear in $\delta t$. Following Ref.~\cite{sinitsyn-13prl}, one can show that in the leading order in deviation from the equilibrium,
\beq
\la \delta { s}^{\alpha}_{{\bf k}_i \rightarrow {\bf k}_j} \ra = \omega_{{\bf k}_i {\bf k}_j} ({\bf s}_{{\bf k}_i} -{\bf s}_{{\bf k}_j})  \delta t ,
\label{avspincur}
\eeq
\beq
{\rm var}\left(\delta { s}^{\alpha}_{{\bf k}_i \rightarrow {\bf k}_j} \right) = \omega_{{\bf k}_i {\bf k}_j} f_D(\epsilon)(1-f_D(\epsilon)) \frac{ \delta t}{2},
\label{varspincur}
\eeq
where
$\omega_{{\bf k}_i {\bf k}_j}$ is the scattering rate between phase space volumes $i$ and $j$, and $f_D(\epsilon)$ is the Fermi-Dirac distribution function at the energy of scattered electrons.  The scattering rate can be estimated quantum mechanically, e.g., by applying the golden rule, as discussed in \cite{sinitsyn-09jcmp}.
Equation~(\ref{avspincur}) can be used to estimate  the standard collision term, $\hat{I}_{col}$, in (\ref{qbe}), while Eq.~(\ref{varspincur}) can be used to obtain Gaussian correlators of the noise term.
Afterwards, the quantum Boltzmann equation acquires the form of a Langevin equation in the phase space, which can be studied using standard techniques that we discussed in the previous section.
We refer the reader to Ref.~\cite{sinitsyn-13prl} for further details.

Here we note that  fluctuations of the spin currents between the phase space volumes appear even when the scattering conserves spins.
It is rather a shot noise type of fluctuations that arise from discreteness of electrons. Such fluctuating spin currents conserve the total spin. However, in presence of the spin orbit coupling, they become responsible for fluctuations of the spin precession in the spin-orbit field, and, therefore, lead to fluctuations of the total spin in the system at a longer time scale $\sim 1/(\lambda_\textnormal{SO} k_F)$.  It is the strength of the quantum Boltzmann equation, in comparison to the Kubo formula, that such a physical interpretation can be developed starting from microscopic physical processes and finishing with the slowest dynamics of the total spin relaxation.

\subsection{Conduction Electrons Beyond Thermodynamic Equilibrium}
Several nonequilibrium effects that can be revealed by SNS of conduction electrons have been discussed
in  \cite{sinitsyn-13prl,sinitsyn-she,golub-noneq} within the kinetic equation approach. The application of an electric
field induces charge currents that influence the spin noise power spectrum
via the spin-orbit coupling \cite{sinitsyn-13prl}. Thus, the
 electric field $E$   acts on conducting electron spins like an in-plane magnetic
   Zeeman field  $\sim \lambda_\textnormal{SO} eE \tau_{\rm tr}$,
 where $\lambda_\textnormal{SO}$ is the strength of
 the  Rashba spin-orbit coupling, $\tau_{\rm tr}$ is the transport lifetime. This effect can be observed in the spin noise power spectrum as
 a shift of the Larmor peak of conduction (but
 not localized) electrons (Fig.~\ref{peak-split}a) and can be used, e.g. to determine
 anisotropy of spin-orbit coupling  (Fig.~\ref{peak-split}b) and separate contributions of
 localized and conducting electrons \cite{sinitsyn-13prl}.  This effect was
suggested as a tool to measure the spin orbit coupling anisotropy.


\begin{figure}
\centerline{\includegraphics[width=0.9\columnwidth]{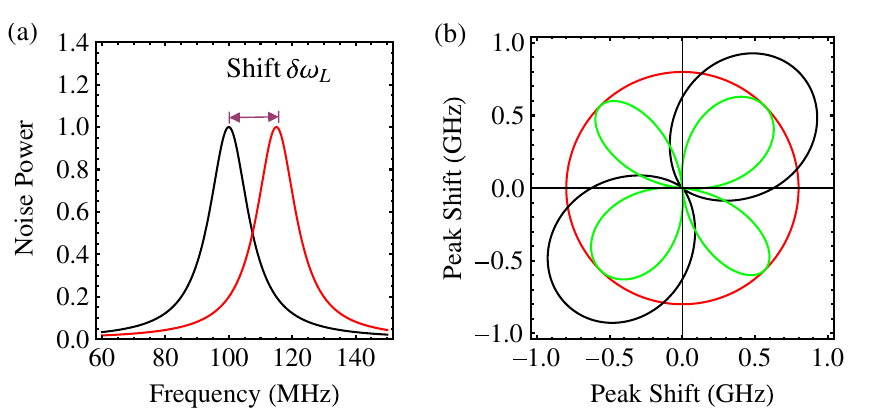}}
\caption{(a) Noise power peak shift in electric
  field and (b) its magnitude vs angle of electric field direction for several different ratios of Rashba and Dresselhaus spin orbit couplings.
    Reprinted figure with permission from F. Li et al.,
  Physical Review Letters 111, p. 067201, 2013 \cite{sinitsyn-13prl}.
  Copyright \copyright (2013) by the American Physical Society.
  }
\label{peak-split}
\end{figure}

Another interesting research direction is to explore the regime of a strong electric field
(above 10 Vcm$^{-1}$) in GaAs.  At such conditions, electrons dissipate energy by exciting localized electrons from the donor band \cite{avalanch,avalanch2}, creating
 considerable correlations and avalanches of electrons in the
 conduction band. Spin noise  in this regime is hard to predict at present, e.g., because
the effect of dynamics of recombination processes on spin relaxation is unclear. Kinetic equation is particularly suitable to study such a strongly nonequilibrium regime theoretically
because it provides self-consistent description even in the cases
when only phenomenological justification exists.  An example of such theoretical studies of spin fluctuations in the streaming regime
of  ballistic electrons can be found in \cite{golub-noneq}. Finally, we point to the recently developed kinetic equation approach based on the Keldysh technique that was applied to nonequilibrium effects in SNS \cite{glazov-keldysh}.

\subsection{Higher Order SNS and Stochastic Path Integral}

The  noise term in  Eq.~(\ref{qbe}) is generally
non-Gaussian and depends  nontrivially on the size of a spin fluctuation, e.g., in the case of many-body exchange interactions.
Calculations of higher order spin correlators from the quantum Boltzmann equation would require the knowledge of
the higher order correlations of  the noise term, such
as $\langle \zeta_{\alpha}(t_3,{\bf r}_3) \zeta_{\beta}(t_2,{\bf r}_2) \zeta_{\gamma}(t_1,{\bf r}_1)
\rangle$. Higher order statistics of microscopic currents has been the subject of research in the context of the electronic shot noise.
One powerful method to obtain statistics of fluctuating {\it charge} currents is
the Levitov-Lezovik formula \cite{ll1,ll2,ll3,ll4,ll5}. However, in our situation the
statistics of spin currents is harder to define since the derivation of the Levitov-Lesovik formula assumes a charge
measurement, i.e. the measurement
operator should commute with the density matrix before and after the
evolution. Hence, extension of the theory of full counting statistics to spin currents
currently remains poorly developed. One possible way around this problem was suggested in \cite{fuxiang-15}, where universality of some shapes of higher order cumulants was found that was justified by the law of large numbers and the fluctuation theorem.

The derivation of the stochastic Boltzmann equation as well as finding its solution in the case of non-Gaussian noise terms require a special approach.
A possible way is to combine Eq.~(\ref{qbe}) with the method of the {\it
   stochastic  path integral}, which has been previously applied to
nonlinear stochastic equations
 \cite{path-int1, sinitsyn-07prl,sinitsyn-09pnas}. Pedagogical introduction to the path integrals with applications to the higher order SNS can be found in  \cite{fuxiang-c4,sinitsyn-13njp}.

The idea is to introduce a generating function that is the sum over all possible stochastic trajectories $\tau$  of the  single particle density matrix $\hat{\rho}$, weighted by a counting field $\chi_c$, e.g.
\begin{equation}
Z(\chi_c)=\sum_{\tau} P_{\tau} e^{\chi_c f_{\tau}},
\label{path-int}
\end{equation}
 where $P_{\tau}$ is the probability of a trajectory $\tau$ and
 $f_{\tau}$ is the counted variable, e.g. the change of the magnetization
 in the observation region produced by a trajectory. Knowing $Z(\chi_c)$ one can calculate an arbitrary correlator of $f_{\tau}$ by
 taking derivatives of $Z(\chi_c)$ at $\chi_c=0$.
 For a mesoscopic number of observable electrons,  $Z(\chi_c)$ can be written in the form
 of a familiar path integral over system variables \cite{path-int1,fuxiang-15}:
 \begin{equation}
Z(\chi_c,T)= \int \textnormal{d}{\bm \rho}(t) \int \textnormal{d}{\bm \chi}(t) e^{\int\limits_0^T \textnormal{d}t\, \{ i \dot{{\bm \rho}} {\bm \chi} +H({\bm \rho}, {\bm \chi}, \chi_c)  \}},
\label{path-int2}
\end{equation}
 where ${\bm \rho}$ is the vector that parametrizes elements of the
 density matrix $\hat{\rho}$ in (\ref{qbe}) and  ${\bm \chi}$ is a
 conjugated variable. The function $H$ plays a similar role to a
Hamiltonian, which is very different from the physical Hamiltonian
$\hat{H}$ in Eq.~(\ref{qbe}). The form of Eq.~(\ref{path-int2}) is the starting point for
 various well justified approximations, for example, the limit of a
 large number of spins corresponds to the semiclassical limit in the
 path integral. This technique can  be used
 to  integrate over fast degrees of freedom and obtain an
 effective significantly simplified description of relatively slow
 processes without the loss of information about their fluctuations,
 including higher order correlations \cite{sinitsyn-09pnas}. Such a time-scale separation   happens typically in applications of Eq.~(\ref{qbe}). Indeed, usually spin relaxation is orders of magnitude slower than the electron scattering times. There is generally a hierarchy of such different time-scales in interacting electron systems, with fast processes eventually influencing the slow spin dynamics. This situation is ideal for application of the technique  \cite{sinitsyn-09pnas}. Finally, we note that the path integral technique can incorporate purely quantum effects within the weak measurement framework, as it is discussed in \cite{jordan-weak1, jordan-weak2}.

\section{Methods III: Discrete Spin Models}

In this section, we review spin noise characteristics of several specific
discrete spin models. These models have attracted the attention from SNS community for their applicability to experimental systems that can be studied by
SNS techniques at frequencies below 1 THz.

\subsection{Symmetric Central Spin Model}

The central spin models describe interactions of a single spin,
which is usually the electron spin-1/2,
with  bath spins, which are usually the nuclear spins (Fig.~\ref{central-spin-fig}).
Recently, new realizations of the central spin model were obtained experimentally based on magnetically doped semiconductor quantum dots that can, potentially, also be probed by SNS \cite{crooker-15nano}.

In the simplest model of this type, it is assumed that  interactions between the central and nuclear bath spins are isotropic and of the same strength. In this case, the model Hamiltonian is given by
\beq
\hat{H}_{\rm cs} =g \sum_{i=1}^N \hat{{\bf S}} \cdot \hat{\bf s}^i,
\label{isotr}
\eeq
where $\hat{{\bf S}} $ is the central spin operator and $\hat{\bf
  s}^i$ is the spin operator of the $i$-th nuclear spin, $N$ is the
number of nuclear spins, and $g$ is the coupling constant. Usually the goal is to find the correlator for the central spin projection along the
$z$-axis,  $\la \hat{S}_z(t)
\hat{S}_z(0) \ra$,  because nuclear spins are optically inactive.

\begin{figure}
\centerline{\includegraphics[width=0.3\columnwidth]{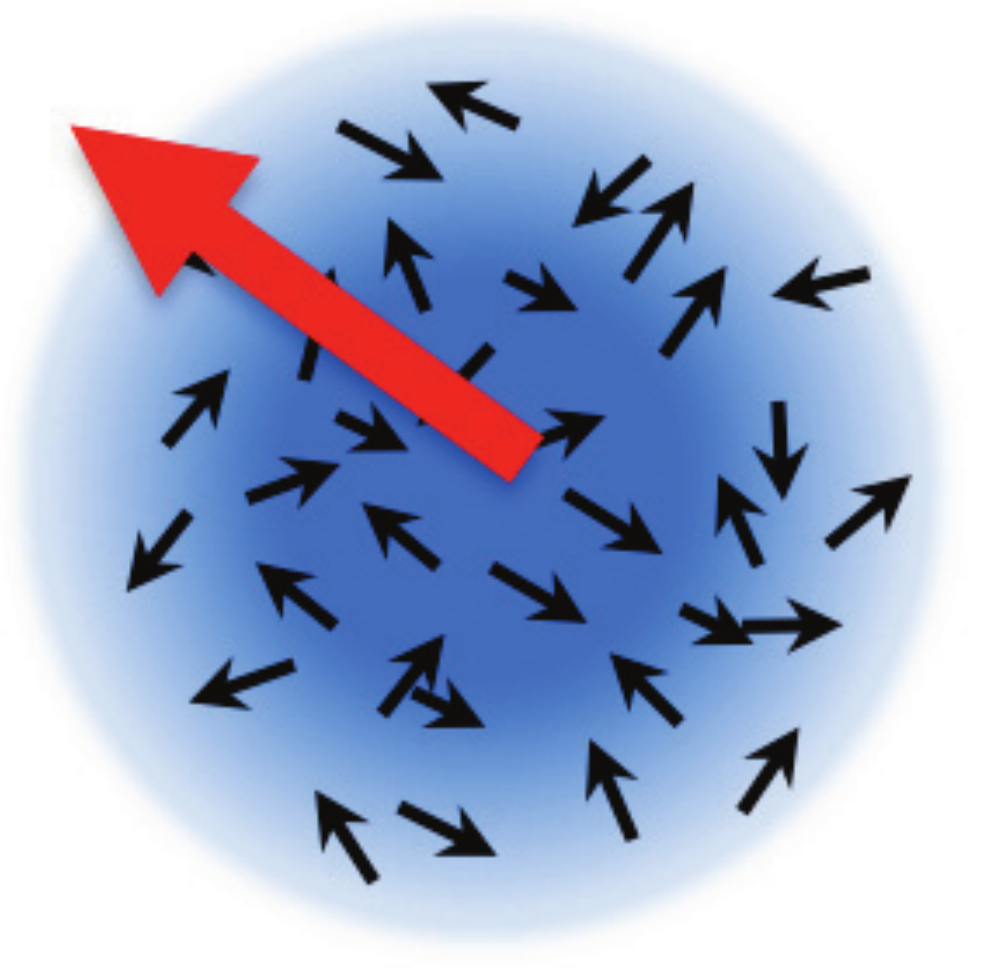}}
\caption{The central spin model describes a single spin interacting
  with a bath of $N$ nuclear spins.
  }
\label{central-spin-fig}
\end{figure}

Equation (\ref{isotr}) model can be solved exactly \cite{central-cold0,central-echo1}. Introducing the spin operator of the nuclear spin bath,
$\hat{\bf I}_N=\sum_{i=1}^N \hat{\bf s}_i$, and the
operator of the total spin, $\hat{\bf I}\equiv \hat{\bf S}+\hat{\bf I}_N$, one
can rewrite the Hamiltonian (\ref{isotr}) as
\beq
\hat{H}_{\rm cs} =\frac{g}{2} \left(  \hat{\bf I}^2 -\hat{\bf I}_N^2
  -\hat{{\bf S}}^2 \right).
\label{isotr1}
\eeq
Taking into account that $\hat{\bf S}^2=3/4$ for the central spin-1/2, and that  the operator $\hat{\bf I}_N^2$
commutes both with  $\hat{\bf I}^2$ and the projection operator of the
total spin along the $z$-axis, $\hat{ I}_z$, the Hamiltonian (\ref{isotr})  can be
diagonalized in the basis of eigenstates of $\hat{\bf I}^2$
and $\hat{ I}_z$. The construction of these eigenstates in terms
of eigenstates of  $\hat{\bf S}$ and  $\hat{\bf s}_i$ is described in textbooks on quantum mechanics \cite{LL-quantum}. Details for the model (\ref{isotr1})  can
be found in \cite{central-echo1}.


Typically, the number of nuclear spins is very large ($N \sim 10^4-10^6$). In the limit of $N\rightarrow \infty$, it was found that  the
quantum-mechanical exact result for the 2nd order spin correlator (found within Eq.~(\ref{isotr}) model) coincides with the prediction of a mean field approach that treats the effect of the nuclear spin bath on the central spin in terms of a constant magnetic field with a magnitude taken from the Gaussian distribution \cite{merkulov}.
According to the full quantum-mechanical solution for $N\gg 1$, at high temperatures, the central spin noise power spectrum has the form shown in Fig.~\ref{broad-dots} \cite{central-echo1}.
This solution does not explain discrepancies  between experimental observations and results of the semiclassical theory \cite{sinitsyn-12prl-2}.

The major difference between the full quantum-mechanical solution and the semiclassical solution  is
the quantum echo effect. The semiclassical theory predicts that after the averaging over all possible
initial states of the bath spins, the central spin polarization relaxes (on average) to $1/3$ of its initial value. The quantum-mechanical model
makes the same prediction, however, for a finite $N$, there is a time $\tau
\sim 1/g$ when the central spin returns to the fully polarized state.  Such a behavior, however, has never been
observed experimentally, which is one more proof that Eq.~(\ref{isotr}) model is too
simplistic to describe  realistic central spin systems.

The exact solution becomes more useful either in the low temperature regime
($k_BT < g$)  or when the nuclear spin bath is strongly polarized. In  such cases, quantum correlations within the
ground state become substantial and the semiclassical solution may become
insufficient. The partition function of Eq.~(\ref{isotr}) model
\cite{central-cold0} shows, for example, that this model does
not have a phase transition at any finite temperature.
In the low temperature regime, spin noise in the central spin
model was discussed in \cite{central-cold1}. The exact solution also
results in some nontrivial features when higher than 2nd order spin correlators are considered \cite{central-echo1}.

\subsection{Disordered Central Spin Problem}
In order to explain the experimentally observable spin relaxation of localized electronic
spins, the central spin model was generalized to include random
coupling effects as follows \cite{merkulov}:
\beq
\hat{H}_{\rm cs} = \sum_{i=1}^N g_i\hat{{\bf S}} \cdot \hat{\bf s}^i,
\label{isotr4}
\eeq
where parameters $g_i$ are taken from some distribution. It turns out
that the model (\ref{isotr4}) is  solvable by the Gaudin anzatz
\cite{Faribault2013,central-gaudin2}, which reduces the complexity
to that of a system of nonlinear algebraic equations. Unfortunately, the time
to solve such equations numerically becomes too long when the number
of nuclear spins is $N \gtrsim 40$. However, it suffices to observe the trends of this model.
Alternatively, the model (\ref{isotr4}) was explored
numerically by simulating the quantum-mechanical evolution directly. However, exact numerical algorithms can treat no more than 20-30 spins \cite{central-echo1,central-numerics1,central-numerics2,Faribault2013}.

Fortunately, it was discovered in \cite{dobrovitski} that a semiclassical
approach that is based on an enforced factorization of the density matrix
produces indistinguishable results for spin relaxation calculations from the results of exact algorithms when the
spin bath is not polarized and the number of spins $N> 20$. This allowed studies of the central spin
relaxation with parameters that are close to realistic, e.g.,
simulations of $N=50000$ nuclear spins were performed in
\cite{alex-nat}.

Numerical simulations have demonstrated that the effect of
randomness of hyperfine couplings on the central spin relaxation is unexpectedly
weak, especially in presence of an out-of-plane anisotropy, which is typical
for hole-doped quantum dots \cite{sinitsyn-12prl-2,central-numerics1}. This fact can be
attributed to the integrability of the model (\ref{isotr4}), as
discussed in \cite{central-numerics2}. Hence, there is a considerable
experimental evidence that the model (\ref{isotr4}) is still insufficient to
describe the spin relaxation in semiconductor quantum dots.
A relatively strong quadrupole interaction (see  Eq.~(\ref{Ham-q}))
should be taken into account to explain some experimental results \cite{Li12,sinitsyn-12prl-2,alex-nat,electron-dot}.

\subsection{Spin Cluster Model}

This model describes a system of mutually interacting  spins subjected to the external magnetic
field applied along the $z$-axis. The system Hamiltonian
is given by \cite{spin-bag2, spin-bag3}
\beq
\hat{H} = g \sum_{i>j=1}^N \hat{\bf s}^i \hat{\bf s}^j +{\bf \Omega} \sum_{i=1}^N {\bf \hat{ s}}^i ,
\label{spin-bag1}
\eeq
where $g$ is the coupling constant,  and ${\bf \Omega}$ is the
external magnetic field. The eigenstates and eigenvalues of this
model (and hence the spin correlators) can be found similarly to these in the symmetric central spin model,
since the Hamiltonian (\ref{spin-bag1}) invovles only the commuting operators $I^2$ and $I_z$,
where $\hat{\bf I}\equiv\sum_{i=1}^N \hat{\bf s}^i$.

Let us consider an insulating semiconductor in the vicinity of transition
to the conduction regime. Imagine that there are
isolated islands constructed of well connected states.
In this case, the electrons can hop
between connected states of the same island and, hence, interact with each other.
In Ref.~\cite{spin-bag1}, Eq.~(\ref{spin-bag1}) model  was used  to mimic effects of the exchange
interaction among electrons inside such an island.

Similarly to the symmetric central spin model, Eq.~(\ref{spin-bag1}) model shows
somewhat trivial behavior in the case of $N \gg 1$ and high
temperatures \cite{spin-bag2, spin-bag3}. The spin noise power spectrum is the same as in a
classical model in which spins rotate around an effective static field
produced by a static spin fluctuation of the order of $\sqrt{N}$. Hence, the
spectrum of Eq.~(\ref{spin-bag1}) model strongly resembles the spectrum
of the central spin model shown in Fig.~\ref{broad-dots}(b).


\begin{figure}
\centerline{\includegraphics[width=0.34\columnwidth]{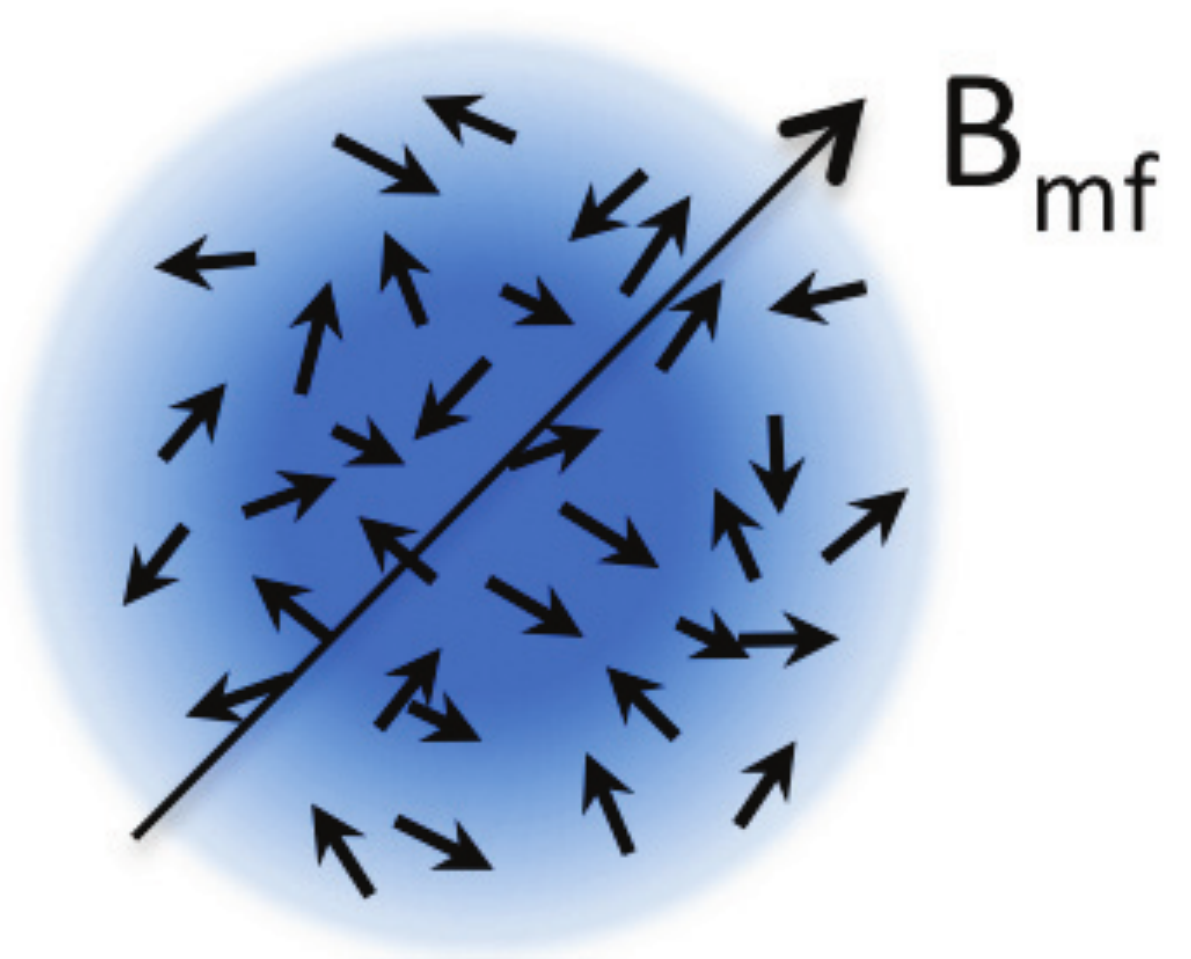}}
\caption{The spin cluster model describes the interaction of $N$ identical
  spins. Each spin couples equally to all other spins. At $N\gg 1$, the
  mean field description in which each spin precesses around an effective magnetic
  feild $B_{\rm mf} \sim g\sqrt{N}$ reproduces the exact spin noise power spectrum. }
\label{spin-bag-fig}
\end{figure}

\subsection{Central Spin Problem at Higher Temperatures}

The Langevin equation can be used to describe the relaxation of a spin localized in
a quantum dot \cite{central-phen1,central-phen2}. In this approach, the nuclear spin
effects can be taken into account by a term describing the coupling of the central spin
to the nuclear Overhauser field, which is defined in Eq.~(\ref{over1}).
Then, the Langevin equation is written as
\beq
{\dot{\vS}} = ({\bf \Omega}+{\bf \Omega}_\textnormal{O}) \times \vS -\Gamma {\bf S}  +\vxi(t),
\label{glazov-1}
\eeq
where ${\bf \Omega}$ corresponds to  the external magnetic field, ${\bf \Omega}_\textnormal{O}$
corresponds to the time-independent component of the Overhauser field, and $\Gamma$ is the relaxation rate.
It is assumed that ${\bf \Omega}_\textnormal{O}$  is taken from the Gaussian distribution with parameters discussed in \cite{merkulov}.

The authors of \cite{central-phen1} have shown that
the noise term $\vxi(t)$ introduces a Lorentzian
broadening of the zero frequency peak such as the one shown in Fig.~\ref{broad-dots}(b). By
increasing the strength of this noise, they observed the disappearance
of the Gaussian peak and transformation of the noise power spectrum
into a single Lorentzian peak centered at zero frequency.

Originally,
Eq.~(\ref{glazov-1}) was proposed to
describe the effect of nuclear spin bath dynamics \cite{central-phen1}. Such an
approach seems, however, to be an oversimplification for such a complex many-body
problem as the central spin interacting with nuclear spins. While the
Overhauser field does have its own dynamics, this dynamics is not independent of the
dynamics of the central spin, what is typically the case in
electron-doped quantum dots. Moreover, even when the feedback from the
central spin can be disregarded, as, e.g., in hole-doped
quantum dots \cite{sinitsyn-12prl-2}, the Overhauser field
fluctuations have a finite  amplitude and correlation time. The form
of this field correlator is also nontrivial. Such details, however, are needed,
e.g., to explain the quick suppression of central spin relaxation by
an out-of-plane magnetic field \cite{Li12}.

Instead, it is expected that Eq.~(\ref{glazov-1}) approach is
applicable to the electron-doped quantum dots at relatively high
temperatures ($T>10$K). In this situation, the phonon-related mechanisms dominate in
spin relaxation. Therefore, the relaxation and noise terms in Eq.~(\ref{glazov-1}) are no longer related to the spin bath physics.
At the same time, the effect of the Overhauser field on the
coherent precession of the central spin cannot be disregarded. As
measurements of spin correlators in electron doped quantum dots have
become available \cite{electron-dot,alex-nat}, an opportunity appears to
test predictions of this Langevin equation approach.

\subsection{Spin Hopping Models}
It was shown theoretically that the electron hopping between quantum dots influences
the spin noise properties \cite{central-hopping1}. In this work, the evolution of electron spin density
${\bf S}_i$ at the site $i$ was described by an equation including a hopping term and a term of
interaction with the local nuclear Overhauser field, ${\bf \Omega}_\textnormal{O}^i$,  \cite{central-hopping1}:
\beq
\frac{\textnormal{d}{\bf S}_i}{\textnormal{d}t}= {\bf \Omega}_\textnormal{O}^i \times {\bf S}_i + \sum_{j} [W_{ij}{\bf S}_j -W_{ji} {\bf S}_i],
\label{hoppingH}
\eeq
where $W_{ij}$ is the hopping rate between sites $j$ and $i$.  In addition to electronic spin dynamics \cite{pershin03a}, such models have been extensively studied, e.g., in the field of molecular spintronics
\cite{dobrovitski-hopping}, so  spin relaxation in  such models is well understood.

When the hopping rate is small compared to $|{\bf \Omega}_\textnormal{O}|$, the electron spin
makes many rotations around the local field before the electron jumps to a different
dot. In this case, the spectrum is an average of single quantum dot spectra
with a Gaussian distribution of $ {\bf \Omega}_\textnormal{O}^i$, as shown in Fig.~\ref{broad-dots}(b).
On the other hand, in the large hopping rate limit,   electron spins experience quickly
fluctuating Overhauser fields. As in the case of the Dyakonov-Perel  spin relaxation,
a randomly fluctuating nuclear field causes an exponential electron spin relaxation \cite{pershin03a}.
Therefore, the noise power spectrum is Lorentzian in
this limit. The width of the Lorentzian peak is inversely proportional to the
hopping rate, i.e. faster transitions lead to a longer spin life-time
(the motional narrowing).  The transition between these two regimes and
implications for the spin noise power spectrum are discussed in more detail in  \cite{central-hopping1}.

We also note that quantum effects at coherent hopping of electrons
between two  singly and doubly charged coupled quantum dots (a quantum dot molecule)
were discussed in \cite{dibyendu}. The Coulomb coupling and hopping are
manifested then in specific noise power spectrum peaks. The authors of
\cite{dibyendu} pointed to an interesting fact: Since different quantum dots
have different optical properties, the  hopping between different
quantum dots, even without any spin relaxation, will produce measurable
fluctuations of the Faraday rotation angle. This effect can be used to
characterize many parameters, which may not be
directly related to spin relaxation in  nanostructures.

\subsection{Spin Diffusion in Disordered Spin Lattices}

The application of SQUID- and cantilever-based magnetometry to disordered interacting spin
systems has raised a theoretical interest to spin noise in systems with a model
Hamiltonian of the type  \cite{PhysRevB.86.134414}
\beq
\hat{H} = \sum_{ij} J_{ij} {\bf S}_{i} \cdot {\bf S}_j,
\label{disorder}
\eeq
where $J_{ij}$ are coupling constants for spins $i$ and $j$. These
constants are often selected randomly based on a distribution
$P(J_{ij})$, which mimics the complex RKKY oscillatory-type interaction
among randomly positioned spins.

A cantilever or SQUID measures   spin polarization in a small region.
Although the Hamiltonian (\ref{disorder}) conserves the total spin
polarization,  spin diffusion between the observation and outside regions leads to
observable fluctuations. Usually, Eq.~(\ref{disorder})-type models displays a power law
($1/\omega^{\alpha}$) dependence in the spin noise power spectrum
defined by a non-universal parameter
$\alpha$. It's a non-trivial task to express $\alpha$ through model parameters.

In two recent theoretical papers \cite{mozyrsky-1,ivar-renorm}, successful approaches to
predict the spin noise power spectrum for Eq.~(\ref{disorder})-type models were developed.
The paper by Sykes {\it et al.} \cite{mozyrsky-1} presents a type of mean field
approach  based on the Bloch-Redfield theory. The authors of \cite{mozyrsky-1}
assume that the dynamics of any spin is influenced
by an effective field from all other environmental spins. An important
addition to the standard mean-field approach is the assumption that
the environmental field has a noisy component with a correlator that can
be determined self-consistently.

The second successful method is a numerical renormalization group approach
\cite{ivar-renorm}. This method starts with identifying a pair of spins with
the strongest effective coupling.  Assuming a static environment, the fast dynamics of this pair
of spins is found analytically including its contribution to the noise power
 spectrum. After that, the selected pair of spins is  replaced by an effective
 single spin with some updated couplings to other spins. This process is
 repeated over and over again so that at each step a new pair of spins with the strongest coupling is selected.
 It was shown in \cite{ivar-renorm} that for a
 disordered 1D chain with random couplings $J_{ij}$, the numerical renormalization group approach
finds the noise power spectra, which are almost
 indistinguishable from these calculated using the exact diagonalization.

The mean field and numerical renormalization group approaches complement
each other. The former is expected to work better for long-range
interactions, while the latter is best suited for short-range couplings.

с\section{Perspectives}


In this Review, we have examined a variety of different applications of SNS
ranging from  conduction electrons to atomic spin systems and quantum dots.
Potentially,  SNS can be applied to any system that demonstrates
measurable Faraday or Kerr rotation effects in the states with finite spin polarization, including cuprates and unconventional
superconductors. However, for some potential applications, the required sensitivity and extension of
the frequency bandwidth to the terahertz
range are yet to be demonstrated.

Continuous instrumentation and methodological improvements of SNS
that we have been observing during the last decade give confidence that
many new applications of SNS will be developed. Along this path,
 SNS should not be perceived as merely an alternative way to determine basic linear
response characteristics. Instead, SNS enables an essentially new
methodology in materials science.  Large streams of information --
gigabytes per second -- can now be processed  to extract  useful spin
correlators. The complexity of higher-order spin
correlators,  which are multidimensional in the frequency  space, creates new challenges of
interpreting large amounts of information. Such new information could be most suitable for simultaneous
multiparameter fits, e.g., to infer the full spin Hamiltonian with
all parameters ``at once" by advanced statistical methods \cite{vapnik} and signal processing \cite{signal-book}.


Below, we review several currently challenging research directions that could be possibly
addressed by SNS within the foreseeable future.


\subsection{Full Counting Statistics}

There is a strong interest in advanced statistical methods to explore
some highly unusual characteristics including the ones that have
never been detected in the condensed matter systems.
For example,  SNS has a potential to investigate the {\it global properties} of
the generating function containing the information about all order spin correlators.
Indeed, since the optical SNS directly measures the phase shift
acquired by the polarized beam, $e^{i\theta_F(t)}$, where  $\theta_F(t)$ is the Faraday rotation angle,
sufficiently long experimental data sets can be used to find the average of any power of this exponent,
$\langle [e^{i\theta_F(t)}]^{\lambda} \rangle$.
Since $\theta_F(t)\sim S_z(t)$,  SNS thus allows
experimental determination of the following generating function:
\begin{equation}
 Z(\lambda,t) = \langle e^{i\lambda \hat{S}_z(t)}e^{-i\lambda \hat{S}_z(0)} \rangle \equiv \langle e^{i\lambda \delta {S}_z(t)} \rangle ,	
\label{gf1}
\end{equation}
where $\hat{S}_z$  is the operator of the total spin polarization along the $z$-axis and $\lambda$ is the {\it counting parameter}.


\subsubsection{Phase Transitions in Rare Event Statistics.}
The possibility to
 observe phase transitions in statistics of rare events attracts a lot
 of theoretical attention \cite{fcs-pt1,ph-tr2,ph-tr3,ph-tr4,liu-LY,sinitsyn-13pre}.  However, these
 critical phenomena, which provide unique opportunities to understand strongly correlated
 systems, remain practically unstudied experimentally.
 Usually, phase transitions at the fluctuation level occur when a system is not at the critical state
 but close to it in some parameter space.

In 1952, Lee and Yang  \cite{LY0,LY1} showed that, in a large class of
interacting spin systems, the partition function becomes
zero at certain points on the complex plane of fugacity or a magnetic field.
In the case of a general Ising ferromagnet, all the Lee-Yang zeros are located
on the unit circle in the complex fugacity plane. When the number of
interacting spins is large, such zeros merge and create a branch cut
(Fig.~\ref{LY}). A nice approach to test the Lee-Yang zeros was proposed
in \cite{liu-LY} showing that a probe spin coherence demonstrates sudden death and
birth at the critical times corresponding to the Yang-Lee singularities.
$Z(\lambda,t)$ given by Eq.~(\ref{gf1}) could be considered as a probe in analogy to the probe spin coherence
discussed in \cite{liu-LY}.

Relation to  phase transitions at fluctuation level can be established when the number of
interacting spins is large. Lee-Yang zeros  then  merge and create a branch cut
(Fig.~\ref{LY}). When $q\equiv e^{i\lambda}$ enters the values at  this cut,
generating functions, such as in Eq.~(\ref{gf1}), may experience a sharp change of behavior \cite{liu-LY}. Critical exponents in the vicinity of this  transition can be derived similarly to how it is done in  applications of Lee-Yang zeros to the theory of classical phase transitions.

In a recent experiment performed by analogy with liquid NMR
quantum computing,  Lee-Yang zeros of the partition function with a
complex fugacity were observed in molecular nuclear spins \cite{Liu-zeros}. In other recent
inspiring papers \cite{flindt-13prl,flindt-13pre}, it was demonstrated theoretically that in order to explore the
phase transitions at the  fluctuation level, unusual rare events do not need to be detected.
Namely, it was shown that phase transitions at the fluctuation level lead to time-dependent
oscillations of potentially measurable lowest cumulants. By tracing
such oscillations, it is possible to explore the discontinuities of
$Z(\lambda,t)$  \cite{flindt-13prl,flindt-13pre}.
Additionally,
it was shown in Ref.~\cite{sinitsyn-zeno} that the well known disappearance of the quantum coherent
oscillations due to strong coupling to a detector, near the onset of the  quantum Zeno effect
\cite{zeno1,zeno2,zeno11,zeno21,zeno31}, can be  manifestation of such a
discontinuity of the generating function of the spin noise statistics.
This proves that, in some sense,  phase transitions at the fluctuation level have been
already observed experimentally, in particular,  in solid state qubits
\cite{zeno-exp1}. The authors of Ref.~\cite{sinitsyn-zeno} also argue
that the quantum Zeno effect can be further explored by SNS.

 \begin{figure}[t]
\centerline{\includegraphics[width=0.9\columnwidth]{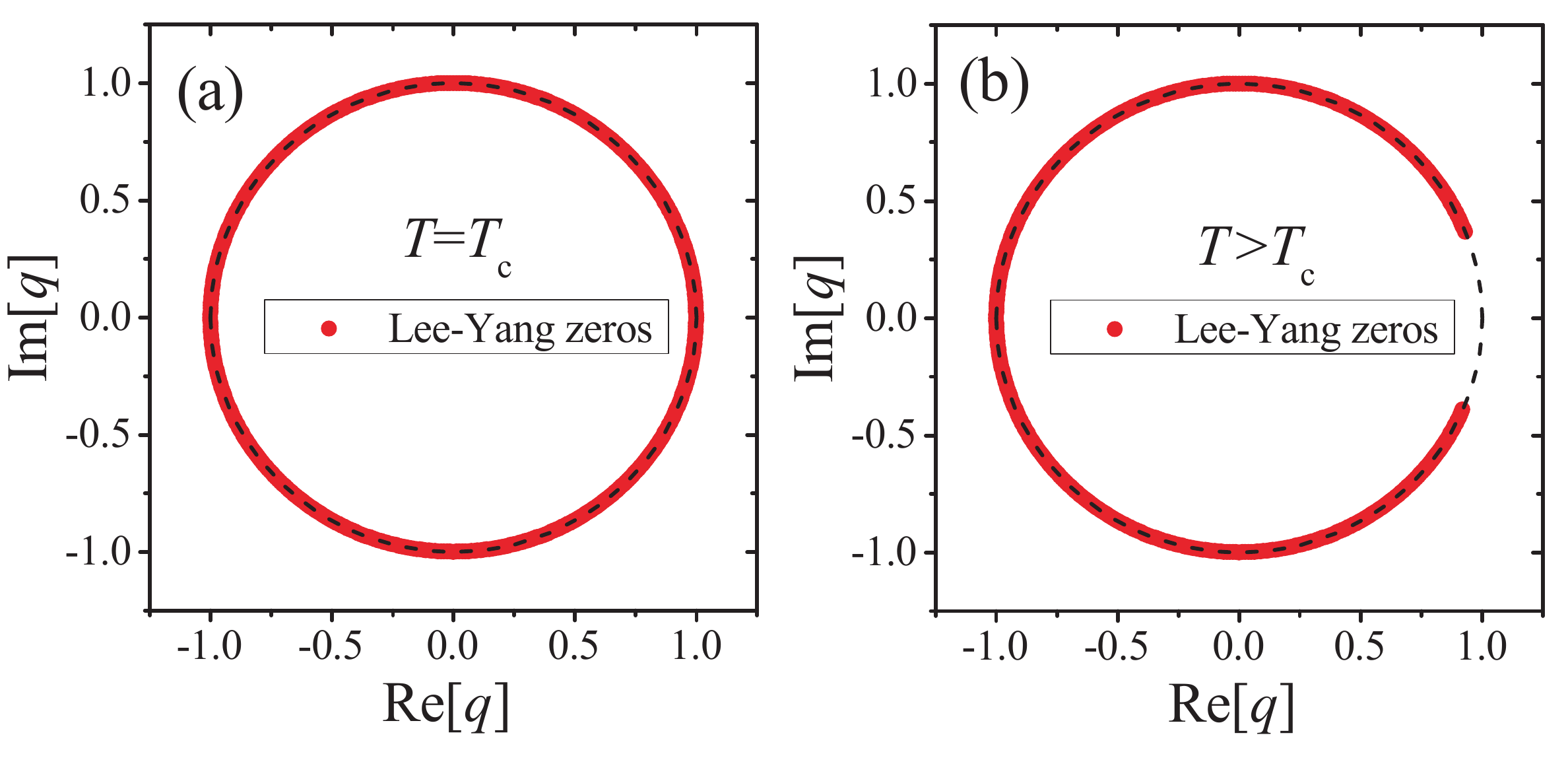}}
\caption{Positions of Lee-Yang zeros on the unit cicrle (a) at the critical temperature,
and (b) for a paramagnetic phase of Ising model. 
}
\label{LY}
\end{figure}

\subsubsection{Quantum Entanglement Entropy and Entanglement Witness.}  It was shown theoretically that the
  counting statistics of spin fluctuations can be used to reconstruct
  the quantum entanglement entropy (QEE) of  spins with their
  environment \cite{klich2}. This work was done in analogy with the demonstration by Klich and Levitov \cite{klich1}
     that the generating function of counting statistics of charge currents is
  directly related to QEE between the leads. The knowledge of QEE can be used to
  determine topological characteristics of correlated electrons and to detect quantum
  phase transitions
  \cite{qee1,qee2,qee3,qee4,qee5,qee6,qee7,qee8,qee9,qee10,qee11}.

 QEE has not been experimentally measured  by SNS techniques yet but there is encouraging experimental progress towards this goal.
Thus, spin noise measurements in atomic vapors have been already used to unveil quantum many-body
 correlations \cite{mitchel-12,mitchel-10,mitchel-11,polzik-01nat}. Moreover, the recently
demonstrated  two-color SNS \cite{two-color-2,two-color-1} now
 allows experimental studies of cross-correlations between different
 spin subsystems, that are of interest in the context of QEE.
These experiments gave birth to hopes that  SNS might eventually be used to study QEE  \cite{qpt}.

While  experimental studies of QEE require considerable advances of the higher order SNS, the quantum entanglement can  already be studied by SNS techniques that probe 2nd order spin correlators.
For example, the spin susceptibility can be used to test the boundaries of the entanglement witness \cite{e-witness1,e-witness2,e-witness3} (if these boundaries are broken then the quantum state of a spin system cannot be represented as a product of different spin states).
It was already shown that the
off-resonant optical pulses could be useful both to create macroscopic entanglement and to detect it by exploring quantum spin fluctuations \cite{polzik-01nat}.

Here, we would like to point out that almost all previous studies of QEE and  susceptibility as an entanglement witness have been focused on static characteristics related to the equal time correlators of variables. The strength of  SNS measurement techniques is rather in resolving {\it temporal} correlations. Hence, the convergence of  SNS with the physics of quantum entanglement will strongly depend on the future theoretical work clarifying the entanglement information encoded in  time-correlators of spin variables. Steps in this direction have  recently been taken \cite{Hauke16a}.

\subsection{Strongly Correlated Spin Systems}

SNS can be particularly useful to study strongly correlated systems because the noise power
spectrum can reveal correlations in a wide range of energy
scales.
 It is likely that some new applications of SNS in this important research area will
emerge in the very near future. Here we briefly discuss possible
research directions.

{\it Magnetic Semiconductors.}  A dilute magnetic
 semiconductor can be created if a few percent of atoms in a paramagnetic semiconductor are
 replaced with magnetic ions. Such materials would be
 especially well-suited for future spintronic
 devices if the mechanisms behind their ferromagnetism were better
 understood. Spin noise studies of such materials have recently became possible \cite{mn-noise}.
 Consider, e.g., the magnetic semiconductor Ga$_{1-x}$Mn$_x$As. In the ferromagnetic phase, it  shows a strong Faraday effect due
to the steady spin polarization of holes \cite{frgamnas}. In the
paramagnetic phase, spins of holes have relatively fast dynamics but close to the phase transition, when the mean spin
polarization is zero on average, spin fluctuations are expected to
be considerable, possibly with a strongly non-Lorentzian shape of the
noise power spectrum due to the critical slow-down of the dynamics and
disorder. One can expect, for example, that there are regions with
accidentally large concentration of Mn ions with relatively strong
ferromagnetic interactions. Such regions may behave as ``quasiparticles
with large spins", whose sizes can be obtained by measuring high order
correlations \cite{sinitsyn-13njp}.

{\it Luttinger Liquid, Kondo Effect, etc.} Recently,
a renormalization group study of spin noise in a Luttinger liquid with
spin-orbit coupling was performed by Sun and  Pokrovsky
\cite{PhysRevB.91.161305}. They have demonstrated the sensitivity of
 spin correlators to phase transitions and optical resonances in the Luttinger liquid
\cite{pokrovsky-13}.
Numerical renormalization group methods were also used in \cite{kondo-noise} to
demonstrate that signatures of the Kondo effect can be observed in SNS
experiments. The two-beam SNS was also proposed to
detect a dynamic localization of quasiparticles in spin chains \cite{roy-15}.
As SNS sensitivity has reached the single spin limit \cite{single-spin},
one can realistically expect the spin noise detection in thin nanowires or from single magnetic impurities.
In fact,  SNS is quite desirable for such applications as it provides a contactless probe that significantly reduces
the sample preparation time and complexity. Moreover,
optical studies can be performed on many nanowires  simultaneously, without the need of their alignment.


{\it Magnetic Films and Micromagnetics.}
There are potential advantages in applying  SNS to ferromagnetic materials and spin
glasses compared to traditional methods, such as the SQUID-based spectroscopy.
First of all, we emphasize that spin noise probed through Kerr rotation fluctuations
can be studied at GHz range sufficient to reveal, for example, magnon interactions in yttrium iron
garnet \cite{magnons}. Moreover, the Kerr rotation can  also be sensitive to the in-plane
magnetization, which can enable other interesting applications, such
as the studies of thermal dynamics of monopoles in magnetic spin ices \cite{nisoli2007ground,nisoli2013colloquium}.

\subsection{Extensions of Spin Noise Spectroscopy}
An experimentally unexplored field is the combination of SNS with other noise
measurement techniques, such as  measurements of the noise of electric
currents. Studies of cross-correlations between spins and other
variables could potentially reveal some important characteristics of
transport or other phenomena.

For example, it was theoretically predicted \cite{sinitsyn-she} that in
paramagnetic semiconductors,
the spin Hall effect \cite{Dyakonov71a,hirsch1999spin,sinitsyn-04prl-2,sinitsyn-04prb-3} creates
cross-correlations between spin noise and transverse voltage fluctuations, which could be used as a probe of the
spin Hall effect. The spin Hall effect is caused by the spin-orbit interaction,
which possible origins in semiconductors include the structure inversion asymmetry \cite{Bychkov84a}, bulk inversion asymmetry \cite{Dresselhaus55a}, or impurities.
Importantly, the spin Hall effect is relatively weak in semiconductors.

\begin{figure}[t]
\centerline{\includegraphics[width=0.7\columnwidth]{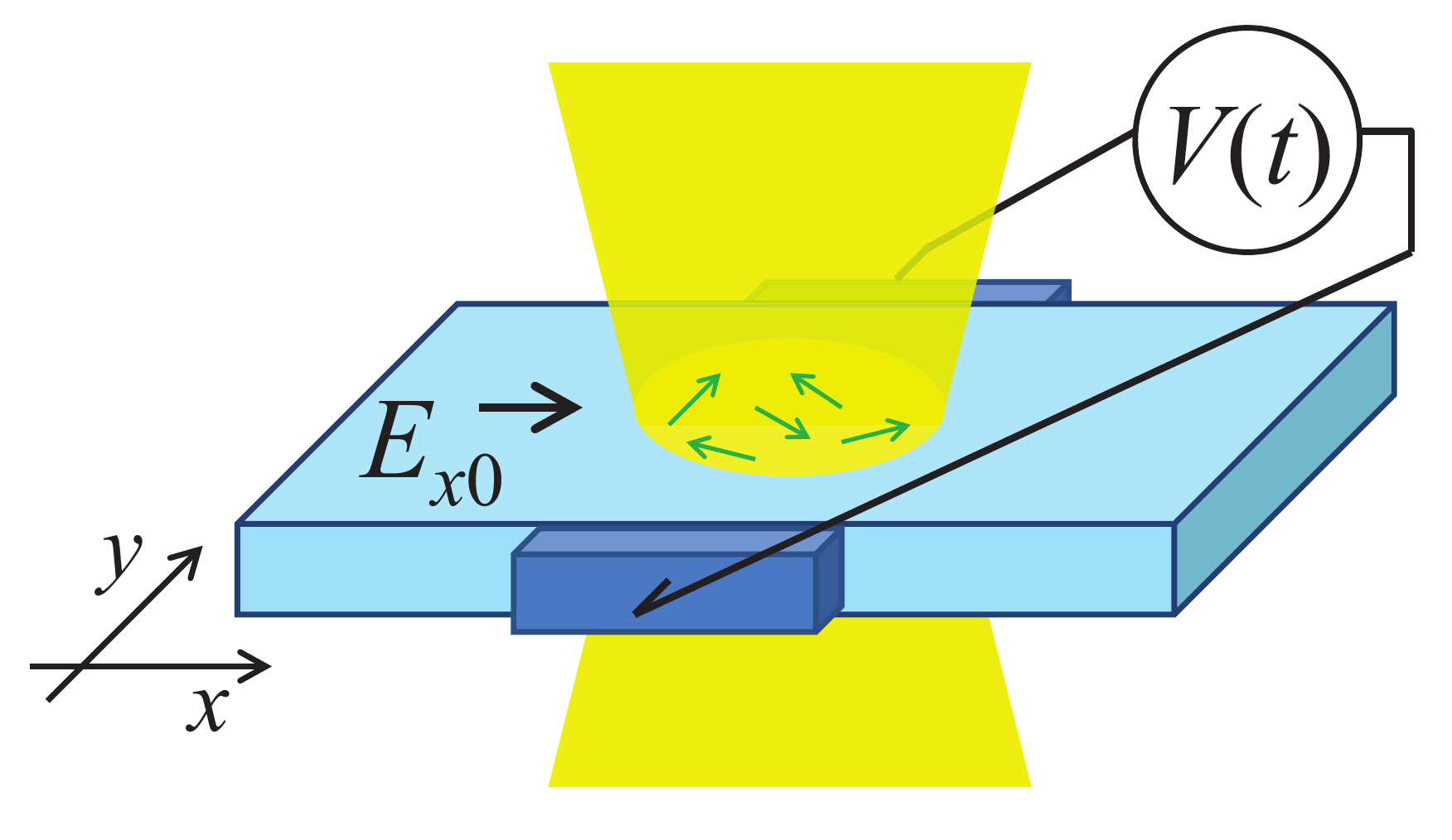}}
\caption{Hybrid spin noise spectroscopy. The spin Hall effect leads to cross-correlations between spin (measured optically)
and transverse voltage (measured electrically) fluctuations.
Reprinted figure with permission from V. A. Slipko et al.,
  Physical Review B 88, 201102(R) (2013) \cite{sinitsyn-she}. Copyright \copyright (2013) by the American Physical Society.}
\label{fig9_2_hybrid}
\end{figure}

The recently suggested method of the {\it hybrid SNS } \cite{sinitsyn-she} provides a pure noise probe of the spin Hall effect.
This approach focuses on correlations between spin (measured optically) and transverse voltage (measured electrically) fluctuations (see Fig.~\ref{fig9_2_hybrid})
and is based on the finding \cite{sinitsyn-she} that, in the presence of the spin Hall effect,  spin fluctuations are dressed by charge dipoles that lead to the transverse voltage fluctuations that are correlated with the local spin noise. The correlation strength is proportional to the spin Hall coefficient  \cite{sinitsyn-she}:
\begin{equation}
\left< S_z(0)V(t)\right> \sim \gamma E_{x0} e^{-\nu_st}. \label{eq_res1}
\end{equation}
Here, $V$ is the transverse voltage, $\gamma$ is the  spin Hall  coefficient describing deflection of spin-up (+) and spin-down (-) electrons, $E_{x0}$ is the longitudinal (applied) electric field and $\nu_s$ is the spin relaxation rate. Additionally, one can show that the transverse voltage-voltage correlation function is quadratic in the spin Hall coefficient, but, due to the smallness of the spin-Hall effect in semiconductors, is likely too small to be observable. We note, however, that the spin Hall effect can also lead to a dc transverse voltage in specially engineered structures \cite{pershin07c,Noh13a}.

\section{Conclusions}
\label{conclusion}


Spin noise spectroscopy has been implemented in a variety of
atomic, molecular, and condensed matter systems. Its advantages
over other competing approaches have been already demonstrated. SNS has
helped to resolve some of the pivotal problems in solid state physics,
such as resolving controversies about the spin relaxation mechanism of a semiconductor quantum dot qubit.

The progress of SNS will certainly depend on
the ability of this technique to explore ever more complex systems and
phenomena. If this progress continues, the  theory of SNS may also
evolve to an essentially novel methodology in materials science that
will operate with unusually high streams of information, quantum entanglement, full counting statistics,
multi-dimensional plots, and large data. The future will put these
expectations to the test.

\section*{Acknowledgments}
Authors thank Luyi Yang, Avadh Saxena, Scott Crooker and Darryl Smith for useful discussion.
The work at LANL was carried out under the auspices of the National Nuclear
Security Administration of the U.S. Department of Energy at Los
Alamos National Laboratory under Contract No. DE-AC52-06NA25396. N.A.S. also thanks the support from the LDRD program at LANL.
Y.V.P. acknowledges the support from the Smart State Center for Experimental Nanoscale Physics at USC.

\section*{References}
\bibliographystyle{unsrt}
\bibliography{queuing-meso}
\end{document}